\documentclass[rmp,aps,twocolumn,nofootinbib]{revtex4}

\usepackage{graphicx}
\usepackage{bm} %bold math

\def\HI{\hbox{H~$\scriptstyle\rm I$}}
\def\HII{\hbox{H~$\scriptstyle\rm II$}}
\def\HeI{\hbox{He~$\scriptstyle\rm I$}}
\def\HeII{\hbox{He~$\scriptstyle\rm II$}}
\def\HeIII{\hbox{He~$\scriptstyle\rm III$}}

\def\HIs{{\rm H\, \scriptstyle I}}
\def\HIIs{{\rm H\, \scriptstyle II}}

\def\HeIs{{\rm He\, \scriptstyle I}}
\def\HeIIs{{\rm He\, \scriptstyle II}}
\def\HeIIIs{{\rm He\, \scriptstyle III}}
\def\Lya{Ly$\alpha\ $}
\def\Lyb{Ly$\beta\ $}

\def\CIIs{\hbox{C~$\scriptstyle\rm II\ $}}

\def\CIVs{\hbox{C~$\scriptstyle\rm IV\ $}}

\def\AlIIs{\hbox{Al~$\scriptstyle\rm II\ $}}

\def\MgIIs{\hbox{Mg~$\scriptstyle\rm II\ $}}

\def\NIs{\hbox{N~$\scriptstyle\rm I\ $}}

\def\NVs{\hbox{N~$\scriptstyle\rm V\ $}}

\def\OVIs{\hbox{O~$\scriptstyle\rm VI\ $}}
\def\OVIIs{\hbox{O~$\scriptstyle\rm VII\ $}}
\def\OVIIIs{\hbox{O~$\scriptstyle\rm VIII\ $}}

\def\SiIIs{\hbox{Si~$\scriptstyle\rm II\ $}}
\def\SiIIIs{\hbox{Si~$\scriptstyle\rm III\ $}}
\def\SiIVs{\hbox{Si~$\scriptstyle\rm IV\ $}}
\def\CII{\hbox{C~$\scriptstyle\rm II$}}
\def\CIII{\hbox{C~$\scriptstyle\rm III$}}
\def\CIV{\hbox{C~$\scriptstyle\rm IV$}}

\def\FeII{\hbox{Fe~$\scriptstyle\rm II$}}
\def\FeIIs{\hbox{Fe~$\scriptstyle\rm II\ $}}

\def\MgII{\hbox{Mg~$\scriptstyle\rm II$}}

\def\NV{\hbox{N~$\scriptstyle\rm V$}}

\def\OI{\hbox{O~$\scriptstyle\rm I$}}
\def\OIII{\hbox{O~$\scriptstyle\rm III$}}

\def\OVI{\hbox{O~$\scriptstyle\rm VI$}}

\def\SiII{\hbox{Si~$\scriptstyle\rm II$}}
\def\SiIII{\hbox{Si~$\scriptstyle\rm III$}}
\def\SiIV{\hbox{Si~$\scriptstyle\rm IV$}}

\def\kms{\>\rm km\>s^{-1}}
\def\s{\mathop{\rm s\,}\nolimits}
\def\m{\mathop{\rm m\,}\nolimits}
\def\J{\mathop{\rm J\,}\nolimits}
\def\ry{\mathop{\rm Ry\,}\nolimits}
\def\gta{\, {}^>_\sim \,}
\def\lta{\, {}^<_\sim \,}
\def\nhat{{\bf\hat n}}
\newcommand{\expd}[1]{\times 10^{#1}}
\ifx\undefined\onethird
 \newcommand{\onethird}{{\frac{1}{3}}}
\else
 \renewcommand{\onethird}{{\frac{1}{3}}}
\fi
\ifx\undefined\onehalf
 \newcommand{\onehalf}{{\frac{1}{2}}}
\else
 \renewcommand{\onehalf}{{\frac{1}{2}}}
\fi

\begin{document}
\title{The Physics of the Intergalactic Medium\footnote{Submitted
to the {\it Reviews of Modern Physics}.}}

\author{Avery A. Meiksin}
\email{A.Meiksin@ed.ac.uk}
\affiliation{School of Physics, University of Edinburgh,
Edinburgh, EH9 3HJ, United Kingdom\\
SUPA\footnote{Scottish Universities Physics Alliance}}

\begin{abstract}
Intergalactic space is filled with a pervasive medium of ionized gas,
the Intergalactic Medium (IGM). A residual neutral fraction is
detected in the spectra of Quasi-Stellar Objects at both low and high
redshifts, revealing a highly fluctuating medium with temperatures
characteristic of photoionized gas. The statistics of the fluctuations
are well-reproduced by numerical gravity-hydrodynamics simulations
within the context of standard cosmological structure formation
scenarios. As such, the study of the IGM offers an opportunity to
probe the nature of the primordial density fluctuations on scales
unavailable to other methods. The simulations also suggest the IGM is
the dominant reservoir of baryons produced by the Big Bang, and so the
principal source of the matter from which galaxies formed. The
detection of metal systems within the IGM shows that it was enriched
by evolved stars early in its history, demonstrating an intimate
connection between galaxy formation and the IGM. The author presents a
comprehensive review of the current understanding of the structure and
physical properties of the IGM and its relation to galaxies,
concluding with comments on prospects for furthering the study of the
IGM using future ground-based facilities and space-based experiments.
\end{abstract}

\maketitle
\tableofcontents

\section{INTRODUCTION}
\label{sec:intro}

According to the Big Bang theory, the primordial hydrogen and helium
created in the Universe first materialized in the form of an extremely
hot ionized gas. By the time the Universe was three hundred thousand
years old, adiabatic expansion cooling brought the temperature of the
primordial plasma down until the hydrogen and helium recombined. The
radiation last scattered at this time appears today as the Cosmic
Microwave Background (CMB). The search for the IGM began well before
the discovery of the CMB with an attempt by
\textcite{1959ApJ...129..525F} to detect the hyperfine 21cm absorption
signature from hydrogen along the line of sight to the extragalactic
radio galaxy Cygnus A. Although no detection was made, combining the
optical depth with the measured temperature of the CMB discovered in
1965 would have been sufficient to exclude the possibility that the
Universe was closed by baryons, with an upper limit on the baryon
density of only 20\% of the closure density required for an
Einstein-deSitter (flat) universe. \footnote{This conclusion requires
making the (at the time) reasonable assumptions that the hydrogen was
all neutral and that the hyperfine structure levels of the hydrogen
were in thermal equilibrium with the CMB.}

Nearly coincident with the discovery of the CMB, however, a
considerably improved measurement of the density of intergalactic
neutral hydrogen was made. Soon after the discovery of the first
Quasi-Stellar Object (QSO) \cite{1963Natur.197.1037H,
1963Natur.197.1040S}, \textcite{1965ApJ...142.1633G} reported a small
decrement in a QSO spectrum shortward of its \Lya emission line.
Attributing the decrement to the \Lya resonance scattering of
radiation from the QSO by intergalactic neutral hydrogen, Gunn and
Peterson demonstrated that the cosmic mass density of neutral hydrogen
was exceedingly smaller than the Einstein-deSitter density. In fact,
it was far smaller than the spatially averaged hydrogen of all the
stars in the Universe. If the Big Bang theory was correct, it meant
either that galaxy formation was an extraordinarily efficient process,
sweeping up all but a tiny residue of the primordial hydrogen, or that
the gas was reionized.

These two themes, the detection of intergalactic gas through the 21cm
signature in the radio or through Lyman resonance scattering lines in
the optical or ultraviolet (UV), continue to dominate studies of the
Intergalactic Medium (IGM). To date, almost all that is known about
the structure of the IGM has been derived from optical and UV
data. This situation is expected to change dramatically in the near
future with the development of low frequency radio arrays like the LOw
Frequency ARray (LOFAR)\footnote{http://www.lofar.org}, the Murchison
Widefield Array
(MWA)\footnote{http://www.haystack.mit.edu/ast/arrays/mwa/}, the
Primeval Structure Telescope (PaST/21CMA)
\footnote{http://web.phys.cmu.edu/$\sim$past/}, the Precision Array to Probe
the Epoch of Reionization (PAPER)
\footnote{http://astro.berkeley.edu/$\sim$dbacker/eor/},
and a possible Square Kilometre Array (SKA).
\footnote{http://www.skatelescope.org} A primary science driver of
these instruments is the direct imaging of the IGM prior to the
completion of the Epoch of Reionization \cite{MMR97}. Most of this
review focuses on the current understanding of the state of the IGM as
determined from optical and UV measurements. The observations have
relied almost exclusively on the spectra of QSOs, although IGM
absorption features have also been detected in the spectra of
Gamma-Ray Bursts (GRBs) \cite{2006PASJ...58..485T}, and indeed played
a key role in establishing the extragalactic character of some bursts
\cite{1997Natur.387..878M}.

Almost immediately after Gunn and Peterson publicized their finding,
it was recognized that individual \Lya absorption features should
appear from neutral hydrogen concentrated into cosmological structures
\cite{1965ApJ...142.1677B, 1967ApJ...149..465W}. Absorption features
had in fact been detected in higher resolution QSO spectra, but these
were identified with intervening ionized metal absorption systems
\cite{1968ApJ...153..689B}, as was expected if galaxies had hot halos
of ionized gas:\ the lines of sight to background QSOs were expected
to pass through such hot halos and intercept any ionized gas clouds
within them \cite{1969ApJ...156L..63B}. The features, however, were
uncomfortably common, hinting at a class of unknown structures not
associated with galaxies. The \Lya resonance line features continued
to prove elusive until 1971, when \textcite{1971ApJ...164L..73L}
recognized several absorption features shortward of the \Lya emission
line of a QSO as \Lya lines\footnote{These are not true absorption
features involving the net destruction of a photon, but the scattering
out of the line of sight of resonance line photons.} arising in a
population of discrete absorption systems also showing metal
features. The \Lya lines form a plethora of distinct absorption
features in the spectra of high redshift QSOs; they are collectively
known today as the \Lya forest.

The properties of the \Lya forest came under increasing scrutiny, with
the first systematic survey conducted several years later by
\textcite{1980ApJS...42...41S}, convincingly demonstrating through the
homogeneity of the observed properties of the absorbers their
intergalactic origin, as opposed to clouds ejected by the QSOs
observed. Although limited by the resolution of the spectrograph, the
measured velocity widths of the \Lya features corresponded to gas
temperatures of a few to several $10^4$~K, characteristic
photoionization temperatures for gas of a primordial composition, for
which there is no significant cooling by metals. The number of
features per comoving length was shown to increase with redshift,
demonstrating that the systems were evolving
\cite{1982ApJ...252...10Y}.

The past decade has witnessed a dramatic improvement in precision
studies of the \Lya forest with the advent of 8--10~m class
telescopes, particularly the Keck 10-m and the 8.2m Very Large
Telescope (VLT). For the first time, the individual absorption
features in the \Lya forest were spectroscopically resolved over their
full range. Velocity widths of $\sim25\kms$ are typical. The neutral
hydrogen column densities of the absorbers range from roughly
$10^{12}-10^{22}\,{\rm cm^{-2}}$. The highest column density systems,
the Damped \Lya Absorbers (DLAs), are of particular interest for
galaxy formation, as they are suspected of containing the neutral gas
that formed the bulk of the stars in present day galaxies.

As the number per length of absorption systems increases along a line
of sight with increasing redshift, so does the mean flux decrement in
a background QSO spectrum due to \Lya scattering. The QSO spectra
measured as part of the Sloan Digital Sky Survey (SDSS)
\footnote{http://www.sdss.org} show a rapid rise in the flux decrement
at $z\gta5.5$, suggesting that the epoch of \HI\ reionization may lie
not far beyond $z\simeq6$. Many of the hydrogen features also show
absorption lines from metals, including carbon, silicon, nitrogen,
oxygen, magnesium, iron and others. The abundances of the metals,
however, are at most about 10\% of solar at low redshifts, and as low
as 1\% at high redshifts, indicating that the absorption systems are
comprised largely of primordial material. The primordial nature of the
gas received further important confirmation in 1994 with the discovery
by \textcite{1994Natur.370...35J} of intergalactic helium using the
{\it Hubble Space Telescope} ({\it HST}) through the detection of
\HeII\ \Lya absorption.

Because the baryons in the IGM are detected only though their
absorption signatures, the physical structures that give rise to the
features must be modeled. Early models characterized the systems as
discrete clouds of gas, with most of the focus on either clouds
pressure-confined by a hot medium \cite{1980ApJS...42...41S}, or
gravitationally confined in a dark matter halo
\cite{1986Ap&SS.118..509I,1986MNRAS.218P..25R}. At the time it was
believed that the absorption systems accounted for only a few percent
of the baryons produced in the Big Bang, much like galaxies, their
visible counterparts. A paradigm shift in the models occurred in the
mid 1990s. Measurements of coincident absorption features along
neighboring lines of sight suggested sizes of tens to hundreds of
kiloparsecs for the absorbers \cite{1992ApJ...389...39S,
1994ApJ...437L..83B, 1997ApJ...491...45D}, much larger than expected
for clouds confined by pressure or dark matter halos. A radical
transformation in the understanding of the nature and structure of the
IGM was initiated by numerical simulations of cosmological structure
formation. Today essentially all of the baryons produced in the Big
Bang are believed to have undergone the same gravitational instability
process initiated by primordial density fluctuations that was
responsible for the formation of galaxies. The computation of the
structure of the IGM has been converted into an initial value problem
similar to that of the CMB fluctuations. Fluctuations in the CMB are
solved for by following the gravitational instability growth of a
postulated spectrum of primordial dark matter density fluctuations.
The growth of structure in the IGM is now treated as the nonlinear
extension of these computations. The result is a network of
filamentary structures, the so-called ``cosmic web''
\cite{1996Natur.380..603B}. The \Lya forest is believed to be a
signature of the cosmic web. Early simulations broadly reproduced the
statistics of the \Lya forest spectacularly well \cite{Cen94,
1995A&A...295L...9P, ZAN95, 1996ApJ...457L..51H, 1997ApJ...485..496Z,
Theuns98}. An immediate conclusion was that at $z\gta1.5$, some 90\%
of the baryons produced in the Big Bang are contained within the IGM,
with only 10\% in galaxies, galaxy clusters or possibly locked up in
an early generation of compact stars.

Soon after the discovery of intervening absorption features, it was
recognized that they provided potentially powerful tests of
fundamental properties of the Universe. The split in the fine
structure lines of the metals was used to set constraints on the
variability of the fine structure constant
\cite{1967ApJ...149L..11B}. A bunching of absorption features near
$z\simeq1.9$ (now known to be fortuitous), gave rise to the
(re)introduction of a cosmological constant to account for the
numerous features as multiple images due to lensing
\cite{1967ApJ...147.1222P}. The expected primordial composition of the
IGM offered the potential of placing constraints on the
photon-to-baryon ratio of the Universe through measurements of the
intergalactic deuterium abundance. More recently, the success of the
models has inspired attempts to exploit the \Lya forest as a new means
to constrain cosmological structure formation models and obtain
stringent constraints on the cosmological parameters.

The description of the IGM by the simulations, however, is far from
complete. There remain many unsolved problems. The current simulations
do not reproduce all the observed properties of the IGM. The
absorption lines are predicted to be substantially narrower than
measured. This likely stems from the principal outstanding missing
piece of physics, the reionization of the IGM. Not only must hydrogen
be ionized, but helium as well. The ionization heats the gas through
the photoelectric effect. Detailed radiative transfer computations are
required to recover the temperatures, for which there is still limited
success. The sources of the reionization and the epochs of
reionization, both of hydrogen and helium, are still not firmly
established. The origin of the metal absorption systems in the diffuse
IGM is still unknown, although it is widely expected they were
deposited by winds from galaxies, possibly driven by intense episodes
of star formation. As such, the metal absorption lines in principle
offer an important means of studying the history of cosmic star
formation. Most fundamentally, the relation of the IGM to the galaxies
that form from it is still mostly unknown, but offers perhaps the most
exciting prospects for new research directions.

The purpose of this review is to describe the progress made in the
understanding of the origin of structure within the IGM, with a view
to presenting the underlying physics that determines the structure. An
understanding of the physics is necessary for future progress. The
past decade has revealed the IGM to be a complex dynamical arena
involving interactions between the IGM gas, galaxies and QSOs. It is
becoming increasingly apparent that the separation of these systems
into distinct and isolated entities is an artificial
construct. Galaxies and QSOs originated from the IGM, and their
radiation and outflows impacted upon it. Any complete understanding of
the origin of these systems requires a unified treatment. In this way,
the IGM resembles the interstellar medium of disk galaxies in which
the gaseous component is intimately linked to the stars and their
evolution and impact. Interpreting the increasingly refined
observations requires detailed modeling, which relies on large-scale
numerical computations involving gas, radiative processes, and
gravity. The physics involved is intermediate in complexity between
that of the CMB and galaxy formation, rendering the IGM a bridge
between these extreme scales of cosmological structure
formation. Unraveling the processes that led to the formation and
structure of the IGM may thus serve as a crucial step in the solution
of the much more involved problem of galaxy formation.

\begin{figure}
\includegraphics[width=8cm, height=6.5cm]{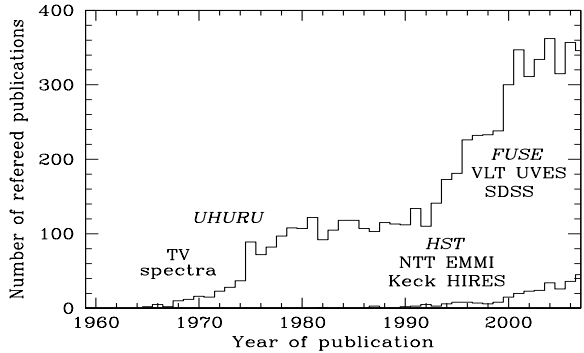}
\caption{Number of refereed papers on the intergalactic medium and QSO
intervening absorption systems published from 1965--2007, as provided
by the ADS abstract service (adsabs.harvard.edu). The sharp rises
follow key developments in astronomical instrumentation and
observations. The lower histogram shows the rise of papers on the
reionization and subsequent ionization state of the IGM.
}
\label{fig:citations}
\end{figure}

A search of the literature for papers on the IGM since the Gunn \&
Peterson (1965) measurement produces close to 6000
references.\footnote{Based on a boolean search of the Astrophysics
Data System abstract service (adsabs.harvard.edu) for all refereed
papers with abstracts or keywords containing ``(intergalactic and
medium) or (quasar and absorption and line).''  Reionization papers
are selected as the subset discussing the reionization of the IGM or
the subsequent ionization structure of the IGM.} While this review
does not have the space to describe the observational methods used to
measure the IGM, it should be recognized that progress in the
understanding of the properties of the IGM is indebted to advances in
observational techniques. This is well illustrated by a plot of the
number of refereed publications in the field against time. Periods of
relatively steady output are punctuated by four leaps. The first
occurs at the end of the 1960s and beginning of the 1970s with the
introduction of image tube spectrographs coupled with integrating
television systems for photon counting \cite{1972ailt.conf..295B,
1972ailt.conf..317M}, which greatly facilitated the taking of
spectra. The next occurs in the mid-1970s with the development of
x-ray astronomy following the launch of the {\it UHURU} satellite in
1970 and the recognition that galaxy clusters contain an extended and
pervasive medium of hot, radiating gas \cite{1973ApJ...185L..13K,
1973ApJ...184L.105L}. Another sharp rise occurs in the early 1990s
following the launch of the {\it Hubble Space Telescope} in 1990, the
installation of EMMI and its echelle spectroscope on the New
Technology Telescope (NTT) \cite{1990Msngr..61...51D}, and the
delivery of the HIRES spectrometer \cite{1994SPIE.2198..362V} to the
Keck telescope. The fourth occurred in 2000 with the introduction of
the UV and Visible Echelle Spectrograph (UVES) to the Very Large
Telescope (VLT) \cite{2000SPIE.4005..121D}, the launch of the {\it Far
Ultraviolet Spectroscopic Explorer} ({\it FUSE})
\cite{2000ApJ...538L...1M}, and the beginning of operations of the
Sloan Digital Sky Survey \cite{2000AJ....120.1579Y}. The latter in
particular triggered a surge of activity in reionization studies
following the discovery of high redshift QSOs ($z>6$) with spectra
indicating a rapid rise in the effective optical depth of the IGM to
\Lya photons, hinting that the Epoch of Reionization was being
approached \cite{2001AJ....122.2850B, 2002AJ....123.1247F}. This is
indicated by a rise in IGM reionization papers in
Fig.~\ref{fig:citations}, a trend which continues today, fueled in
part by the growing interest in the influence of reionization on the
CMB fluctuations measured by the {\it Wilkinson Microwave Anisotropy
Probe} ({\it WMAP}) \cite{Kogut03}. The next major leap may well come
from the anticipated radio measurements.

The rapidly rising tide of IGM studies has brought along with it
several reviews. This is fortunate, as it is impossible in a single
review to cover all areas thoroughly today. Early reviews of the QSO
absorption line literature, now largely historical, were provided by
\textcite{1976ARA&A..14..307S} and
\textcite{1981ARA&A..19...41W}. \textcite{1998ARA&A..36..267R} and
\textcite{Bechtold03} provide reviews of the \Lya forest that are
still largely up-to-date in the sense that most of the topics
currently engaging the community are treated, with developments since
mostly improvements in accuracy and in the details of the numerical
models. Reviews of the low redshift IGM are provided by
\textcite{2003ASSL..281....1S} and
\textcite{2006pces.conf..111S}. \textcite{2005ARA&A..43..861W} have
reviewed the current understanding of Damped \Lya Absorbers, an
absorber class of special concern as it represents the closest link to
the gaseous component forming present day galaxies. A new series of
reviews followed the recent explosion of activity on the reionization
of the IGM in anticipation of the detection of the Epoch of
Reionization (EoR) through its high redshift 21cm signature. An early
review in this area is by \textcite{2001ARA&A..39...19L}. Since then,
current observational and theoretical aspects of reionization have
been exhaustively covered by \textcite{2006ARA&A..44..415F},
\textcite{2006PhR...433..181F}, and \textcite{2007RPPh...70..627B}.

Rather than repeating the wide range of IGM phenomenology already
covered by previous reviewers, I concentrate here on the physics
underlying the structure of the IGM. One aim is to describe the
physical underpinnings of current numerical simulations as required
for future simulations to further progress. As observations are
crucial for constraining any models, I begin by giving a broad
overview of the current observational situation. The next section
describes the physics of ionization equilibrium, followed by a
discussion of the UV metagalactic background that maintains the
ionization of the IGM. A brief review of early models of the \Lya
forest absorbers is then presented, followed by a discussion of
current numerical models. The reionization of the IGM is then
summarized along with means for its detection. This is followed by a
discussion of the connection between galaxies and the IGM before
concluding with observational and theoretical prospects for the
future.

Unless stated otherwise, the cosmological parameters $\Omega_m=0.3$
and $\Omega_v=0.7$ are assumed for, respectively, the ratios of the
present day matter density and the present day vacuum energy density
to the current Einstein-deSitter density $\rho_{\rm
crit}(0)=3H_0^2/8\pi G$, a Hubble Constant of $H_0=100h\,{\rm
km\,s^{-1}\,Mpc^{-1}}$ with $h=0.7$ (${\rm Mpc} = {\rm
Megaparsec}\simeq 3.0856\times10^{22}\,{\rm m}$), and a baryon density
of $\Omega_bh^2=0.022$. These values are consistent within the errors
with the current best estimates for a flat Universe based on CMB
measurements using Year-5 {\it WMAP} data of $\Omega_m=0.279\pm0.014$,
$\Omega_v=0.721\pm0.015$, $h=0.701\pm0.013$, and
$\Omega_bh^2=0.02265\pm0.00059$ \cite{2008arXiv0803.0547K}, or
intergalactic $D/H$ measurements, giving $\Omega_bh^2=0.021\pm0.001$
\cite{2006ApJ...649L..61O}.

\section{Observations}
\label{sec:obs}

\subsection{Resonance absorption lines}
\label{subsec: absorption-lines}

The IGM is detected through the absorption features it produces in the
spectrum of a background bright source of light (typically a QSO).
The production of the absorption features is governed by the equation
of radiative transfer through the IGM, conventionally expressed in
terms of the specific intensity of a background radiation source.

The specific intensity $I_\nu({\bf r},t,\nhat)$ is defined as the rate
at which energy crosses a unit area per unit solid angle per unit time
as carried by photons of energy $h_{\rm P}\nu$ traveling in the
direction $\nhat$ relative to some fiducial direction. The equation of
radiative transfer for $I_\nu({\bf r},t,\nhat)$ in a medium with
attenuation coefficient $\alpha_\nu({\bf r},t,\nhat)$ is
\begin{eqnarray}
\frac{1}{c}\frac{\partial I_\nu({\bf r},t,\nhat)}{\partial t}+\nhat\cdot
{\bf \nabla}I_\nu({\bf r},t,\nhat) = \nonumber\\
-\alpha_\nu({\bf r},t,\nhat)
I_\nu({\bf r},t,\nhat) + j_\nu({\bf r},t,\nhat).
\label{eq:tdRT}
\end{eqnarray}
Because of its use below, a radiative source term has been included
through the {\it emission coefficient} $j_\nu({\bf r},t,\nhat)$, which
describes the local specific luminosity per solid angle per unit
volume emitted by sources. Normally the random orientation of atoms
will ensure $j_\nu$ is isotropic, but not always, as for instance if
the source includes a scattering term, so that a dependence on $\nhat$
is included to account for the possibility of anisotropic sources. A
central source like a QSO or galaxy will in fact generally radiate
anisotropically.

In general, the attenuation of the radiation field is due both to the
absorption of photons and their scattering out of the beam. The
attenuation coefficient is then
\begin{equation}
\alpha_\nu({\bf r},t,\nhat) = \rho({\bf r},t)\kappa_\nu({\bf r},t,\nhat)
+ n({\bf r},t)\sigma_\nu({\bf r},t,\nhat),
\label{eq:atten}
\end{equation}
where $\rho({\bf r},t)$ is the mass density of the medium,
$\kappa_\nu({\bf r},t,\nhat)$ is the opacity, which expresses the
absorption cross section per unit mass, $n({\bf r},t)=\rho({\bf
r},t)/\bar m$ is the number density of scattering particles of mean
mass $\bar m$, and $\sigma_\nu({\bf r},t,\nhat)$ is the scattering
cross section. In general, the scattering could arise from more than
one type of particle, in which case $n\sigma_\nu$ would be replaced by
a sum over particle species $i$ of density $n_i$ and cross section
$\sigma_\nu(i)$. In a static medium, $\alpha_\nu({\bf r},t,\nhat)$ is
generally isotropic, but not always. As an example, the alignment of
atoms in a strong magnetic field would absorb anisotropically. In a
moving medium, an anisotropic contribution to the attenuation will be
produced by the dependence of Doppler scattering on the relative
motion of the radiation and the fluid.

The value of $I_\nu$ at any given time $t$ and position $s$ along the
direction $\nhat$ will be given by any incoming intensity $I_\nu^{\rm
inc}$ at position $s_0$ at time $t_{\rm ret}=t-(s-s_0)/c$, attenuated
by intervening material at positions $s^\prime$ at the retarded times
$t^\prime_{\rm ret}=t-(s-s^\prime)/c$, along with contributions from
sources at positions $s^{\prime\prime}$ that emitted at the retarded
times $t^{\prime\prime}_{\rm ret}=t-(s-s^{\prime\prime})/c$, followed
by attenuation. The formal solution to Eq.~(\ref{eq:tdRT}) is then
\begin{eqnarray}
I_\nu(s,t) = \left(I_\nu^{\rm inc}\right)_{s_0,t_{\rm ret}}
\exp\left[-\int_{s_0}^s\,ds^\prime\,
(\alpha_\nu)_{s^\prime,t_{\rm ret}^\prime}
\right]\nonumber\\
+\int_{s_0}^s\,ds^{\prime\prime}\,
\left(j_\nu\right)_{s^{\prime\prime},t_{\rm ret}^{\prime\prime}}
\exp\left[-\int_{s^{\prime\prime}}^s\,ds^\prime\,
(\alpha_\nu)_{s^\prime,t_{\rm ret}^\prime}
\right]
\label{eq:tdRT-sol}
\end{eqnarray}
as may be verified by direct substitution. In a uniformly expanding
(or contracting) medium described by the scale factor $a(t)$,
Eq.~(\ref{eq:tdRT}) must be modified to
\begin{eqnarray}
\frac{1}{c}\frac{\partial I_\nu({\bf r},t,\nhat)}{\partial t}+
\frac{1}{c}\frac{\dot a}{a}\left[3I_\nu({\bf
r},t,\nhat)-\nu\frac{\partial I_\nu({\bf
r},t,\nhat)}{\partial\nu}\right]\nonumber\\ + \nhat\cdot {\bf
\nabla}I_\nu({\bf r},t,\nhat) = -\alpha_\nu({\bf r},t,\nhat)
I_\nu({\bf r},t,\nhat) + j_\nu({\bf r},t,\nhat),
\label{eq:tdRT-exp}
\end{eqnarray}
where the frequency redshifts according to $\nu\propto a(t)^{-1}$. The
solution becomes
\begin{eqnarray}
I_\nu(s,t) = \left(I_{\nu_0}^{\rm inc}\right)_{s_0,t_0}
\left[\frac{a(t_0)}{a(t)}\right]^3\exp\left[-\int_{s_0}^s
\,ds^\prime\,
(\alpha_{\nu^\prime})_{s^\prime,t^\prime}
\right]\nonumber\\
+\int_{s_0}^s\,ds^{\prime\prime}\,
\left(j_{\nu^{\prime\prime}}\right)_{s^{\prime\prime},t^{\prime\prime}}
\left[\frac{a(t^{\prime\prime})}{a(t)}\right]^3
\exp\left[-\int_{s^{\prime\prime}}^s\,ds^\prime\,
(\alpha_{\nu^\prime})_{s^\prime,t^\prime}
\right],
\label{eq:tdRT-exp-sol}
\end{eqnarray}
where $\nu_0=\nu a(t)/a(t_0)$, $\nu^\prime=\nu a(t)/a(t^\prime)$, and
$\nu^{\prime\prime}=\nu a(t)/a(t^{\prime\prime})$.

In the case of no intermediate sources, $j_\nu=0$ and the received intensity
depends only on how the incident intensity is modified by the intervening
gas. This is the situation for a single background source such as a QSO.
The Lyman absorption features arise from the scattering of resonance
line photons received from the background QSO through a medium of
scatterers with number density $n({\bf r},t)$. The frequency dependent
scattering cross section for a scatterer at rest is given by the
Lorentz profile\footnote{SI units for electrostatic quantities throughout
the text may be accommodated by including the square-bracketed
quantity $[1/4\pi\epsilon_0]$, where $\epsilon_0$ is the vacuum
permittivity. Without the factor, the expressions correspond to
the forms appropriate for cgs units.}
\begin{equation}
\sigma_\nu=\left(\frac{\pi e^2}{m_e c}\right)\left[\frac{1}{4\pi\epsilon_0}
\right]f_{lu}
\frac{(\Gamma_{ul}/4\pi^2)}{(\nu-\nu_{lu})^2+(\Gamma_{ul}/4\pi)^2},
\label{eq:Lorentz}
\end{equation}
where $\nu_{lu}$ is the resonance line frequency, corresponding to the
wavelength $\lambda_{lu}$, of the transition between an upper level
broadened by radiation damping to a sharp lower level (the ground
state), $\Gamma_{ul}=(g_lf_{lu}/g_u)[1/4\pi\epsilon_0] 8\pi^2e^2/(m_e
c\lambda_{lu}^2)$ is the damping width of the upper level, where $g_l$
and $g_u$ are the respective statistical weights of the lower and
upper levels, $f_{lu}$ is the upward oscillator strength, $e$ is the
electric charge of an electron, and $m_e$ the electron
mass.\footnote{The Lorentz profile neglects the frequency dependence
of the scattering cross-section far from line center. It does not, for
example, recover the low frequency Rayleigh limit ($\sigma_\nu\propto
\nu^4$). The profile, however, is fully adequate for all practical
applications to the IGM. Small deviations may become detectable for
broad absorption features such as Damped \Lya Absorbers at high
spectral resolution \citep{2003ApJ...594..637L}. Such deviations,
however, would prove difficult to distinguish from the effects of
density inhomogeneities in the absorbing gas and the associated gas
velocities, except possibly in a statistical sense over many systems.}
The cross section is normalized according to
\begin{equation}
\sigma\equiv\int_{-\infty}^\infty\, d\nu\, \sigma_\nu =
\left(\frac{\pi e^2}{m_e c}\right)\left[\frac{1}{4\pi\epsilon_0}\right]f_{lu}.
\label{eq:sigmanu-norm}
\end{equation}
For $\lambda_{lu}$ in m,
$\Gamma_{ul}=6.670\times10^{-5}(g_lf_{lu}/g_u)\lambda_{lu}^{-2}$. For
the Lyman series,
\begin{equation}
g_1f_{1n}=\frac{2^9n^5(n-1)^{2n-4}}{3(n+1)^{2n+4}},
\label{eq:gfLy}
\end{equation}
where $n$ is the quantum number of the upper state
\cite{1935MNRAS..96...77M}. For the \Lya transition, $g_l=2$, $g_u=6$
and $f_{lu}=0.4162\cdots$. For hydrogen Ly$\alpha$,
$\lambda_{lu}=1215.67$\AA\ and $\Gamma_{ul}=6.262\times10^8\,{\rm
s}^{-1}$. For \HeII\ Ly$\alpha$, $\lambda_{lu}$ is smaller by a factor
of 4, and $\Gamma_{ul}$ larger by a factor of 16.

In general, the atoms will not be at rest. At the very least they will
generally undergo thermal motions described by a Maxwellian velocity
distribution corresponding to their temperature $T$. They may
additionally take part in an ordered flow of velocity ${\bf v}$, but
this may be accounted for by shifting the line center frequency $\nu$
to $\nu(1-v_{||}/c)$, to first order accuracy in $v/c$, where $v_{||}$
is the line-of-sight velocity. In addition, there may be a disordered
turbulent, or so-called {\it microturbulent}, component in some
collapsed or shocked regions. Ignoring non-thermal velocities, the
profile including thermal motions is found by convolving
Eq.~(\ref{eq:Lorentz}) with a Maxwellian. The result, which neglects
additional frequency dependences far from line center, is the Voigt
function
\begin{equation}
H(a,x)=\frac{a}{\pi}\int_{-\infty}^\infty\,dy
\frac{e^{-y^2}}{(x-y)^2+a^2},
\label{eq:phiV}
\end{equation}
where $a=\Gamma_{ul}/(4\pi\Delta\nu_D)$ is the ratio of the damping
width to the Doppler width, and $x=(\nu-\nu_{lu})/\Delta\nu_D$ is the
frequency shift from line center in units of the Doppler width
$\Delta\nu_D=\nu_{lu} b/c$ with Doppler parameter
\begin{equation}
b=\left(\frac{2k_{\rm B}T}{m_a}\right)^{1/2},
\label{eq:bparam}
\end{equation}
where $m_a$ is the mass of the scattering particle and $k_{\rm B}$ is
Boltzmann's constant. A kinematic component such as microturbulence is
often accounted for by adding in quadrature the characteristic
kinematic velocity to the thermal component of the Doppler parameter
\begin{equation}
b= (b_{\rm th}^2 + b_{\rm kin}^2)^{1/2},
\label{eq:bbparam}
\end{equation}
where $b_{\rm th}$ is the thermal contribution from
Eq.~(\ref{eq:bparam}). An expansion approximation for the Voigt
function in $a$ is provided by \textcite{1948ApJ...108..112H}. Fast
methods for arbitrary $a$ are described by
\textcite{Humlicek:JQSRT21:306}, \textcite{Humlicek:JQSRT27:437} and
\textcite{1993JQSRT..50..635S}. Various evaluation methods are
compared by \textcite{Schreier:JQSRT48:743}.

The resonance line scattering cross section is
\begin{equation}
\sigma_\nu = \left(\frac{\pi e^2}{m_e c}\right)
\left[\frac{1}{4\pi\epsilon_0}\right]f_{lu}\varphi_V(a,\nu)=
\sigma\varphi_V(a,\nu),
\label{eq:sigmaV}
\end{equation}
where
\begin{equation}
\varphi_V(a,\nu)=\frac{1}{\Delta\nu_D}\phi_V(a,x)
=\frac{1}{\pi^{1/2}\Delta\nu_D}H(a,x)
\label{eq:Vprof}
\end{equation}
is the Voigt profile, normalized to
$\int_0^\infty\,d\nu\,\varphi_V(a,\nu)=1$. For hydrogen \Lya,
$a\simeq0.0472T^{-1/2}$. For \HeII\ Ly$\alpha$, $a$ is larger by a
factor of 8.

The intensity of the background QSO is attenuated by the factor
$e^{-\tau_\nu}$, where Eq.~(\ref{eq:tdRT-exp-sol}) gives
\begin{equation}
\tau_\nu = \int_{s_0}^s ds^\prime\, n(s^\prime,t^\prime)\sigma_{\nu^\prime}.
\label{eq:taunu}
\end{equation}
For a homogeneous and isotropic background Universe expanding with
scale factor $a(t)$, radiation emitted by the source at time $t_0$ and
rest frame frequency $\nu_0>\nu_{lu}$, where $\nu_{lu}$ is the
resonance line frequency of a (rest frame) Lyman transition, will be
scattered by the medium at epoch $t^\prime$ given by $\nu^\prime=\nu_0
a(t_0)/a(t^\prime)=\nu_{lu}$. Thus all of the received QSO intensity
will be attenuated at the time $t$ at all observed frequencies $\nu$
in the range $\nu_{lu}>\nu>\nu_{lu}a(t_0)/a(t)$, or wavelengths
$\lambda$ in the range $\lambda_{lu}<\lambda<\lambda_{lu}a(t)/a(t_0)$,
where $\lambda_{lu}$ is the (rest frame) wavelength of the particular
Lyman transition, producing a trough in the spectrum of the QSO. This
is the Gunn-Peterson effect \cite{1965ApJ...142.1633G,
1965Natur.207..963S}. Intervening metal absorption, assuming solar
abundances distributed uniformly throughout intergalactic space, would
produce similar troughs \cite{1964AZh....41..801S,
1965SvA.....8..638S}. The corresponding optical depth is given by
noting that $ds^\prime=cdt^\prime=cH^{-1}(t^\prime)
da(t^\prime)/a(t^\prime)$, where $H(t^\prime)= \dot
a(t^\prime)/a(t^\prime)$ is the Hubble parameter at time
$t^\prime$. Changing the integration variable to
$a^\prime=a(t^\prime)$ and setting $z=\nu_{lu}/\nu - 1$, it follows
from Eq.~(\ref{eq:taunu}) that, taking $\langle n\rangle$ to denote
the mean number density of particles in the lower state,
\begin{eqnarray}
\tau_\nu&=&c\int\,\frac{da^\prime}{a^\prime}
H(a^\prime)^{-1}\sigma_{\nu/a^\prime-\nu_{lu}}
\langle n(a^\prime)\rangle\nonumber\\
&=&\frac{g_u}{g_l}\frac{1}{8\pi}\frac{\lambda_{lu}^3\Gamma_{ul}
\langle n(z)\rangle}{H(z)}\nonumber\\
&=&1.191\times10^4\frac{(1+z)^3}{[\Omega_m(1+z)^3+\Omega_K(1+z)^2
+\Omega_v]^{1/2}}\nonumber\\
&&\times\left(\frac{f_{lu}\lambda}{506.0\AA}\right)
\left(\frac{g_u/g_l}{3}\right)
\left(\frac{\langle n\rangle}{\langle n_{\rm H}\rangle}\right),
\label{eq:tauGP}
\end{eqnarray}
after noting from Eq.~(\ref{eq:sigmanu-norm}) that
$\sigma=(g_u/g_l)\Gamma_{ul}\lambda_{lu}^2/8\pi$.  The Hubble
parameter $H(z)$ at redshift $z$ is related to the value of the Hubble
constant today $H_0$ by $H(z) = H_0E(z)$, where $E(z) =
[\Omega_m(1+z)^3 + \Omega_K(1+z)^2 + \Omega_v]^{1/2}$
\cite{1980ApJS...42...41S, Peebles93}. Here, $\Omega_m$ is the mass
density of the Universe expressed as a fraction of the Einstein-de
Sitter critical density, and $\Omega_K$ and $\Omega_v$ are similar
fractions arising from the curvature and vacuum energy contributions,
respectively. They satisfy the identity $\Omega_m+\Omega_K+\Omega_v =
1$. In the last line of Eq.~(\ref{eq:tauGP}), the result was scaled to
the present day mean cosmic hydrogen density $n_{\rm H}(0)=\rho_{\rm
H}(0)/m_{\rm H}=(1-Y)\Omega_b \rho_{\rm crit}(0)/m_{\rm
H}\simeq0.186\,{\rm m^{-3}}$, where $m_{\rm H}$ is the mass of a
hydrogen atom and assuming a baryon density of $\Omega_bh^2=0.022$
(see \S~\ref{subsubsec:deuterium} below) and cosmic helium mass
fraction $Y\simeq 0.2477\pm0.0029$ \cite{2007ApJ...666..636P}, and to
a Hubble Constant of $H_0=70\,{\rm km\,s^{-1}\,Mpc^{-1}}$
(\S~\ref{subsubsec:evolution} below). Eq.~(\ref{eq:tauGP}) is the
inverse of the Sobolev parameter for a homogeneously and isotropically
expanding medium, and was first derived in the context of cosmological
hydrogen \Lya photon scattering by \textcite{1959ApJ...129..536F}.

\begin{figure*}
\includegraphics[width=16cm]{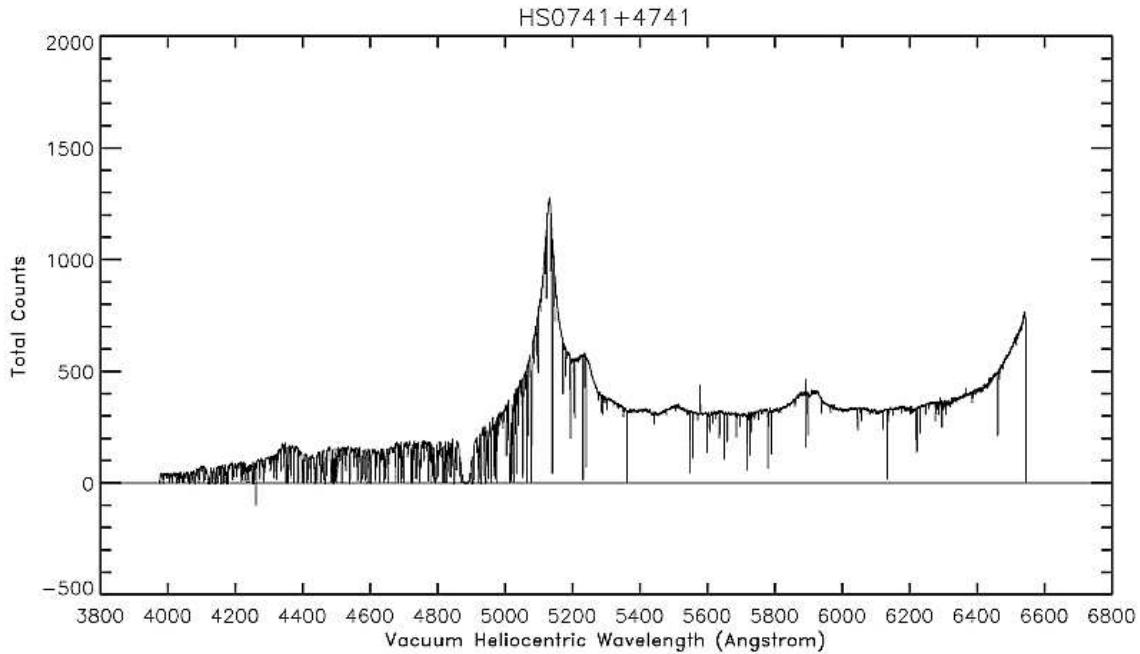}
\caption{The \Lya forest measured in the spectrum of the $z=3.22$
quasar HS$0741+4741$. The spectrum was obtained using the HIRES
spectrograph on the Keck 10~m telescope. The integration time was 4.0
hours with a spectral resolution of $R=37,000$ covering 3900\AA\ to
6600\AA\ with a signal-to-noise ratio of 145 per resolution
element. The sharp peak at $\lambda=5130$\AA\ is the \Lya emission
line of the quasar. The \Lya forest lies towards the blue side of the
feature. A Damped \Lya Absorber at $z=3.02$ appears at $\lambda\simeq4890$\AA\
\cite{2001ApJS..137...21P}. Intervening \CIVs systems give rise to
rarer absorption features longward of the emission line
\cite{2006AJ....131...24S}. (Figure courtesy of A. Songaila.)
}
\label{fig:LyaF}
\end{figure*}

Observations show that in fact the IGM is not homogeneous, but that
the baryons have coalesced, creating a fluctuating density field
$n(s,t)$. The discreteness of the absorption lines measured in the
spectra of high redshift QSOs suggests they originate in distinct
localized regions. The resulting collection of absorption features,
shown in Fig.~\ref{fig:LyaF}, is known as the \Lya forest. Denoting
the centers of the absorbing regions by the positions $s_i$, the
optical depth becomes $\tau_\nu = \sum_i\tau_\nu(i)$, where
\begin{equation}
\tau_\nu(i) = \int_{s_i-\Delta s/2}^{s_i+\Delta s/2} ds^\prime\,
n(s^\prime,t_i)\sigma_{\nu^\prime}.
\label{eq:taunu-i}
\end{equation}
Here, the integral has been localized to a region of width $\Delta s$
around $s_i$, and $\nu^\prime= \nu a(t)/a(t_i)$, where $t_i$
corresponds to the epoch of coordinate $s_i$ in an expanding
universe. The absorption features thus probe discrete spatial
structures in the IGM. Expressing the optical depth as a sum of
discrete absorption systems is, however, an approximation, as the
intergalactic gas forms an evolving spatial continuum. Gas distributed
over a wide spatial range, and even at different epochs, can therefore
affect the same observed frequency $\nu$. However, for $z\lta3.5$,
observations show the discrete absorber approximation provides a
reasonably good description of the hydrogen \Lya
forest. Eq.~(\ref{eq:taunu-i}) may be further simplified to
\begin{equation}
\tau_\nu = \pi^{1/2}\tau_0 \langle \phi_V(a,x)\rangle,
\label{eq:taunu-line}
\end{equation}
where
\begin{equation}
\tau_0 = \frac{N\sigma\lambda}{\pi^{1/2}b} = \frac{\sqrt{\pi}e^2} {m_e
c}\left[\frac{1}{4\pi\epsilon_0}\right]\frac{N}{b}\lambda f_{lu}
\label{eq:tau0}
\end{equation}
is the optical depth at line centre. Here, $N$ is the column density
$\int \,ds^\prime\, n(s^\prime,t^\prime)$, and $\langle
\phi_V(a,x)\rangle$ is averaged over the line of sight, weighted by
the density. This may differ from $\phi_V(a,x)$ when the temperature
or large-scale macroscopic velocity field varies along the line of
sight. Tabulations of $\lambda f_{lu}$ and related atomic data for
resonance lines for a variety of elements are available in the
literature \cite{1991ApJS...77..119M, 2000ApJS..130..403M,
2003ApJS..149..205M}. For hydrogen \Lya,
\begin{equation}
\tau^{\rm HI}_0\simeq0.38\left(\frac{N_{\rm HI}}{10^{13}\,{\rm cm}^{-2}}\right)
\left(\frac{b}{20\kms}\right)^{-1}.
\label{eq:tau0-HI}
\end{equation}

\begin{figure}
\includegraphics[width=9cm, height=9cm]{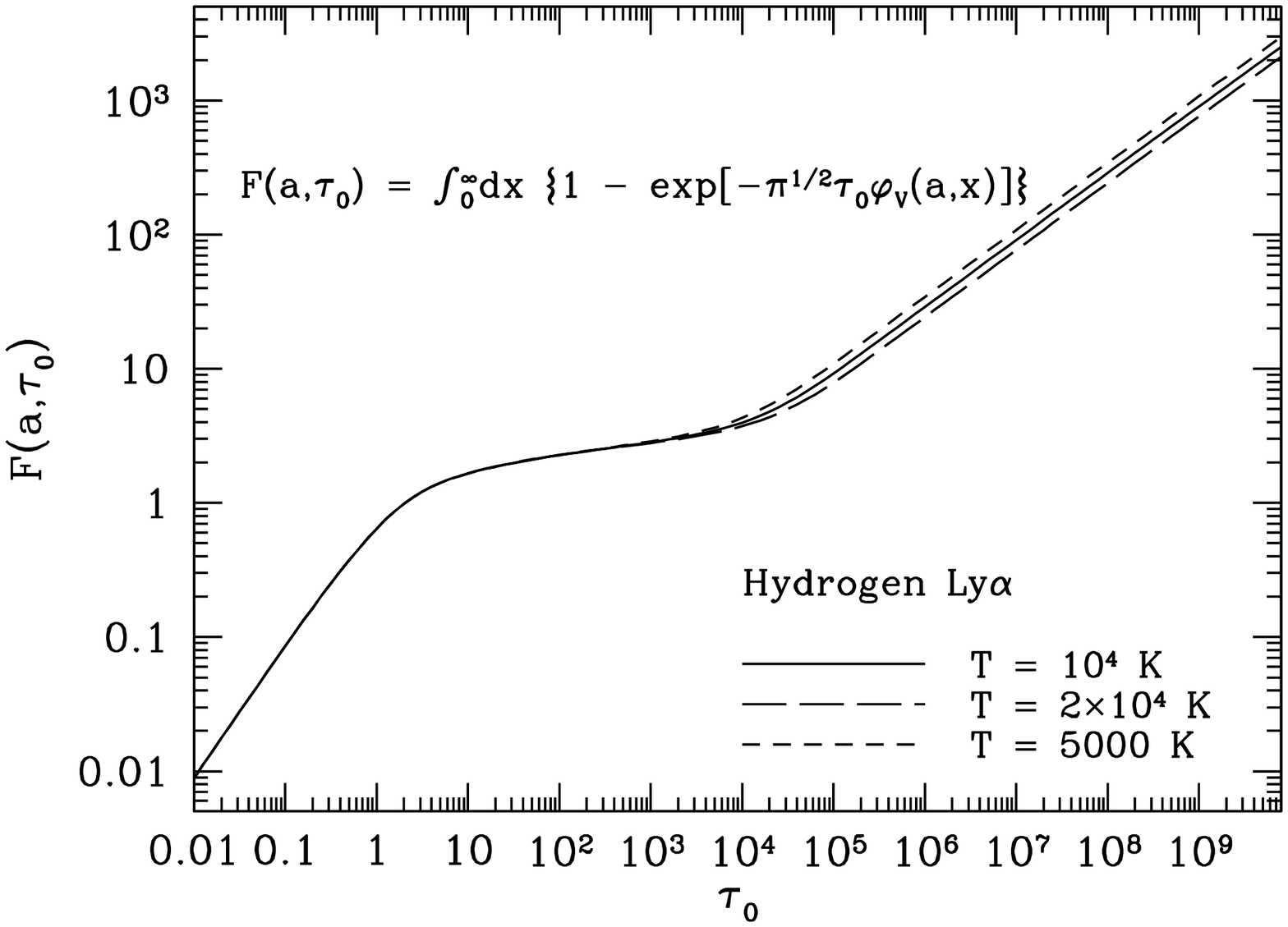}
\caption{The function $F(a,\tau_0)$ describing the equivalent
width of an absorption feature. Curves are shown for hydrogen
\Lya at three temperatures. Except for large values of $\tau_0$, the
curves are independent of $a$.}
\label{fig:Ftau}
\end{figure}

At redshifts $z>3.5$, the blending of absorption lines makes it
increasingly difficult to attribute the absorption to distinguishable
systems. By $z>5.5$, the individual lines have essentially all merged,
forming in effect a Gunn-Peterson trough (though still not
corresponding to a completely neutral IGM). Absorption features at
intermediate redshifts not easily deblended into subcomponents may be
characterized by an equivalent width $w_\lambda$. This is defined to
be the width, expressed in units of wavelength, a square-well
absorption feature with zero flux at its bottom must have to match the
integrated area of the detected feature under the continuum:
\begin{equation}
w_\lambda = \int_0^\infty\,d\lambda \left(1 - e^{-\tau_\nu}\right) =
\frac{\lambda^2}{c}\int_0^\infty\,d\nu \left(1 - e^{-\tau_\nu}\right).
\label{eq:w}
\end{equation}
For a Voigt profile, the equivalent width is related to the column
density through the line-center optical depth according to
\begin{equation}
\frac{w_\lambda}{\lambda}=2\frac{b}{c}F(a,\tau_0),
\label{eq:wFtau0}
\end{equation}
where
\begin{equation}
F(a,\tau_0)=\int_0^\infty\,dx\,
\left\{1-\exp\left[-\pi^{1/2}\tau_0\phi_V(a,x)\right]\right\}.
\label{eq:Ftau0}
\end{equation}
The relationship between the equivalent width and the column density
is known as the ``curve of growth.''  For $\tau_0<10^4$, the
absorption profiles are well-approximated as Doppler in shape,
corresponding to $\phi_V(a,x)\simeq\pi^{-1/2}e^{-x^2}$, and
$F(a,\tau_0)$ may be denoted more simply by $F(\tau_0)$. The function
$F(\tau_0)$ may be expressed as the power series
\cite{1968S&SS....7..365M}
\begin{equation}
F(\tau_0)=\frac{\sqrt\pi}{2}\sum_{n=1}^\infty\,(-1)^{n-1}
\frac{\tau_0^n}{n!\sqrt{n}}.
\label{eq:Ftau0Dlin}
\end{equation}

For very optically thin lines ($\tau_0<<1$),
$F(\tau_0)\simeq(\sqrt{\pi}/2)\tau_0$, and the equivalent width is
related linearly to the column density,
\begin{equation}
\frac{w_\lambda}{\lambda}=\frac{\pi e^2}{m_e c^2}
\left[\frac{1}{4\pi\epsilon_0}\right]N\lambda f_{lu}
\simeq8.85\times10^{-13}N\lambda f_{lu},
\label{eq:wlin}
\end{equation}
where $N\lambda$ has the dimension ${\rm cm}^{-1}$. Such features are
said to lie on the ``linear'' part of the curve of growth.

For $6<\tau_0<300$, better than 1\% accuracy is provided by the
asymptotic series \cite{1968S&SS....7..365M}
\begin{equation}
F(\tau_0)\sim\left(\ln\tau_0\right)^{1/2}\left[1+
\frac{0.2886}{\ln\tau_0}-\frac{0.1335}{(\ln\tau_0)^2}+\cdots\right].
\label{eq:Ftau0Dlog}
\end{equation}
For \HI\ \Lya at a temperature $T=10^4$~K, $F(a,\tau_0)$ deviates from
the logarithmic approximation Eq.~(\ref{eq:Ftau0Dlog}) by less than
10\% for $2<\tau_0<5000$. Such features are said to lie on the
``logarithmic'' part of the curve of growth. They are also referred to
as ``saturated'' lines, since an increase in the column density has
only a small change on the shape of the absorption feature. The
measurement of accurate column densities is notoriously difficult for
these features, necessitating the search for higher order resonance
lines on a more linear part of the curve of growth if accurate column
densities are desired. On the other hand, the equivalent width is
nearly directly proportional to the Doppler width, allowing an
accurate determination of the Doppler parameter.

For very large values of $\tau_0$, the Lorentzian radiation damping
wings of the Voigt profile dominate the line profile, which in this
limit is well approximated by $\phi_V(a,x)\simeq a/(\pi x^2)$.
In this case, the equivalent width is given approximately by
\begin{equation}
\frac{w_\lambda}{\lambda}\simeq\frac{2}{c}\left(\lambda^2N\sigma
\frac{\Gamma_{ul}}{4\pi}\right)^{1/2} = 2\pi^{1/4}\frac{b}{c}(a\tau_0)^{1/2}.
\label{eq:Ftau0sqrt}
\end{equation}
Because of the square-root dependence on column density, this limit is
referred to as the ``square-root'' part of the curve of growth. While
the independence of the equivalent width on the Doppler width of the
feature leaves $b$ poorly determined, the stronger dependence on the
column density permits a more accurate determination of the column
density from the equivalent width, or from line-profile fitting, than
is possible for systems on the logarithmic part. The square-root
approximation Eq.~(\ref{eq:Ftau0sqrt}), however, is accurate only for
very large values of $\tau_0$. It is more than 25\% too low for
$\tau_0<10^4$. Better than 1\% accuracy is achieved for
$\tau_0>2\times10^5$.

\subsection{Absorption line properties}
\label{subsec:absprop}

\begin{table*}
\caption{Summary of absorption line system properties}
\begin{ruledtabular}
\begin{tabular}{ccccccccc}
&\multicolumn{2}{c}{Line parameters}
&\multicolumn{4}{c}{Physical characteristics}&\multicolumn{2}{c}
{$\frac{dN^d}{dz}=N_0(1+z)^\gamma$}\\
Absorber class&$N_{\rm HI}$~(cm$^{-2}$)&$b^a\, (\rm km\,s^{-1})$
&$n^b\,({\rm m^{-3}})$&$T^b\,({\rm K})$&Size~(kpc)
&$[M/H]^c$&$N_0$&$\gamma$\\ \hline
Ly$\alpha$ forest &$\lta10^{17}$ & 15--60 & $0.01-1000$
&$5000-50000$ &15--1000(?) & -3.5 -- -2 &6.1&2.47\\ LLS
&$10^{17}-10^{19}$ & $\sim15$ & $\sim10^3-10^4$ &$\sim30000$ &--&
-3 -- -2 &0.3&1.50\\ Super LLS &$10^{19}-2\times10^{20}$ & $\sim15$ &
$\sim10^4$ &$\sim10000$ &--& -1 -- +0.6 &0.03&1.50\\ DLA
&$>2\times10^{20}$ & $\sim15$ & $\sim10^7$; $\sim10^4$ &$\sim100$;
$\sim10000$&$\sim10-20$(?)& -1.5 -- -0.8 &$\sim0.03$&$\sim1.5$\\
\label{tab:absprops}
\end{tabular}
\end{ruledtabular}
\begin{tabular}{l}
\multicolumn{1}{l}{$^a$Approximate ranges. Not well determined for most
Lyman Limit Systems and super Lyman Limit Systems.}\\
\multicolumn{1}{l}{$^b$Values not well constrained by direct observations.}\\
\multicolumn{1}{l}{$^c$Approximate metallicity range, expressed as a
logarithmic fraction of solar:\ $[M/H] =
\log_{10}(M/H)-\log_{10}(M/H)_\odot$.}\\
\multicolumn{1}{l}{$^d$For the following \HI\ column density and
redshift ranges. For the \Lya forest:\ $13.64<\log_{10}N_{\rm HI}<17$
and $1.5<z<4$;}\\ \multicolumn{1}{l}{for Lyman Limit Systems:\
$\log_{10}N_{\rm HI}>17.2$ and $0.32<z<4.11$. The same evolution rate
is adopted for super Lyman}\\ \multicolumn{1}{l}{Limit Systems. The
evolution rate of Damped \Lya Absorbers over the range $2<z<4$ is
consistent with that of}\\ \multicolumn{1}{l}{Lyman Limit Systems, but
poorly constrained by observations.}
\end{tabular}
\end{table*}

\subsubsection{Physical properties of absorption line systems}
\label{subsubsec:physprops}

The absorption features comprising the \Lya forest are broadly
classified into three main types:\ \Lya forest systems, Damped \Lya
Absorbers (DLAs), and Lyman Limit Systems (LLSs). The classification
is based primarily on the physical origin of the features. The \Lya
forest systems, by far the most common, are generally well-fit by
Doppler line profiles. The much rarer DLAs have sufficiently high
hydrogen column densities that they show the radiation damping wings
of the Lorentz profile, and require the Voigt line profile for
accurate fitting. The intermediate LLSs have a sufficiently large
column density to absorb photons with energies above the photoelectric
edge, or Lyman limit. The classifications are not strictly
exclusive. Damped absorbers of course produce Lyman Limit
Systems. Lower column density Lyman Limit Systems will produce a \Lya
forest feature. For convenience, the features, however, are treated as
distinct, corresponding typically to the column density ranges shown
in Table~\ref{tab:absprops}. Recently, a subclass of super Lyman Limit
Systems (sLLSs) has been introduced, corresponding to systems with
$10^{19}<N_{\rm HI}<2\times10^{20}\, {\rm cm^{-2}}$
\cite{2005ARA&A..43..861W}.\footnote{Although
the SI system is used throughout this review, column densities are
expressed with dimensions of ${\rm cm^{-2}}$ to be consistent with the
convention in the literature.} These
systems are convenient for statistical studies since their column
densities are more easily determined than their lower column density
counterparts because the damping wings begin to appear. They are
sometimes also referred to as sub-damped absorbers, but there is an
important physical distinction between them and the general class of
DLAs:\ the hydrogen in the DLAs is essentially neutral, while the
sLLSs are sufficiently penetrated by the UV metagalactic background as
to be partially ionized. It is therefore reasonable to distinguish the
DLAs as an entirely unique class of absorber.

\begin{figure}
\includegraphics[width=8cm, height=8cm]{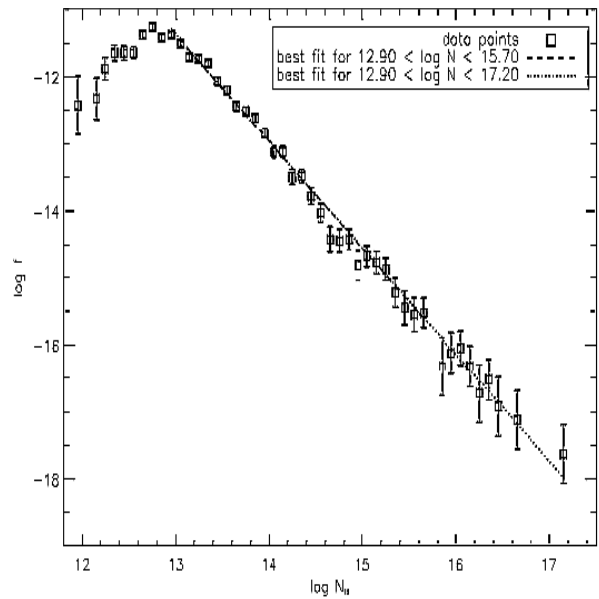}
\caption{Distribution of \Lya forest column densities with fits for
different ranges, for absorption systems in the redshift range
$0.5<z<1.9$. For $12.90<\log_{10}N_{\rm HI}<15.70$,
$\beta=1.60\pm0.03$. For $12.90<\log_{10}N_{\rm HI}<17.20$,
$\beta=1.59\pm0.02$. From \textcite{2006A&A...458..427J}.
(Figure reproduced by permission of Astronomy \& Astrophysics.)
}
\label{fig:dNdNHI}
\end{figure}

\begin{figure}
\includegraphics[width=8cm, height=8cm]{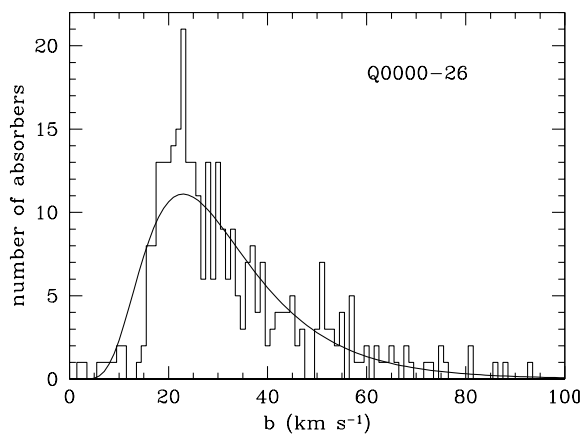}
\caption{Distribution of \Lya forest Doppler parameters for Q0000--26
\cite{1996ApJ...472..509L}. A lognormal distribution (solid line)
provides a statistically acceptable fit to the distribution.
}
\label{fig:bdist}
\end{figure}

A great number of surveys of QSO intervening absorption systems have
been carried out, building up a large inventory of their statistical
properties. Systematic surveys of the \Lya forest were conducted by
\textcite{1979ApJ...234...33W}, \textcite{1980ApJS...42...41S},
\textcite{1994ApJS...91....1B}, \textcite{1995AJ....110.1526H},
\textcite{1996ApJ...468..121S}, \textcite{2002MNRAS.335..555K},
\textcite{2002ApJS..140..143B}, and \textcite{2004AJ....128.1058T}.
Surveys of Lyman Limit Systems were conducted by
\textcite{1989ApJS...69..703S}, \textcite{1995ApJ...440..435L} and
\textcite{2007ApJ...656..666O}.  Surveys of Damped \Lya Absorbers were
conducted by \textcite{1986ApJS...61..249W},
\textcite{1991ApJS...77....1L}, \textcite{1993ApJS...84....1L},
\textcite{1994AJ....108...44L}, \textcite{1995ApJ...440..435L},
\textcite{1995ApJ...454..698W}, \textcite{2000ApJ...543..552S},
\textcite{2001A&A...379..393E}, \textcite{2002A&A...383...91E},
\textcite{2002ApJ...566...51C}, \textcite{2005A&A...440..499A},
\textcite{2005MNRAS.362..549S}, \textcite{2006ApJ...636..610R} and
\textcite{2006ApJ...646..730J}. A radio survey of intervening
absorption systems was carried out by
\textcite{2007ApJ...654L.111G}. Surveys for galaxies associated with
intervening absorption systems were conducted for galaxies associated
with diffuse \Lya absorption systems at $z<1$ by
\textcite{1995ApJ...442..538L}, \textcite{2005ApJ...629L..25C},
\textcite{2006MNRAS.367.1261M}, and \textcite{2006MNRAS.367.1251R},
with Damped \Lya Absorbers by \textcite{2003ApJ...597..706C},
\textcite{2005ApJ...621..596C}, and \textcite{2006ApJ...652..994C},
and with metal absorption systems at $z<1$ by
\textcite{2005ApJ...620...95T}, \textcite{2006ApJ...643..680P},
\textcite{2006ApJ...641..217S} and \textcite{2008AJ....135..922K}, and
at $z>1$ by \textcite{2005ApJ...629..636A}.

The neutral hydrogen column densities of the absorbers are measured to
range from roughly $10^{12}-10^{22}\,{\rm cm^{-2}}$. Lower column
density systems may exist, but are difficult to detect. An upper
cut-off at $3-5\times10^{21}\,{\rm cm^{-2}}$ is suggested by
\textcite{2005ApJ...635..123P}. It was early recognized by
\textcite{1987ApJ...321...49T} that the distribution function of the
column densities is a near perfect power law, $dN/dN_{\rm HI}\propto
N_{\rm HI}^{-\beta}$, with $\beta=1.5-1.7$ \cite{1987ApJ...321...49T,
1995AJ....110.1526H,2002MNRAS.335..555K}. A recent determination for
absorption systems in the redshift range $0.5<z<1.9$ is shown in
Fig.~\ref{fig:dNdNHI} \cite{2006A&A...458..427J}. Although there may
be small deviations from a perfect power law
\cite{1993ApJ...416..137G,MM93,1993MNRAS.262..499P}, the nearness to a
single power law over such an enormous dynamic range strongly suggests
a single formation mechanism.

The measured Doppler velocities $b$ range over about $10<b<100\,{\rm
km\,s^{-1}}$, with the vast majority concentrated between $15-60\kms$
\cite{1985ApJ...292...58A, 1991ApJ...371...36C, 1992ApJ...390..387R,
1995AJ....110.1526H, 1996ApJ...472..509L, 1997AJ....114....1K,
1997ApJ...484..672K}. A typical distribution is shown in
Fig.~\ref{fig:bdist}, using the line list for Q0000--26 from
\textcite{1996ApJ...472..509L}. Temperatures may in principle be
inferred from Eq.~(\ref{eq:bparam}), but doing so is hampered by two
difficulties:\ 1.\ the systems may be broadened by a kinematic
component and 2.\ the absorption features may be blends of more than a
single system. Evidence for kinematic broadening is found when metal
features are also detected (see below). In general, there is no unique
fit to an absorption feature, particularly in the presence of
blending:\ several statistically acceptable fits are possible
\cite{1997ApJ...484..672K}, and these will change as the
signal-to-noise ratio or spectral resolution changes
\cite{1993MNRAS.260..589R}. Indeed, Eq.~(\ref{eq:taunu}) shows that
each absorption feature itself may be regarded as the blending of an
infinity of smaller features. It is only because of clumpiness of the
IGM that the features may be localized (Eq.~[\ref{eq:taunu-i}]), yet
internal structure is still visible when multiple narrower metal
features are detected in a single \Lya system.

\begin{table*}
\caption{Lognormal fits to the Doppler parameter distributions
measured for the \Lya forest seen in the spectra of
Q0000--26 \cite{1996ApJ...472..509L}, HS~1946$+$7658
\cite{1997ApJ...484..672K}, and Q0014$+$813, Q0302--003,
Q0636$+$680 and Q0956$+$122 \cite{1995AJ....110.1526H}.
The fits are represented in the form $f(b)=\exp(-0.5\xi^2)/\sqrt{2\pi}$,
where $\xi=\gamma+\delta\ln(b)$.
}
\begin{ruledtabular}
\begin{tabular}{rllllll}
QSO name & $z\, {\rm range}$ & no. lines & $\langle b\rangle$ ($\kms$) & $\gamma$ & $\delta$ & $p_{\rm KS}$ \\
 \hline
 Q0000--26      & $3.42 < z < 3.98$ & 334 & 32.8 & $-7.15$ & 2.13 & 0.36 \\
 Q0014$+$813    & $2.70 < z < 3.20$ & 262 & 33.9 & $-8.86$ & 2.58 & 0.99 \\
 Q0302--003     & $2.62 < z < 3.11$ & 266 & 33.9 & $-8.75$ & 2.56 & 0.97 \\
 Q0636$+$680    & $2.54 < z < 3.03$ & 312 & 29.3 & $-7.24$ & 2.23 & 0.30 \\
 Q0956$+$122    & $2.62 < z < 3.11$ & 256 & 31.6 & $-8.05$ & 2.42 & 0.53 \\
 HS~1946$+$7658 & $2.42 < z < 2.98$ & 399 & 32.5 & $-7.18$ & 2.16 & 0.50 \\
 \hline
\label{tab:bfit}
\end{tabular}
\end{ruledtabular}
\end{table*}

Broadening is also expected from the line finding and fitting
procedure. The systematic influence of procedures used to locate and
fit the absorption lines on the resulting distributions of Doppler
parameters has received limited attention. The potential usefulness of
the Doppler parameter distribution for extracting physical information
about the IGM, such as its temperature distribution, merits further
study of the effect of fitting algorithms on the derived
distribution. Artificially large $b$-values, for example, have been
reproduced through Monte Carlo simulated spectra with much narrower
gaussian distributions for the $b$-values \cite{1996ApJ...472..509L,
1997ApJ...484..672K}. The Monte Carlo models assume Poisson
distributed line centers. Allowing for clustering of the line centers
(see below) may lead to even broader distributions.

The measured distributions may be fit by a lognormal function
\cite{1997ApJ...485..496Z, MBM01}. As an illustration, the best
fitting lognormal function $f(\xi)=\exp(-0.5\xi^2)/\sqrt{2\pi}$, where
$\xi=-7.15+2.13\ln(b)$, provides a statistically acceptable
description of the $b$ distribution of Q0000--26
\cite{1996ApJ...472..509L} for systems lying between the restframe
\Lyb and \Lya wavelengths of the QSO, excluding a small region
possibly influenced by the proximity effect (see \S~\ref{subsubsec:
proximity}). The fit distribution is shown in Fig.~\ref{fig:bdist}.
The KS test yields a probability for obtaining as large a difference
as found between the predicted and measured cumulative distributions
of $p_{\rm KS}=0.361$. A lognormal distribution will result when the
error on a quantity is proportional to the magnitude of the
quantity. For example, consider an absorption system with an intrinsic
Doppler parameter $b$. If the iterations of the nonlinear fitting
procedure of the remaining features in the spectrum each perturb the
first feature by an amount $\Delta b\propto b$, then $\Delta\ln b$
will change by a randomly distributed amount. Fitting several features
will then result in a sum of random changes to $\Delta\ln b$. By the
central limit theorem, a normal distribution for $\ln b$ will result.
A comparison between the measured $b$-values and their errors from
\textcite{1996ApJ...472..509L} shows a positive correlation. The
Spearman rank-order correlation coefficient is $r_s=0.31$, with
$p(>r_s)=9.3\times10^{-9}$. A similar result is found for
HS~1946$+$7658 \cite{1997ApJ...484..672K}, for which a correlation
between the Doppler parameters and their errors is found with
$r_s=0.28$ and $p(>r_s)=7.4\times10^{-9}$. A lognormal distribution
again provides an acceptable fit to the Doppler parameter
distribution. The results for this and several other distributions
measured in high resolution Keck spectra are shown in
Table~\ref{tab:bfit}. In all cases, a lognormal distribution provides
a good fit. Although not conclusive, these results suggest that the
measured Doppler parameters may in part be induced by a
lognormal-generating stochastic process. It is noted that the process
need not arise entirely from the line-fitting, but could also
originate from the physical processes that gave rise to the structures
that produce discrete absorption features.

Uncertainty in the origin of the Doppler parameter distribution leaves
the physical interpretation of the Doppler parameters with some
ambiguity. Although their relation to the gas temperature is not
straightforward, the Doppler parameters may usually be used
legitimately to set upper limits on the gas temperature. (Even an
upper limit is not always guaranteed, since noise affecting a weak
line may leave it too narrow, or even produce an artificial narrow
line.) For pure Doppler broadening, the range $20-60\kms$ corresponds
to temperatures of $2.4\times10^4-3.8\times 10^4$~K, the range
expected from photoionization at the moderate overdensities expected
for the absorbers \cite{Meiksin94,1997MNRAS.292...27H}. Cooler
temperatures are possible, however, particularly if the gas has been
expanding sufficiently fast for adiabatic cooling to be appreciable.

The element of indeterminacy in the measurements of the line
parameters has made it difficult to search for evolution in the
distribution of Doppler parameters. Evolution is especially
interesting in the context of late \HeII\ reionization, which may have
occurred at $z\simeq3-3.5$. Evolution of the Doppler parameter
distribution, however, could also result if aborption systems with
different physical characteristics dominate the detected and fit
absorption lines at different epochs. Numerical simulations suggest
the systems that give rise to a given \HI\ column density range shift
to structures of different gas densities, and therefore different gas
temperatures, and different sizes as the Universe expands and as the
intensity of the photoionizing UV background evolves. The difficulty
of interpreting any change in the Doppler parameter distribution in
terms of sudden heating is exacerbated by an increase in line blending
with increasing redshift. Using an analysis based on the lower cut-off
in the Doppler parameter distribution, \textcite{Schaye00} report
evolution in the inferred IGM gas temperature over the range
$2.0<z<4.5$, peaking at $z\simeq3$, consistent with the onset of
\HeII\ reionization at this redshift. Similar results are obtained
from a separate analysis by
\textcite{2000ApJ...534...41R}. \textcite{2002A&A...383..747K} also
suggest evolution in the gas temperature based on an increase in the
lower cut-off of the Doppler parameter distribution from $z=3.8$ to
$z=3.3$. But the values for the cut-off they derive are discrepant
with those of \textcite{Schaye00}, which casts uncertainty on either
interpretation. \textcite{2002A&A...383..747K} show there is
considerable sample variance in the cut-off at the same redshift, and
attribute the discrepancy with \textcite{Schaye00} to this
variance. On the other hand, \textcite{2007ApJ...658..680L} find no
significant evolution in the $b$-distribution over the redshift range
$1.5<z<3.6$ using the same sample as \textcite{2002MNRAS.335..555K}.
This result is confirmed by the nearly identical parameters obtained
for the best fit lognormal distributions between the highest and
lowest redshift samples in Table~\ref{tab:bfit}.
\textcite{2007ApJ...658..680L} do argue for the presence of a second
broader population at $z\lta0.4$. \textcite{2006A&A...458..427J} find
no evidence for evolution in the Doppler parameters over the redshift
range $0<z<2$.

Damped \Lya Absorbers show a more complex physical structure.
Measurements of the 21cm hyperfine absorption signature produce a
range of spin temperatures, from values as low as $T_S\simeq200$~K to
lower limits of $T_S\gta9000$~K, with the lower values occurring only
at $z\lta3$. \textcite{2003A&A...399..857K} provide a tabulation from
the literature. The presence of high ionization metal species show
there is a second warmer component present as well, with
$T\simeq10^4$~K (see below). \textcite{2005ARA&A..43..861W} suggest a
two-component interstellar medium consisting of a warm neutral medium
(WNM) at $T\simeq8000$~K in pressure equilibrium with a cold neutral
medium (CNM) at $T\simeq100$~K.

A summary of the absorption line properties, as well as of the
physical characteristics discussed below, is provided in
Table~\ref{tab:absprops}.

\subsubsection{Evolution in the number density of the \Lya forest}
\label{subsubsec:evolution}

\begin{figure}
\includegraphics[width=8cm, height=6cm]{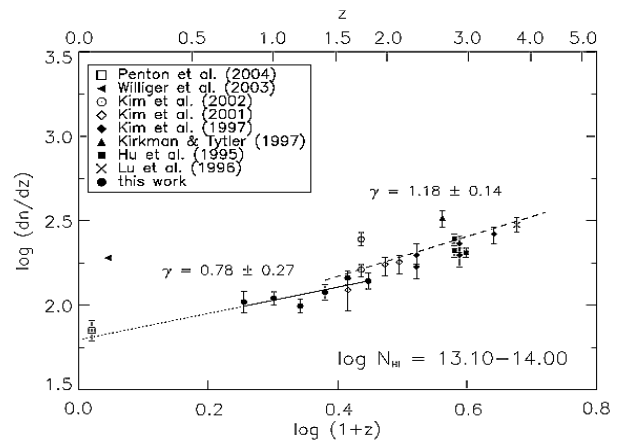}
\caption{Evolution of the number density of \Lya systems with $13.10 <
\log_{10} N_{\rm HI} < 14.00$, including best-fits to the solid and
open circles over the ranges indicated, as represented by the solid
and dashed lines, respectively, taking $dN/dz\propto (1+z)^\gamma$.
The dotted line extrapolates the lower redshift fit to $z=0$. From
\textcite{2006A&A...458..427J}. (Figure reproduced by permission of
Astronomy \& Astrophysics.)
}
\label{fig:dNdz_low}
\end{figure}

\begin{figure}
\includegraphics[width=8cm, height=6cm]{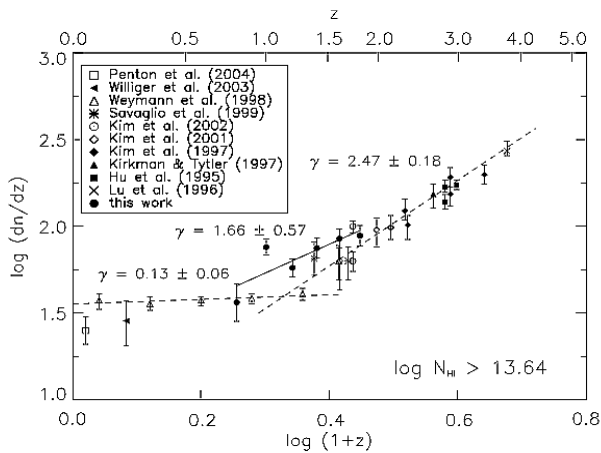}
\caption{Evolution of the number density of \Lya systems with
$\log_{10} N_{\rm HI} > 13.64$, including best-fits to the solid and
open circles over the ranges indicated, as represented by the solid
and dashed lines, taking $dN/dz\propto(1+z)^\gamma$. The additional
low redshift dashed line is the best fit to the open triangles. The
open square, left-pointing solid triangle, and stars correspond to
systems with $\log_{10} N_{\rm HI}>14.00$.  From
\textcite{2006A&A...458..427J}.  (Figure reproduced by permission of
Astronomy \& Astrophysics.)
}
\label{fig:dNdz_high}
\end{figure}

The number of absorption systems per unit redshift increases with
redshift. Some evolution is expected as a result of the expansion of
the Universe. For a proper number density $n_a(z)$ of absorption
systems at redshift $z$, with proper cross section $\sigma_a(z)$,
the expected number of absorbers per unit proper length is
$dN/dl_p=n_a(z)\sigma_a(z)$. The proper line element is related to
redshift according to $dl_p/dz = c/[H(z)(1+z)]$, where $H(z)=H_0E(z)$
is the Hubble parameter (see \S~\ref{subsec: absorption-lines} above).
For a flat Universe ($\Omega_K=0$) and standard cosmological parameters,
$E(z)\simeq0.55(1+z)^{3/2}[1+2.3/(1+z)^3]^{1/2}$. The evolution in the
number density is then given by
\begin{equation}
\frac{dN}{dz} \simeq
(2100\,{\rm Mpc})n_{a, {\rm c}}(z)\sigma_a(z)(1+z)^{1/2}
\biggl[1+\frac{2.3}{(1+z)^3}\biggr]^{-1/2},
\label{eq:dNdz}
\end{equation}
where $n_{a, {\rm c}}(z)=n_a(z)(1+z)^{-3}$ is the comoving number
density of systems. For a constant comoving number density and fixed
proper cross-section, only moderate evolution is expected,
$dN/dz\propto(1+z)^{1/2}$. At low redshifts, this gives a reasonable
description of the evolution. Using {\it Hubble Space Telescope}
observations, \textcite{1998ApJ...506....1W} find for all system with
equivalent widths above 0.24\AA, $dN/dz =
(32.7\pm4.2)(1+z)^{0.26\pm0.22}$ for $z<1.5$. Subsequent higher
spectral resolution surveys using {\it HST} have extended the
statistics to systems with equivalent widths below 0.1\AA\ at $z<0.1$
\cite{2000ApJ...544..150P, 2004ApJS..152...29P}. The large survey of
\textcite{2008ApJ...679..194D} obtains $dN/dz=129^{+6}_{-5}$ down to
0.030\AA\ at $z<0.4$ with $\langle z\rangle\simeq0.14$.

For $z>1.5$, however, the number density evolves considerably faster
than the constant comoving rate \cite{1987ApJ...321...69T}.
\textcite{2002MNRAS.335..555K} obtain $dN/dz=6.1(1+z)^{2.47\pm0.18}$
for systems with column densities in the range $13.64<\log_{10} N_{\rm
HI} <17$. This corresponds to considerable evolution in the product
$n_{a, {\rm c}}(z)\sigma_a(z)$. A smooth transition is found over the
redshift range $0.5<z<1.9$ by \textcite{2006A&A...458..427J}, with a
dependence of the evolution rate on column density. High column
density systems corresponding to Lyman Limit Systems evolve somewhat
more slowly than lower column density
systems. \textcite{1995ApJ...444...64S} find for systems with
$\tau_L>1$, $dN/dz=0.25^{-0.10}_{+0.17}(1+z)^{1.50\pm0.39}$ over the
redshift range $0.32<z<4.11$. Based on a larger, higher redshift
sample, \textcite{2003MNRAS.346.1103P} find somewhat more rapid
evolution, with
$dN/dz=0.07^{-0.04}_{+0.13}(1+z)^{2.45^{+0.75}_{-0.65}}$ over the
redshift range $2.4<z<5$. Damped \Lya Absorbers show evolution
comparable to that of the Lyman Limit Systems, with $dN/dz$ increasing
by about a factor of 2 from $z=2.5$ to $z=4$
\cite{2005ApJ...635..123P}.

As will be discussed in \S~\ref{sec:simuls} below, numerical
simulations show that the structure of the IGM evolves, although in
the comoving frame the structure is remarkably stable. Because $dN/dz$
is fit for a fixed \HI\ column density range, the diminishing gas
density towards decreasing redshifts will result in fewer systems
satisfying the column density threshold, so that $dN/dz$ will decrease
towards decreasing redshift. Evolution in the ionizing background will
also affect the number of absorption systems lying above the column
density threshold, and this is a major factor in the evolution of
$dN/dz$. The slowdown at $z<1.5$ in the evolution is in fact
attributed predominantly to a reduction in the intensity of the
ionizing background:\ as the ionizing rate decreases, fewer systems
will slip below the column density threshold than under pure density
evolution. As a result, the decrease in $dN/dz$ towards lower $z$
slows down. The difference in the rate of evolution between low column
density and high column density systems found by
\textcite{2006A&A...458..427J} (compare Figs.~\ref{fig:dNdz_low} and
\ref{fig:dNdz_high}), shows that structural evolution in the IGM must
also play a role.

\subsubsection{Characteristic sizes and spatial correlations}
\label{subsubsec:sizes}

Multiple QSOs in proximity to each other on the sky provide an
invaluable means for estimating the physical sizes of the regions in
the IGM that give rise to the absorption features. Especially useful
have been multiply imaged lensed quasars. Using high resolution
spectra of a pair of images of a lensed QSO with a separation of
3~arcsecs, \textcite{1995A&AS..113..199S} detected a large number of
absorbers coincident between the two lines of sight. All the clearly
detected coincident systems show comparable equivalent widths between
the lines of sight, suggesting the same physical structures are being
probed. These authors set $2\sigma$ lower limits of about 70~kpc for
weak absorption systems ($\log_{10} N_{\rm HI}<14$), and of about
15~kpc for strong systems ($\log_{10} N_{\rm HI}>15$), assuming the
clouds are spherical. For disk-like systems, the lower limits are
about twice as large.

Other efforts resulted in similar lower limits
\cite{1990MNRAS.242..544M,1994ApJ...437L..83B, 1992ApJ...389...39S}.
Multiple lines of sight with wider separations of about 1 arcminute
suggest still larger sizes of ~0.3--1~Mpc \cite{1998ApJ...494..567D,
1998A&A...339..678D, 1998A&A...334L..45P,1998ApJ...502...16C}, and
possibly as large as 40 Mpc (comoving) \cite{2000ApJ...532...77W}.
\textcite{1995A&AS..113..199S} report a likely size of about
10--30~kpc for a single Damped \Lya Absorber, comparable to the lower
limit obtained by direct \HI\ imaging measurements of a separate
system by \textcite{1989ApJ...341..650B}.

A difficulty with interpreting the larger Megaparsec-scale sizes for
the \Lya forest systems is that the equivalent widths of systems with
common redshifts along the neighboring lines of sight do not always
match well over these scales. An alternative interpretation of the
coinciding redshifts is that they indicate underlying spatial
correlations between the absorption systems rather than their spatial
coherence. Measurements of the auto-correlation function between
absorption systems along single lines of sight suggest correlations on
scales of a few hundred $\kms$ scales \cite{1995ApJ...448L..85M,
1995MNRAS.273.1016C, 1995AJ....110.1526H}. An anti-correlation at
separations of $500-1000\kms$ \cite{1995ApJ...448L..85M,
1995MNRAS.273.1016C, 1995AJ....110.1526H} has been disputed by
\textcite{1997AJ....114....1K} on the basis of the limited lines of
sight analysed, but the issue appears not fully resolved. In the
presence of correlations, the error on the correlation function
depends on the 3 and 4-pt functions, which are not well-measured. A
more recent analysis at $z<2$ detects only weak positive correlations
on the few hundred $\kms$ scale, and no clear anti-correlation
\cite{2006A&A...458..427J}. Transverse positive spatial correlations
over comparable scales at $\langle z\rangle=2.14$ were obtained by
\textcite{1998ApJ...502...16C} and at $0.4<z<0.9$ by
\textcite{2006AJ....132.2046P}. Coherence in absorption lines along
lines of sight separated by $\sim2h^{-1}$~Mpc at $2.6<z<3.8$ has been
detected by \textcite{2008AJ....136..181C}.

A characteristic size of 70~kpc for systems with $N_{\rm
HI}\simeq10^{14}\,{\rm cm^{-2}}$ suggests a large fraction of the
baryons are contained within the \Lya forest \cite{MM93}. Taking the
size to correspond to a characteristic line-of-sight thickness, the
corresponding \HI\ number density is $n_{\rm
HI}\simeq5\times10^{-4}\,{\rm m^{-3}}$. Ionization equilibrium in a UV
background producing an ionization rate of $\Gamma_{\rm
HI}\simeq\Gamma_{\rm HI, -12}10^{-12}$ \HI\ atoms per second
(\S~\ref{subsubsec: sources} below), gives a neutral fraction of
$x_{\rm HI}\simeq (1.2n_{\rm HI}\alpha_A/\Gamma_{\rm HI})^{1/2}
\simeq1.5\times10^{-5}N^{1/2}_{\rm HI, 14}T_4^{-0.37}\Gamma_{\HI,
-12}^{-1/2}$, where $\alpha_A\simeq4\times10^{-19}\,{\rm
m^3\,s^{-1}}T_4^{-0.75}$ is the radiative recombination rate for gas
at a temperature $T_4=T/10^4$~K and $N_{\rm HI, 14}=N_{\rm
HI}/10^{14}\,{\rm cm^{-2}}$. The ratio of the baryonic mass density of
the \Lya forest to the critical Einstein-deSitter density $\rho_{\rm
crit}(z)=\rho_{\rm crit}(0)H(z)^2/ H_0^2$ is
\begin{eqnarray}
\Omega_{\rm Ly\alpha} &=& \frac{1.4m_{\rm H}}{\rho_{\rm crit}}
\int dN_{\rm HI}\frac{\partial^2{\cal N}}{\partial N_{\rm HI}\partial z}
\frac{N_{\rm HI}}{x_{\rm HI}}\biggl(\frac{dl_p}{dz}\biggr)^{-1}\nonumber\\
&=& 1.4m_{\rm H}\frac{8\pi G}{3cH(z)}(1+z)
\int dN_{\rm HI}\frac{\partial^2{\cal N}}{\partial N_{\rm HI}\partial z}
\frac{N_{\rm HI}}{x_{\rm HI}}\nonumber\\
&\simeq&3.0\times10^{-5}N_0h^{-1}\Omega_m^{-1/2}T_4^{0.37}
\Gamma^{1/2}_{\rm HI, -12}\nonumber\\
&\times&(1+z)^{\gamma-1/2}
\ln\biggl(\frac{N_{\rm HI, max}}{N_{\rm HI, min}}\biggr)\nonumber\\
&\simeq&0.06T^{0.37}_4\Gamma^{1/2}_{\rm HI, -12},
\label{eq:OmLya}
\end{eqnarray}
for $dN/dz=N_0(1+z)^\gamma$ with $N_0=6.1$ and $\gamma=2.47$, and
$dN/dN_{\rm HI}\propto N_{\rm HI}^{-1.5}$ with $N_{\rm HI,
min}=10^{13.64}\,{\rm cm^{-2}}$ and $N_{\rm HI, max}=10^{17}\,{\rm
cm^{-2}}$, evaluated at $z=3$ with $\Omega_m=0.3$ and $h=0.7$. This is
comparable to estimates for the baryon density of the Universe (see
\S~\ref{subsubsec:deuterium} below). The result is not conclusive,
however. Since the size estimates are in the transverse direction, a
much smaller value for $\Omega_{\rm Ly\alpha}$ is possible if the
structures are flattened, down by the aspect ratio of the thickness to
breadth of the systems \cite{1995MNRAS.275L..76R}.

\subsubsection{Deuterium absorption systems}
\label{subsubsec:deuterium}

High column density \Lya absorption systems offer the opportunity to
measure the primordial deuterium abundance ${\rm D/H}$, a sensitive
indicator of the cosmic baryon density, or equivalently the nucleon to
photon ratio $\eta$, according to Big Bang nucleosynthesis. To date, a
relatively limited number of accurate determinations of $\eta$ from
the deuterium abundance have been possible. A line of sight must meet
several criteria to be useful for obtaining a clean measurement
\cite{2003ApJS..149....1K}:\ the hydrogen column density must be
sufficiently large that deuterium is detectable; the velocity width of
the system must be sufficiently narrow so as not to encroach on the
$82\kms$ offset deuterium feature; the structure of the system must
not be so complicated that a weak interloping \HI\ feature could be
mistaken for deuterium; and the background QSO must be sufficiently
bright to obtain a high signal-to-noise ratio measurement. These
combined criteria whittle down the number of acceptable QSO
lines-of-sight to about 1\% of those available at $z=3$.

Despite the difficulty, several accurate determinations have been made
\cite{1998ApJ...499..699B, 1998ApJ...507..732B, 2001ApJ...552..718O,
2001ApJ...560...41P, 2003ApJS..149....1K, 2002ApJ...565..696L,
2004MNRAS.355.1042C, 2006ApJ...649L..61O}. The current best estimate
for ${\rm D/H}$ is $\log_{10} {\rm D/H} = -4.55\pm0.04$
\cite{2006ApJ...649L..61O}. Using standard Big Bang nucleosynthesis
models, this translates to a baryon density parameter of
$\Omega_bh^2=0.0213\pm0.0013\pm 0.0004$, where the $1\sigma$ errors
refer to the error on the ${\rm D/H}$ measurement and the
uncertainties in the nuclear reaction rates, respectively. This
corresponds to a nucleon to photon ratio of
$\eta=(5.8\pm0.3)\times10^{-10}$. This compares favorably to the
determination from the 5-year {\it WMAP} measurements of the CMB
fluctuations, combined with Baryonic Acoustic Oscillations and Type Ia
supernovae constraints, of $\eta\simeq(6.2\pm0.2)\times10^{-10}$
\cite{2008arXiv0803.0547K}.

\subsubsection{Helium absorption systems}
\label{subsubsec:helium}

The UV spectral capability of the {\it Hubble Space Telescope} made it
possible for the first time to measure intergalactic \HeII\ \Lya
$\lambda304$. Using the spectrum of Q0302$-$003 at $z_{\rm
em}\simeq3.286$ observed by the Faint Object Camera,
\textcite{1994Natur.370...35J} found $\tau_{\rm HeII}>1.7$ (90\%
confidence) at $z=3.2$, revised to a 95\% confidence interval of $1.5
< \tau_{\rm HeII}<3.0$ by \textcite{1997AJ....113.1495H} and to
$\tau_{\rm HeII} > 4.8$ by \textcite{2000ApJ...534...69H} after more
detailed modeling of the absorption trough. Because the same sources
which ionize the \HI\ should also ionize \HeI, virtually all the
intergalactic helium will be in the form of either \HeII\ or \HeIII,
depending on whether or not there were sources hard enough to fully
ionize helium.

Remarkably, using the spectrum of HS~1700$+$6416 ($z_{\rm
em}\simeq2.7$) observed by the {\it Hopkins Ultraviolet Telescope}
({\it HUT}), \textcite{1996Natur.380...47D} measured the more moderate
optical depth of $\tau_{\rm HeII}\simeq1.00\pm0.07$ over the somewhat
lower redshift range $z=2.2-2.6$. This led to the speculation that
\HeII\ was ionized around $z\simeq3$. Subsequent measurements of the
spectrum of HE~2347$-$4342 ($z_{\rm em}\simeq2.885$) have
shown the \HeII\ \Lya optical depth to be very patchy at $z\lta3$,
as if a second reionization epoch were being observed, that of \HeII\
\cite{1997A&A...327..890R}.

Observations of PKS~1935$-$692 ($z_{\rm em}\simeq3.18$) using {\it
HST} \cite{1999AJ....117...56A} and of HE~2347$-$4342 using the {\it
Far Ultraviolet Spectroscopic Explorer} have detected a \HeII\ \Lya
forest \cite{2001Sci...293.1112K, Zheng04, 2004ApJ...600..570S},
permitting more detailed assessments of the fluctuations in the \HeII\
to \HI\ column densities. Measurements of the \HeII\ \Lya forest also
provide an opportunity to probe the underlying IGM velocity
field. Using combined {\it FUSE} and VLT data, \textcite{Zheng04} were
able to identify isolated absorption systems in underdense regions
with common \HI\ \Lya and \HeII\ \Lya features. A comparison between
the fit Doppler parameters gives the ratio $b_{\rm HeII}/b_{\rm
HI}=0.95\pm0.12$. According to Eq.~(\ref{eq:bbparam}), this indicates
a velocity field in underdense regions with kinematic motions that
dominate the thermal.

Subsequent measurements continue to confirm the patchy nature of the
\HeII\ \Lya optical depth, including {\it HST} observations of
HE~2347$-$4342 \cite{2002ApJ...564..542S}, SDSS~J2346$-$0016 ($z_{\rm
em}\simeq3.50$) \cite{2004AJ....127..656Z}, and Q1157$+$3143 ($z_{\rm
em}\simeq3.01$)\cite{Reimers05}, and a {\it FUSE} observation of
HS~1700$+$6416 \cite{2006A&A...455...91F}. As some of these references
point out, and as will be discussed in
\S~\ref{subsubsec:He-constraints} below, the interpretation in terms
of reionization is not unambiguous:\ the patchiness may also arise
from fluctuations in the radiation field.

\subsubsection{Metal absorption systems}
\label{subsubsec:metals}

The earliest identifications of intervening absorption systems were
for elements heavier than helium, so called metals
\cite{1968ApJ...153..689B}. Common are carbon, nitrogen, silicon and
iron, but to date the list extends much further, including oxygen,
magnesium, neon, and sulfur \cite{1980ApJS...42...41S,
1987ApJ...315L...5M, 1995AJ....109.1522C, 1996A&A...312...33K,
1996AJ....112..335S, 1998AJ....115...55L, 1999ApJ...520..456E,
1999Ap&SS.269..201E, 2000ApJ...541L...1S, 2002ApJ...578..737S,
2004ApJ...606...92S, 2004ApJS..153..165R, 2005ApJ...626..776S}. The
metals provide invaluable insight into the structure of the IGM in
several different ways. The widths of metal absorption features allow
direct estimates of the temperature of the IGM and its small-scale
velocity structure. The narrow widths of \CIVs features were used by
\textcite{1984ApJ...280L...1Y} to demonstrate the absorbers had the
characteristic temperatures of photoionized gas rather than
collisionally ionized. The metal systems probe the impact star
formation has had on the IGM, as the metals were most likely
transported into the IGM through galactic winds, or introduced {\it in
situ} by local small-scale regions forming first generation,
Population III, stars. In the nearby Universe, they indicate the
presence of shocked gas, as may accompany the formation of galaxy
clusters. As such, the metals in principle document the history of
cosmic structure and star formation. Ratios of the metal column
densities may be used to constrain the spectral shape of the
metagalactic ionizing UV background, which in turn puts constraints on
the possible sources of the background and their relative
contributions. Lastly, metal absorption lines offer a unique
opportunity to search for variability in the constants of nature.

Tables of the resonance lines most easily detectable in intervening
absorption systems are provided by \textcite{1989ApJS...69..701M} and
\textcite{1995A&A...294..377V}. Searches for intervening metal
absorbers have been facilitated by the exploitation of atomic line
doublets for some of the stronger species. The most common doublets
used are \MgIIs$\lambda\lambda2796, 2803$~\AA\ and
\CIVs$\lambda\lambda1548, 1551$~\AA. These are among the strongest
lines detected in the \Lya forest. They also have the important
advantage of wavelengths redward of Ly$\alpha$, placing them outside
the \Lya forest and avoiding this potential source of
confusion. Searches for \MgIIs absorption systems are by far the most
common because of their established connection to galaxies. \MgIIs
absorber surveys were carried out by \textcite{1987ApJ...322..739L},
\textcite{1987ApJS...64..667T}, \textcite{1988ApJ...334...22S},
\textcite{1992ApJS...80....1S}, \textcite{1992A&A...262..401B},
\textcite{1993AJ....105.2054A}, \textcite{1993A&A...279...33L},
\textcite{1999ApJS..120...51C}, \textcite{2005ApJ...628..637N},
\textcite{2006MNRAS.367..945Y}, \textcite{2006ApJ...640...81L},
\textcite{2006ApJ...643...75N}, \textcite{2006ApJS..165..229F} and
\textcite{2007ApJ...660.1093N}. Surveys of \CIVs absorption systems
were conducted by \textcite{1988ApJS...68..539S} and
\textcite{2002AJ....123.1847M}. Absorption features by \OVIs
$\lambda\lambda1032, 1038$\AA\ are more difficult to identify since
they lie within the \Lya forest, and so are not easily distinguished
from the forest. Searches have been successful, however, and have
moreover discovered a second IGM environment. While at high redshifts
the \OVIs systems probe the component of the IGM containing \CIVs
\cite{1996ApJ...460..584B, 2002ApJ...573..471B, 1997ApJ...489L.123K},
at low redshifts they are associated with a warm-hot intergalactic
medium (WHIM) \cite{2000ApJ...542...42T, 2002ApJ...564..631S,
2004ApJS..153..165R, 2005ApJ...624..555D, 2008ApJ...679..194D,
2008ApJS..177...39T}, that is most likely collisionally ionized rather
than photoionized. A search for \OVIs absorption systems as an
indicator of a WHIM at high redshift was conducted by
\textcite{2002ApJ...578..737S}. Efforts are currently underway to
detect an intergalactic x-ray absorption line forest using the {\it
XMM-Newton} and {\it Chandra} x-ray satellites
\cite{2005ApJ...629..700N, 2006ApJ...652..189K,
2007ApJ...656..129R}. Detections of low redshift \OVIIIs have been
reported by \textcite{2002ApJ...572L.127F, 2007ApJ...670..992F} and
\textcite{2007ApJ...665..247W}.

Attempts to infer the origin of the metals, whether distributed by
galactic outflows or introduced locally by Population III stars, have
led to various efforts to estimate the range in ambient metallicities
and to determine their spatial distribution \cite{1986A&A...169....1B,
1994ApJ...436...33B, 1995AJ....109.1522C, 1996AJ....112..335S,
1998Natur.394...44C, 1998AJ....115.1725B, 2000AJ....120.1175E,
2001ApJ...561L.153S, 2004A&A...419..811A, 2005AJ....130.1996S,
2006AJ....131...24S, 2006ApJ...638...45P, 2006ApJ...637..648S,
2006MNRAS.365..615S}. Updated solar abundances useful for modeling the
abundances of the metal absorption systems are provided by
\textcite{1998SSRv...85..161G}, \textcite{2001AIPC..598...23H},
\textcite{2001ApJ...556L..63A, 2002ApJ...573L.137A} and
\textcite{2003astro.ph..2409A}. Early attempts to tease out weak metal
features were based either on a spectral shift-and-stack approach to
generate a high signal-to-noise ratio composite \CIVs line
\cite{1998astro.ph..2189L} or through a pixel-by-pixel statistical
analysis of the spectra \cite{1996AJ....112..335S,
1998Natur.394...44C}. A comparison of these two techniques is provided
by \textcite{2000AJ....120.1175E}.

An alternative is to perform long integrations of bright QSOs to
obtain high resolution, high signal-to-noise ratio spectra. With
signal-to-noise ratios of 200--300 per spectral resolution element,
\textcite{2004ApJ...606...92S} detected metal lines in 70\% of the
\Lya forest systems. Absorption systems with \HI\ column densities as
low as $N_{\rm HI}>10^{14}\,{\rm cm}^{-2}$ show \CIVs features, while
systems with $N_{\rm HI}>4\times10^{13}\,{\rm cm}^{-2}$ show \OVIs
features. Although the inferred metallicities are model dependent
(fixed in part by the assumed spectrum and intensity of the ionizing
background), the measurements suggest metallicities as low as
$3\times10^{-4}$ that of the Sun. No evidence for a metallicity floor
could be discovered. The metallicity distribution inferred is
consistent with a lognormal distribution with a mean of 0.006 solar.

The mechanism and degree of metal mixing in the diffuse IGM are
unclear. While the mixing of metals in stellar interiors by convection
and diffusion may produce a homogeneous composition, preserved in
supervovae ejecta and winds, it is far from clear that the mixing of
the ejecta will result in uniform metallicity. The mixing process is
still poorly understood in the Interstellar Medium, and much more so
in the IGM. Just as stirring a drop of milk in a cup of coffee will
result in the mixing and intermingling of the two fluids, so may the
stirring produced by dynamical instabilities such as the
Kelvin-Helmholtz and Rayleigh-Taylor mix metal enriched stellar ejecta
with the surrounding primordial gas. Insufficient mixing, however,
rather than a uniform mixture, will instead result in striations or
patches of high metallicity gas entrained within the primordial gas of
essentially zero metallicity. \textcite{2007MNRAS.379.1169S} find
evidence for just such inadequate mixing for diffuse absorption
systems (with a median upper limit of 13.3 on $\log_{10} N_{\rm HI}$),
at $z\simeq2.3$. By combining \CIVs measurements with upper limits on
\HI, \CIII, \SiII, \SiIII, \NV\ and \OVI, they place robust constraints
on the metallicities and physical properties of the \CIVs absorption
systems, with a median lower limit on the metallicities of ${\rm
[C/H]}>-0.42$ and a median upper limit on the absorber sizes of
$R\lta1.5$~kpc. They obtain typical median lower and upper limits on
the hydrogen densities of $n_{\rm H}>100\,{\rm m^{-3}}$ and $n_{\rm
H}<1000\,{\rm m^{-3}}$, a range that suggests a cloud size of about
100~pc. The clouds, however, would likely be transient, either quickly
dispersing if freely expanding or shorn apart by dynamical
instabilities if moving relative to a confining medium, on a timescale
of about $10^7$~yrs. \textcite{2007MNRAS.379.1169S} arrive at a
picture in which new clouds are continuously created through dynamical
or thermal instabilities. In this picture, nearly all the metals in
the IGM could be processed through just such a cloud phase. The
dispersed clouds would retain their coherence as small patches, but of
too low column density to be detectable. The resulting IGM metallicity
would persist as patchy, not mixed.

\begin{figure}
\includegraphics[width=8cm, height=6cm]{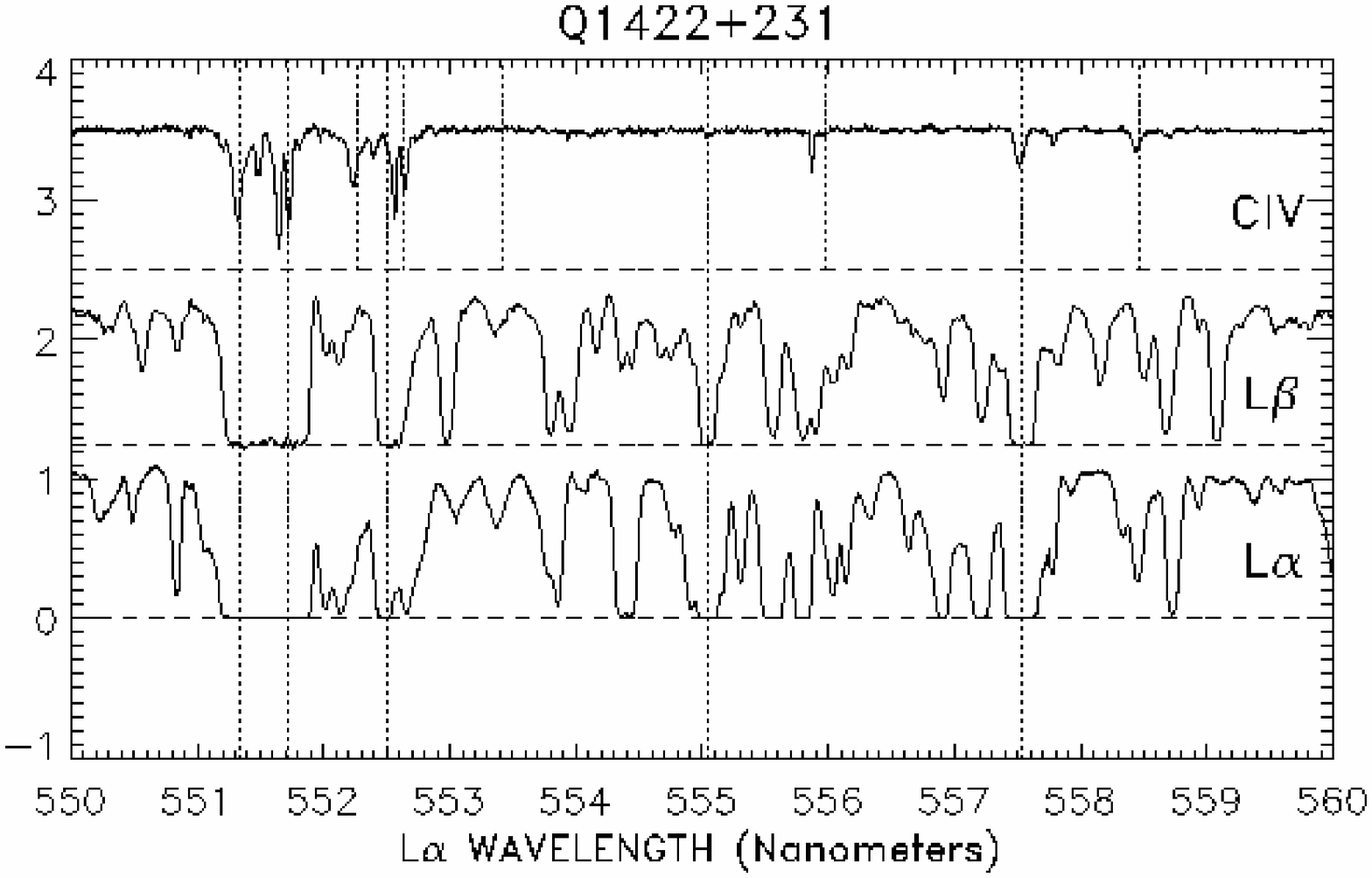}
\caption{Portion of the \Lya forest in the spectrum of Q$1422+231$ at
$z=3.6$ (lower panel) with the corresponding \Lyb systems (middle
panel) and $\lambda1548$\AA\ systems of the \CIVs$\lambda\lambda1548,
1551$~\AA\ doublet (upper panel) shifted to coincide. Absorption
systems saturated in both \Lya and \Lyb are indicated by the long
vertical dotted lines. The short dotted lines show where the
corresponding \CIVs $\lambda1551$\AA\ line should occur for these
systems. From \textcite{1996AJ....112..335S}.
(Figure reproduced by permission of the AAS.)
}
\label{fig:civhi}
\end{figure}

Matching the metal absorption lines to the corresponding \HI\ \Lya
features frequently reveals multicomponent substructure obscured by
the width of the \Lya feature. An example from QSO~$1422+231$ is shown
in Fig.~\ref{fig:civhi}. The \CIVs systems show evidence for spatial
clustering on scales of up to $\sim300\kms$ \cite{1996ApJ...467L...5R,
1996AJ....112..335S, 1998A&AS..127..217D, 2003ApJ...597L..97P}. This
is consistent with a $2\sigma$ lower limit of $\sim30$~kpc for the
clustering of \CIVs and \MgIIs systems obtained by
\textcite{1995A&AS..113..199S} based on a statistical analysis of
coincident metal systems in neighboring lines of sight to a lensed
QSO. An analysis of a triply imaged quasar at $z=3.9$ by
\textcite{2004A&A...414...79E} shows that high ionization systems like
\CIVs and \SiIVs absorbers retain coherence over scales of
$\sim100$~kpc. By contrast, low ionization systems like \MgIIs and
\FeIIs absorbers show coherence only over much smaller scales. Assuming
a spherical geometry, they are estimated to have sizes of 1.5--4.4~kpc
(95\% c.i.). Using similar measurements, \textcite{2001ApJ...554..823R}
find \CIVs systems at $1.5\lta z\lta3$ are coherent over scales of
about 300~pc, with column densities along neighboring lines of sight
differing by 50\% for separations of about 1~kpc. Detectable levels
of \CIVs extend over scales of several kiloparsecs.

Damped \Lya Absorbers show a wide range of metals with both low and
high ionization states, including boron, carbon, nitrogen, oxygen,
magnesium, aluminum, silicon, phosphorus, sulfur, chlorine, argon,
titanium, chromium, manganese, iron, cobalt, nickel, copper, zinc,
germanium, arsenic and krypton \cite{1996ApJS..107..475L,
2005ARA&A..43..861W, 2006A&A...445...93D}. The metallicities evolve
from about 0.03 solar at $z\simeq4$ to 0.15 solar at $z\simeq1$, with
few above 0.3 solar \cite{2005ARA&A..43..861W}.  No DLA is found with
a metallicity smaller than 0.0025 solar, distinguishing them as a
population from the \Lya forest. By contrast, super Lyman Limit
Systems are found to have metallicities ranging from 0.1 solar to
super-solar abundances with as much as 5 times solar in zinc
\cite{2007ApJ...661...88K}. As both these classes of systems are
likely related to galaxies, further discussion of them is deferred to
\S~\ref{sec:abs-gal}.

The presence of the metal lines in absorbers of all \HI\ column
densities in principle offers an opportunity to obtain a nearly
model-free measurement of the gas temperature. Allowing for kinetic
broadening of the Doppler parameters following Eq.~(\ref{eq:bbparam}),
the measurements of the Doppler parameters from features corresponding
to two or more elements permits the thermal and kinematic broadening
components to be separated.  (In principle, instrumental broadening
could be incorporated into the kinematic term.) Given total Doppler
parameter measurements $b_i$ and $b_j$ for two species $i$ and $j$ of
masses $m_i$ and $m_j$, the temperature may be solved for as
\begin{equation}
T=\frac{m_i m_j}{2k_{\rm B}}\frac{b_i^2-b_j^2}{m_j-m_i}.
\label{eq:Tfromb}
\end{equation}

Applying this method to \CIVs and \SiIVs features detected in three
high redshift QSOs, and assuming that the \CIVs and \SiIVs features
trace the same systems with the same temperature and velocity
structure, \textcite{1996ApJ...467L...5R} obtain a typical temperature
of $T\approx38000$~K and typical kinetic motions of $b_{\rm
kin}\approx6\kms$ for a select subsample of especially well-fit
features. The large \CIVs column densities of the systems (typically
exceeding $10^{12}\,{\rm cm^{-2}}$, M. Rauch, personal communication),
suggest large \HI\ column densities of $N_{\rm
HI}>10^{15}\,{\rm cm^{-2}}$ \cite{2004ApJ...606...92S}.

A few other \Lya systems selected for quiescent velocity fields as
part of a program of deuterium abundance measurements yield somewhat
cooler temperatures. The Doppler parameters of \CIVs and \SiIVs
measured by \textcite{2000ApJ...529..655K} in a system with
$\log_{10}N_{\rm HI}=16.7$ yields a temperature of
$T\approx30000-32000$~K. For a Lyman Limit System with
$\log_{10}N_{\rm HI}=17.1$ at $z=0.7$ \cite{1999AJ....117...63T}, the
Doppler parameter measured for \MgIIs compared with that of \HI\ gives
$T\approx31000$~K and $b_{\rm kin}\approx13\kms$, assuming the
hydrogen and metals sample the same velocity field, which is not an
obvious assumption to make. For a super Lyman Limit System with
$\log_{10} N_{\rm HI}=19.42$, \textcite{2001ApJ...552..718O} find from
a simultaneous fit to \OI, \NIs and \HI\ $T=1.15\pm0.02\times10^4$~K
and $b_{\rm kin}=2.56\pm0.12\kms$. In all cases, the temperatures
obtained are consistent with photoionization heated gas. Although
direct gas density measurements are not possible, models generally
suggest that higher column density systems correspond to higher
densities, especially at these high values. The temperatures support
the expected trend of decreasing temperature with increasing density
for sufficiently dense absorption systems \cite{Meiksin94}.

The small kinetic contributions to the line broadening are in contrast
to what is found for \HeII\ absorption systems, which appear to be
dominated by kinetic broadening \cite{Zheng04}. The \HeII\ features,
however, were chosen for their clear isolation in the spectrum, and so
may probe physically underdense regions, in contrast to the higher
column density systems above. An attempt to estimate temperatures from
a larger sample of \HeII\ and \HI\ systems has not been able to
produce secure values, but nonetheless shows that only $\sim45\%$ of
the lines are dominated by kinematic broadening
\cite{2007A&A...463...69F}.

Intergalactic metal absorption systems may be used to probe the time
variability of fundamental physical quantities over cosmological
timescales. The excitation of ionic fine-structure levels by the CMB
allows metal lines in the \Lya forest to be exploited for measuring the
evolution of the CMB temperature \cite{1968ApJ...152..701B}. Because
of the possible influence of other excitation mechanisms, the
measurements are strictly speaking upper limits. Existing excitation
temperature measurements are $T=7.4\pm0.8$~K at $z=1.776$
\cite{1994Natur.371...43S}, $T=7.9\pm1.0$~K at $z=1.9731$
\cite{1997ApJ...474...67G}, $T=15.7\pm3.5$~K at $z=2.34$
\cite{2001ApJ...547L...1G}, and $T=12.1_{-3.2}^{+1.7}$~K (95\% c.i.)
at $z=3.025$ \cite{2002A&A...381L..64M}. These values are all
consistent with a linearly evolving CMB temperature of
$T(z)=T_0(1+z)$, expected for a homogeneous and isotropic expanding
universe, with $T_0=2.725$~K \cite{1999ApJ...512..511M}.

Detected resonance line multiplets have been used to place constraints
on the time variation of fundamental
constants. \textcite{1967ApJ...149L..11B} used the fine structure
wavelength splittings of the \SiIIs lines near $\lambda1260$\AA\ and
$\lambda1527$\AA\ and the \SiIVs lines near $\lambda1394$\AA\ to show
that the fine structure constant ($\alpha=[1/4\pi\epsilon_0]e^2/\hbar
c$) varies by less than 5\% between $z=0$ and $z=1.95$. Tentative
evidence for an evolving fine structure constant between $1<z<1.6$ of
$\Delta\alpha/\alpha\simeq(-1.9\pm0.5)\times10^{-5}$ was reported by
\textcite{1999PhRvL..82..884W} using measurements of \MgIIs and \FeIIs
transitions. The value was later revised to
$\Delta\alpha/\alpha=(-5.7\pm1.1)\times10^{-6}$ over the redshift
range $0.2<z<4.2$\cite{2004LNP...648..131M}, suggestive of a positive
detection of a variation. There is by no means concensus on the
detection. \textcite{2004A&A...417..853C} and
\textcite{2004PhRvL..92l1302S} report
$\Delta\alpha/\alpha=(-0.6\pm0.6)\times10^{-6}$ over the redshift
range $0.4<z<2.3$. Their analysis method has been contested by
\textcite{2007PhRvL..99w9001M} and \textcite{2008MNRAS.384.1053M}, and
defended by \textcite{2007PhRvL..99w9002S}. In an independent effort
using a modified technique designed to eliminate some systematics,
\textcite{2007A&A...466.1077L} obtain
$\Delta\alpha/\alpha=(5.4\pm2.5)\times10^{-6}$ between $z=0$ and
$z=1.7$. Including the hyperfine splitting of the hydrogen 21cm line
permits constraints on the variation of $x\equiv \alpha^2 g_p\mu$,
where $\mu$ is the electron to proton mass ratio and $g_p$ is the
proton gyromagnetic ratio. The current best limits are, for a linear
fit over the redshift range $0.2<z<2.4$, $\dot x/x =
(-1.43\pm1.27)\times10^{-15}\,{\rm yr}^{-1}$
\cite{2005PhRvL..95d1301T}. Combining \HI\ 21cm data and OH 18cm data,
\textcite{2005PhRvL..95z1301K} obtain a limit on the effective
combination $F\equiv g_p(\alpha^2/\mu)^{1.57}$ of $\Delta
F/F=(0.44\pm0.36^{\rm stat}\pm1.0^{\rm syst})\times10^{-5}$ between
$z=0$ and $z\approx0.7$.

\subsection{Direct flux statistics}
\label{subsec:fluxstats}

While the absorption line parameters relate most directly to the
physical characteristics of the absorption systems, measuring similar
parameters in numerical simulations can be very computationally
demanding, particularly when building up sufficient statistics to make
such a comparison meaningful. More straightforward comparisons may be
made using the flux statistics directly. The statistical measures in
most use are the mean transmitted flux, the transmitted flux per pixel
probability distribution, the flux power spectrum and its Fourier
transform, the flux autocorrelation function.  An additional analysis
based on wavelets permits a quick means of gauging the distribution of
widths of the absorption features.

\subsubsection{Mean transmitted flux}
\label{subsubsec:transmission}

By far the simplest statistic is the mean transmitted flux through the
IGM. \textcite{1982ApJ...255...11O} introduced two measurements,
$D_A$, the mean flux of a background source removed by scattering
between the \Lya and \Lyb transitions in the restframe of the QSO (but
avoiding emission line wings), and $D_B$, the mean flux removed
between \Lyb and the Lyman edge. These definitions are
straightforwardly generalized to the fraction of flux removed between
each successive pair of Lyman transitions. Removing the effect on
higher orders by lower orders, however, is not straightforward but may
be accomplished in a statistical sense \cite{1993ApJ...414...64P}. The
measurement $D_A$ may be evaluated as $D_A=\langle 1 -
f_i/c_i\rangle$, where $f_i$ is the measured flux in pixel $i$, $c_i$
is the corresponding predicted continuum of the background object,
normally a QSO, and the average is carried out over a given wavelength
(or corresponding redshift) interval between \Lya and \Lyb in the
restframe of the QSO. Considerable care must be taken in measuring the
contribution at a given redshift from \Lya alone. Systematics that
would bias the result if not corrected for include:\ 1.\ a low \HI\
fraction near the QSO due to its own ionizing intensity (the
``proximity effect''), 2.\ contamination by metal absorption systems,
3.\ the large stochastic influence of Lyman Limit Systems and Damped
\Lya Absorbers, the latter of which produce a divergent mean
\cite{MW04}, and 4.\ evolution of the absorption in the region
measured, including over individual pixels of the spectrum. A detailed
discussion is provided by \textcite{2004ApJ...617....1T}.

For a spectrum that may be decomposed into individual absorption
features with equivalent widths $w({\bf x})$ depending on some set of
parameters ${\bf x}$, such as column density or Doppler parameter, the
mean of the transmitted flux ${\cal T_\nu}\equiv\exp(-\tau_\nu)$ is
given by
\begin{equation}
\langle{\cal T_\nu}\rangle=\langle \exp(-\tau_\nu)\rangle = \exp(-\tau_l),
\label{eq:transmission-lines}
\end{equation}
with the effective optical depth $\tau_l$ due to the lines given by
\begin{eqnarray}
\tau_l &=& \int\, d{\bf x} w({\bf x})
\frac{d{\cal N}({\bf x})}{d\lambda}\nonumber\\
&=& \frac{1+z}{\lambda_0}\int\, dw \frac{\partial^2{\cal N}}{\partial w\,
\partial z}w,
\label{eq:taul}
\end{eqnarray}
where $d{\cal N}({\bf x})/d\lambda$ is the number of systems per unit
rest wavelength per unit volume $d{\bf x}$ in parameter space
\cite{1948ApJ...108..276S, 1993ApJ...414...64P}, here represented in
terms of the number per (proper) equivalent width per unit redshift
$\partial^2{\cal N}/{\partial w\,\partial z}$. The variance $\langle
{\cal T}^2\rangle - \langle {\cal T}\rangle^2$ of the transmission
depends on the details of the line profiles. Expressions for idealized
examples are provided by \textcite{1968S&SS....7..365M} and
\textcite{1993ApJ...414...64P}.

It is instructive to consider a comparison between Eq.~(\ref{eq:taul})
and the Gunn-Peterson optical depth (\ref{eq:tauGP}). Using
Eq.~(\ref{eq:dNdz}) for the number density evolution given by
$dN/dz=\int\,dw {\partial^2{\cal N}}/{\partial w\, \partial z}$ and
Eq.~(\ref{eq:wlin}) for systems on the linear part of the
curve-of-growth, $\tau_l$ may be recast in the form of the
Gunn-Peterson optical depth
\begin{equation}
\tau_l=\frac{3}{8\pi}\frac{\lambda_{lu}^3\Gamma_{ul}
\langle n(z)\rangle}{H(z)},
\label{eq:taul-GP}
\end{equation}
for a mean distributed scattering gas density $\langle n(z)\rangle=
8\pi f_{lu} n_a(z)\sigma_a(z){\bar N}(z) = 8\pi f_{lu}Q_a(z){\bar
n}(z)$, where ${\bar N}={\bar n}L$ is the mean column density of the
absorption systems with mean atomic density $\bar n$ and line of sight
thickness $L$, and $Q_a=n_a\sigma_a L$ is the porosity of the
absorbers, equivalent to the spatial filling factor of the absorption
systems for $Q_a<1$. For a column density distribution varying like
$dN/dN_{\rm HI}\propto N_{\rm HI}^{-1.5}$, the effective absorber
optical depth (excluding DLAs) is dominated by saturated systems, with
$F(\tau_0)\approx 1-2$. In this case, Eq.~(\ref{eq:wFtau0}) gives
$w/\lambda_0=2(b/c)F(\tau_0)\approx 3b/c$. If the absorber velocity
widths scale like the Hubble expansion across them, $b\approx fHL$,
Eq.~(\ref{eq:taul}) gives instead of Eq.~(\ref{eq:taul-GP}),
\begin{equation}
\tau_l \simeq 2Q_a(z)f(z)F(\tau_0) \simeq 3Q_a(z)f(z).
\label{eq:taul-sat}
\end{equation}

In practice the effective optical depth will scale somewhat between
the linear and saturated line limits. It will not in general scale
like the Gunn-Peterson optical depth Eq.~(\ref{eq:tauGP}), as is often
asserted in the literature. For $dN/dz\propto(1+z)^\gamma$,
Eq.~(\ref{eq:taul}) instead gives the dependence
$\tau_l\propto(1+z)^{1+\gamma}$. This is corroborated by estimates of
the mean \HI\ transmission due to \Lya scattering from all components,
whether or not they may be modeled as individual Voigt profile
absorption systems. More generally, the mean transmission may be
expressed as
\begin{equation}
\langle{\cal T_\nu}\rangle \equiv \exp(-\tau_{\rm eff}).
\label{eq:transmission}
\end{equation}
From an analysis of \Lya scattering by the IGM in the redshift range
$2.5<z<4.3$ using moderate resolution spectra,
\textcite{1993ApJ...414...64P} obtained $\tau_{\rm
eff}=A(1+z)^{1+\gamma}$ with $A=0.0175-0.0056\gamma\pm0.0002$ and
$\gamma=2.46\pm0.37$. The evolution exponent $\gamma$ agrees closely
with the finding of \textcite{2002MNRAS.335..555K} for $dN/dz$
above. A similar result was obtained by \textcite{2001A&A...373..757K}
using high resolution spectra:\ $A=0.0144-0.00471\gamma$ and
$\gamma=2.43\pm0.17$ (as inferred by
\textcite{MW04}). \textcite{2002MNRAS.335..555K} give the slight
variation based on an expanded data set of $A=0.0032\pm0.0009$ and
$\gamma=2.37\pm0.20$. Because the implied probability distribution for
$\langle {\cal T}\rangle$ is skewed, the statistical expectation value
for the mean transmission must be computed from these expressions for
$\tau_{\rm eff}$ using Monte Carlo integration \cite{MW04}. A good fit
to the evolution of $\tau_{\rm eff}$ based on the data compiled by
\textcite{MW04}, extended to lower redshifts using the data of
\textcite{2007MNRAS.376.1227K}, is given by
\begin{eqnarray}
\tau_{\rm eff} &=& 0.0164(1+z)^{1.1};\quad 0 < z < 1.2\\
&=& 0.00211(1+z)^{3.7};\quad 1.2 < z < 4\nonumber\\
&=& 0.00058(1+z)^{4.5};\quad 4 < z < 5.5\nonumber
\label{eq:taul_ev}
\end{eqnarray}
\cite{2006MNRAS.365..807M}. The evolution is found to vary from
relatively slowly at low redshifts, as expected for discrete saturated
features, to the more rapid evolution of the Gunn-Peterson effect at
high redshifts as the porosity of the absorption systems approaches
unity and increasingly diffuse gas dominates the effective optical
depth.

A dip in the evolution of $\tau_{\rm eff}$ at $z\simeq3.2$ was
reported by \textcite{2003AJ....125...32B} based on a sample of 1061
moderate-resolution QSO spectra measured by the SDSS. The feature was
interpreted as evidence for sudden heating due to the onset of \HeII\
reionization. This interpretation is not supported by recent
simulations \cite{2008arXiv0807.2799M}. The feature was not confirmed
by \textcite{2006ApJS..163...80M} using an alternative analysis method
applied to 3035 SDSS QSO spectra. Using a sample of 86 high-resolution
high signal-to-noise ratio spectra, however,
\textcite{2008ApJ...681..831F} find a feature similar to that reported
by \textcite{2003AJ....125...32B}. The existence of the feature
appears still not to be entirely secure, but tentatively suggests that
a physical change in the state of the IGM occurred at $z\simeq3.2$,
possibly reflecting a change in the \HI\ ionization rate, perhaps
owing to the growing influence of QSOs (\S~\ref{subsubsec: sources}).

\subsubsection{Statistics based on pixel fluxes}
\label{subsubsec:PFS}

More detailed transmission information may be extracted from the use
of the full spectrum, including the probability distribution of the
transmission per pixel, the flux power spectrum, the autocorrelation
function of the transmission, and a wavelet decomposition. Because
these statistical measures are best interpreted in conjunction with
numerical simulations, they are discussed in more detail in
\S~\ref{sec:simuls}. Here they are only defined.

The pixel transmission distribution (PTD) and its variants, like the
transmission flux probability distribution function (TFPDF)
\cite{2000ApJ...543....1M}, simply quantify the frequency of
occurrence of a given flux value in each pixel of the spectrum,
normalized by the continuum, often expressed in frequency bins. It was
introduced by \textcite{1991ApJ...376...33J} to search for underlying
intergalactic absorption in excess of the amount accounted for by
discrete absorption systems. Several tabulations are available in the
literature:\ \textcite{1997ApJ...489....7R},
\textcite{2000ApJ...543....1M}, \textcite{2007ApJ...662...72B},
\textcite{2007MNRAS.374..206D} and \textcite{2007MNRAS.376.1227K}.

Second order statistics include the flux autocorrelation function
and the flux power spectrum, in analogy to similar statistical
descriptions of the clustering of galaxies. The flux transmission
autocorrelation function was introduced by \textcite{1994ApJ...423...73Z}
to investigate properties of the \Lya forest. For a velocity
separation $v$, it is defined by
\begin{equation}
\xi_{\rm F}(v) = \langle f(v^\prime)f(v^\prime-v)\rangle/\langle f\rangle^2-1,
\label{eq:xiF}
\end{equation}
where $f(v)$ denotes the pixel flux at velocity $v$, and $\langle
f\rangle$ is the measured mean flux at the relevant
redshift.\footnote{This differs somewhat from the definition of
\textcite{1994ApJ...423...73Z}, who in place of $\langle f\rangle^2$
in the denominator, used the more general form $\langle
f(v^\prime)\rangle\langle f(v)\rangle$. In most of the literature, it
is assumed the mean flux changes negligibly over the velocity
separations for which it is computed. This will break down for
sufficiently large velocity separations that cosmological expansion
becomes a factor, but in this case the physical interpretation of the
autocorrelation function becomes less transparent as it would
correlate transmissions at different cosmological epochs.} Several
estimates of $\xi_{\rm F}$ are provided in the literature:\
\textcite{1994ApJ...423...73Z}, \textcite{2000MNRAS.311..657L},
\textcite{2000ApJ...543....1M}, \textcite{2002ApJ...581...20C},
\textcite{2004ApJ...613...61B}, \textcite{2006MNRAS.372.1333D},
\textcite{2006ApJS..163...80M} and \textcite{2007MNRAS.376.1227K}. The
results generally show significant correlations over scales of a few
Megaparsecs (comoving). Similar to the flux autocorrelation function
is the flux cross-correlation function along neighboring lines of
sight to QSOs near each other on the sky \cite{2000MNRAS.311..657L,
2004ApJ...613...61B, 2006MNRAS.372.1333D}. Cross-correlations of
similar strength to the autocorrelations are found, extending over
similar spatial scales.

The flux power spectrum has received considerable attention as a means
of constraining cosmological parameters
\cite{1998ApJ...495...44C,2001ApJ...552...15H}. It is a measure of the
variance in the amplitude of the Fourier transform coefficients of the
transmitted flux. Defining $\delta f=f-\langle f\rangle$ and its
Fourier transform
\begin{equation}
{\hat{\delta f}}(k)=\frac{1}{\Delta v}\int_{-\Delta v/2}^{\Delta
v/2}\,dv \delta f(v)e^{ikv},
\label{eq:FTflux}
\end{equation}
over the velocity interval $\Delta v$, the flux power spectrum is
\begin{equation}
P_{\rm F}(k)=(\Delta v)\langle\vert{\hat{\delta f}}\vert^2
\rangle/\langle f\rangle^2.
\label{eq:PFk}
\end{equation}
The dimensionless form $kP_{\rm F}(k)/\pi$ is
often more convenient. It forms a Fourier transform pair with the
flux autocorrelation function through
\begin{equation}
\frac{kP_{\rm F}(k)}{\pi} =
\frac{k}{\pi}\int_{-\Delta v/2}^{\Delta v/2}\,dv\xi_F(v)e^{ikv}.
\label{D2FxiF}
\end{equation}
The reader should be aware that several different conventions for
$P_{\rm F}(k)$ exist in the literature, some without a mean flux
normalization, some differing by a factor of $2\pi$, and some
referring to an effective 3D flux power spectrum extracted from the 1D
flux power spectrum in a manner analogous to the relation between the
1D and 3D mass power spectra. Many tabulations of $P_{\rm F}(k)$ are
available:\ \textcite{1998ApJ...495...44C},
\textcite{1999ApJ...520....1C}, \textcite{2000ApJ...543....1M},
\textcite{2002ApJ...581...20C} and \textcite{2004MNRAS.347..355K,
2004MNRAS.351.1471K}. A substantial effort was made to measure $P_{\rm
F}(k)$ from the spectra of high redshift quasars discovered by the
Sloan Digital Sky Survey \cite{2006ApJS..163...80M}.

A shortcoming of the Fourier transform is that it loses spatial
information. In the case of a spectrum showing the \Lya forest, the
positions of individual absorption features are lost. A wavelet
transform is analogous to a Fourier transform, but retains local
information, giving in effect a description of the power on different
scales as a function of position. A wavelet is a square-integrable
function, $\psi(x)$, defined in real space such that $\psi_{jk}\equiv
2^{j/2}\psi(2^j x - k)$ (where $j$ and $k$ are integers) forms an
orthonormal basis for the set of square-integrable functions. It
satisfies $\int_{-\infty}^{\infty}dx\,\psi(x)=0$, and is normally
concentrated near $x=k2^{-j}$. The wavelet coefficients are defined by
$w_{jk}=\int\, dx f(x)\psi_{jk}(x)$; the original function may be
recovered through $f(x)=\sum_{j,k} w_{jk}\psi_{jk}(x)$. The wavelet
transform effects a localized multiscale rendition of the
data. Although not commonly used, wavelets are in principle a powerful
tool for analysing the \Lya forest, producing a statistical
description similar to the distribution of $b$-values. The discrete
wavelet transform, like the fast Fourier transform, may be computed
extremely rapidly, permitting a large number of spectra to be analyzed
much more quickly than by performing a full absorption line
analysis. This is an advantage over full Voigt profile line fitting
for assessing the match of simulations to observations
\cite{2000MNRAS.314..566M, 2000MNRAS.317..989T, MW01, MBM01, Theuns02,
2002ApJ...564..153Z}.

\section{Ionization Equilibrium}
\label{sec:ionization}

\subsection{Ionization}
\label{subsec:ionization}

\subsubsection{Hydrogen and helium}
\label{subsubsec:pi-HHe}

The absorption signatures of the IGM are produced by atoms and ions
with bound electrons, either through resonance scattering or
photoelectric absorption. The amount of scattering or absorption
depends on the ionization structure of the IGM. In this section, the
photoionization of hydrogen and helium is discussed. In hot
environments like galaxy clusters, collisional ionization can
dominate. Collisional ionization equilibrium may play a particularly
important role in the physics of the WHIM \cite{1999ApJ...514....1C,
  1999ApJ...511..521D, 2001ApJ...552..473D, 2006ApJ...650..560C} and
is treated more generally in the section below on metal ionization.

Photons with energies exceeding the ionization potential of a
bound electron in a hydrogen atom will ionize the neutral hydrogen
atoms in the IGM at the rate per neutral atom
\begin{equation}
\Gamma_{\rm HI} = c\int_{\nu_{\rm T}}^\infty\,d\nu
\frac{u_\nu}{h_{\rm P}\nu}a_\nu^{\rm HI},
\label{eq:GHI}
\end{equation}
where $a_\nu^{\rm HI}$ is the hydrogen photoelectric cross section,
$\nu_{\rm T}$ is the threshold frequency required to ionize hydrogen
(the Lyman limit), and $u_\nu$ is the specific energy density of the
ambient radiation field. The specific energy density is related to the
specific intensity of the radiation field $I_\nu({\bf r},t,{\hat{\bf
n}})$ by $u_\nu=4\pi J_\nu/c$, where $J_\nu({\bf
r},t)=(1/4\pi)\oint\,d{\bf\Omega} I_\nu({\bf r},t,{\hat{\bf n}})$ is
the angle-averaged specific intensity.

Free electrons will be radiatively captured by the protons at the rate
per proton $n_e\alpha_{\rm HII}(T)$, where $\alpha_{\rm HII}$, often
referred to as the Case A radiative recombination coefficent
$\alpha_A(T)$, is the total rate coefficient for radiative capture
summed over recombinations to all energy levels. Expressed in terms of
the ionization fractions $x_{\rm HI}=n_{\rm HI}/ n_{\rm H}$ and
$x_{\rm HII}=n_{\rm HII}/ n_{\rm H}$, where $n_{\rm H}= n_{\rm HI} +
n_{\rm HII}$ is the total hydrogen number density, the ionization rate
equations become
\begin{eqnarray}
\frac{dx_{\rm HI}}{dt} &=& -x_{\rm HI}\Gamma_{\rm HI}
+x_{\rm HII}n_e\alpha_A(T), \nonumber\\
\frac{dx_{\rm HII}}{dt} &=& -\frac{dx_{\rm HI}}{dt}.
\label{eq:phionizeH}
\end{eqnarray}
The time-derivatives are lagrangian, so that Eqs.~(\ref{eq:phionizeH})
are valid in the presence of velocity flows.

In equilibrium, the neutral fraction will be
\begin{equation}
x_{\rm HI}^{\rm eq}=\frac{n_e\alpha_A(T)}{\Gamma_{\rm HI} + n_e\alpha_A(T)}.
\label{eq:phionizeHI-eq}
\end{equation}
For a pure hydrogen gas, $n_e=n_{\rm HII}$, and the rate equation
for $x_{\rm HI}$ becomes
\begin{equation}
\frac{dx_{\rm HI}}{dt}-t_{\rm rec}^{-1}x_{\rm HI}^2
+(\Gamma_{\rm HI} + 2t_{\rm rec}^{-1})x_{\rm HI}-t_{\rm rec}^{-1}=0,
\label{eq:phionize-xiHI}
\end{equation}
where $t_{\rm rec}=1/(n_{\rm H}\alpha_A)$ is the total radiative
recombination time for the gas.
The equilibrium neutral fraction may be expressed as
\begin{equation}
x_{\rm HI}^{\rm eq}=\left(\frac{1}{2}\phi + 1\right)
-\left[\left(\frac{1}{2}\phi + 1\right)^2-1\right]^{1/2},
\label{eq:phionize-xiHI-eq}
\end{equation}
where $\phi=\Gamma_{\rm HI}t_{\rm rec}$ denotes the number of
photoionizations per neutral atom over a recombination time. It is
instructive to consider the case of constant $\phi$.
Eq.~(\ref{eq:phionize-xiHI}) then has the general solution
\begin{equation}
x_{\rm HI} = x_{\rm HI}^{\rm eq} -
\frac{\eta}{\left(1 - \delta_0^{-1}\eta\right)\exp(\eta\tau) - 1},
\label{eq:phionizeH-sol}
\end{equation}
where $\eta = \phi + 2 - x_{\rm HI}^{\rm eq}$, $\tau=t/t_{\rm rec}$,
and $\delta_0 = x_{\rm HI}(0) - x_{\rm HI}^{\rm eq}(0)$ is the initial
deviation of the neutral hydrogen fraction from its equilibrium
value. For a fast photoionization rate ($\phi\gg1$), $x_{\rm HI}^{\rm
eq}\simeq \phi^{-1}\ll 1$ and $x_{\rm HI}\simeq x_{\rm HI}^{\rm eq} +
\delta_0 \exp(-\Gamma_{\rm HI}t)$. An initial deviation $\delta_0$
from equilibrium decays exponentially. The neutral fraction approaches
its equilibrium value on the photoionization rate timescale
$\Gamma_{\rm HI}^{-1}$, regardless of whether equilibrium is
established through net photoionizations or net recombinations.
Ionization equilibrium will be maintained provided any change to the
gas (such as source variability or the expansion of the gas) occurs on
a time scale long compared with the photoionization time scale. By
contrast, for a slow photoionization rate $\phi\ll1$, $x_{\rm HI}^{\rm
eq}\simeq 1$ and the neutral fraction approaches equilibrium on the
recombination timescale, $x_{\rm HI}\simeq 1 + \delta_0\exp(-t/t_{\rm
rec})$ for $\vert\delta_0\vert\ll1$ ($\delta_0<0$).

The photoionization equations for helium are analogous to
Eqs.~(\ref{eq:phionizeH}). Denoting the respective \HeI, \HeII, and
\HeII\ fractions by $x_{\rm HeI}$, $x_{\rm HeII}$ and $x_{\rm HeIII}$,
they are given by

\begin{eqnarray}
\frac{dx_\HeIs}{dt}  &=& -x_\HeIs \Gamma_\HeIs + x_\HeIIs n_e\alpha_\HeIIs, \nonumber \\
\frac{dx_\HeIIs}{dt}&=& -\frac{dx_\HeIs}{dt} - \frac{dx_\HeIIIs}{dt}, \nonumber \\
 \frac{dx_\HeIIIs}{dt}&=&  x_\HeIIs \Gamma_\HeIIs - x_\HeIIIs n_e\alpha_\HeIIIs, \label{eq:phionizeHe}
\end{eqnarray}
where $\alpha_\HeIIs$ and $\alpha_\HeIIIs$ are the total recombination
rates to all levels of \HeI\ and \HeII, respectively, and
$\Gamma_\HeIs$ and $\Gamma_\HeIIs$ are the respective photoionization
rates. The total electron density is $n_e = n_{\rm HII} + n_{\rm HeII}
+ 2n_{\rm HeIII}$.

\begin{table*}
\caption{Case A recombination ($\alpha_i$ [$\m^3 \s^{-1}$]) and recombination cooling ($\beta_i$ [$\J \m^3 \s^{-1}$]) coefficients.}
\begin{ruledtabular}
\begin{tabular}{rl}
 $\alpha_\HIIs$  & $2.065\expd{-17} T^{-\onehalf} \left(6.414 - \frac{1}{2}\ln\, T
                   + 8.68\expd{-3} T^\onethird \right)$ \\
 $\alpha_\HeIIs$ & $3.294\expd{-17} \left\{\left(\frac{T}{15.54}\right)^\onehalf
                  \left[1+\left(\frac{T}{15.54}\right)^\onehalf \right]^{0.309}
		  \left[ 1+\left(\frac{T}{3.676\expd{7}}\right)^\onehalf
		        \right]^{1.691}
		  \right\}^{-1}
                  +1.9\expd{-9} 
	          \left( 1 + 0.3 e^{\frac{-9.4\expd{4}}{T}} \right)
	          e^{\frac{-4.7\expd{5}}{T}}  T^{-\frac{3}{2}}$ \\
 $\alpha_\HeIIIs$ & $8.260\expd{-17} T^{-\onehalf} \left( 7.107 
                     - \frac{1}{2}\ln\,T
                    + 5.47\expd{-3} T^\onethird \right)$ \\
\hline
 $\beta_\HIIs$   & $2.851\expd{-40} T^{\onehalf} \left(5.914 -\frac{1}{2}\ln\, T
                   + 0.01184 T^{\onethird} \right)$ \\
 $\beta_\HeIIs $ & $1.55\expd{-39} T^{0.3647}
                 +1.24\expd{-26} 
	          \left( 1 + 0.3 e^{\frac{-9.4\expd{4}}{T}} \right)
	          e^{\frac{-4.7\expd{5}}{T}}  T^{-\frac{3}{2}}$ \\
 $\beta_\HeIIIs$ & $1.140\expd{-39} T^\onehalf \left( 6.607 - \frac{1}{2}\ln\, T
                   + 7.459\expd{-3} T^\onethird \right)$ \\
\hline
\label{table.RecombCoef}
\end{tabular}
\end{ruledtabular}
\end{table*}

The Case A radiative recombination coefficient for hydrogenic ions is
provided by \textcite{Seaton59}, from which the coefficients for
$\alpha_\HIIs(T)$ and $\alpha_\HeIIIs(T)$ may be derived. (In general,
the radiative recombination coefficient $\alpha_Z(T)$ for a hydrogenic
ion of atomic number $Z$ is related to that of hydrogen by
$\alpha_Z(T)=Z\alpha_\HIIs(T/Z^2)$.) For elements He and beyond, the
recombination coefficient is the sum of two rates. One is the usual
radiative capture term, in which $e^{-}+X^{n+1}\rightarrow X^n+\gamma$
for a species $X$ with $n$ electrons. The other arises from
dielectronic recombination, in which a fast incoming electron excites
an ion, which becomes trapped in a weakly bound autoionizing state
upon capturing the electron, then decays by the emission of a
resonance line photon:\ $e^{-}+X^{n+1}\rightarrow {X^n}^* \rightarrow
X^n+h_{\rm P}\nu_{ul}$. The probability for emission of the photon is
suppressed by the factor $\exp(-h_{\rm P}\nu_{ul}/k_{\rm B}T)$; usually
excitation is followed by the reverse reaction. A form for the
radiative recombination rate coefficient contribution to
$\alpha_\HeIIs(T)$ for \HeII\ $\rightarrow$ \HeI\ is provided by
\textcite{VF96}, while a form for the dielectronic term is provided by
\textcite{AP73}. The recombination coefficients are listed in
Table~\ref{table.RecombCoef}. Recombination to \HeI\ may also occur
through charge transfer, ${\rm He}^+ + {\rm H}^0 \rightarrow {\rm
He}^0 + {\rm H}^+ + \Delta E$, where $\Delta E$ is the energy
defect. Where hydrogen is highly ionized, this is a negligible effect
compared with radiative recombinations. For partially ionized
hydrogen, however, as for example within a Damped \Lya Absorber, or
the IGM prior to its complete reionization, the effect may play an
important role. The rate coefficient is provided by
\textcite{1996ApJS..106..205K}. The reverse process, charge transfer
ionization, is suppressed by a Boltzmann factor and so is normally
negligible except at high temperatures.

\begin{table*}
\caption{Ionization cross-section parameters used in
Eq.~(\ref{eq.CrossSection}). From \textcite{Osterbrock89}.}
\begin{ruledtabular}
\begin{tabular}{rllll}
 & $a_T [\m^2]$ & $\nu_T$ [Hz] & $\beta$ & $\s$ \\
 \hline
 $a_\HIs$   & $6.30\expd{-22}$ & $3.282\expd{15}$ & 1.34 & 2.99 \\
 $a_\HeIs$  & $7.83\expd{-22}$ & $5.933\expd{15}$ & 1.66 & 2.05 \\
 $a_\HeIIs$ & $1.58\expd{-22}$ & $1.313\expd{16}$ & 1.34 & 2.99 \\
 \hline
\label{table.CrossSection}
\end{tabular}
\end{ruledtabular}
\end{table*}

The exact photoionization cross section for hydrogenic atoms with
atomic number $Z$ in the ground state is \cite{Heitler}
\begin{equation}
a_Z(\nu)=\frac{A_0}{Z^2}\left(\frac{\nu_T}{\nu}\right)^4
\frac{e^{4-(4{\rm tan^{-1}}\epsilon)/\epsilon}}{1-e^{-2\pi/\epsilon}},
\label{eq:sigmaHZ}
\end{equation}
for $h_{\rm P}\nu>h_{\rm P}\nu_T=13.60Z^2$eV, the threshold energy for
ionization, and where $A_0 = (2^9 \pi / 3 e^4) \alpha \pi a_0^2=
6.30 \times 10^{-22}\,{\rm m}^2$, where
$a_0=[4\pi\epsilon_0] \hbar^2/ m_e e^2$ is the Bohr radius,
$\alpha=[1/4\pi\epsilon_0] (e^2/ \hbar c)$ is the fine structure
constant, and $\epsilon=(\nu/\nu_T-1)^{1/2}$. For most practical
applications, however, an approximate form is preferred.  The
ionization cross sections are well-approximated by the form given by
\textcite{Osterbrock89},
\begin{equation}
\label{eq.CrossSection}
a(\nu) = a_T \left[ \beta
\left(\frac{\nu}{\nu_T}\right)^{-s} +
(1-\beta)\left(\frac{\nu}{\nu_T}\right)^{-s - 1}\right] .
\end{equation}
The parameters for hydrogen and helium are listed in
Table~\ref{table.CrossSection}. For results requiring very high
accuracy, more precise fits to the photoionization cross sections are
provided by \textcite{1996ApJ...465..487V}, including results for an
exhaustive list of metals.

\subsubsection{Metals}
\label{subsubsec:pi-metals}

The large number of ionization states of metals results in more
complex processes involved in their photoionization. These include
autoionization and Auger electron production in addition to
dielectronic recombination. In an autoionization, an excited ion with
one or two electrons lying in the photoionization continuum is
produced following a photoionization event in which a deep shell
electron is ejected. The excited ion spontaneously ionizes through a
radiationless transition and the ejection of multiple Auger
electrons. Although there has been considerable progress in accurate
estimates of all the relevant rate coefficients, many remain poorly
determined either experimentally or theoretically, requiring
interpolation (and sometimes extrapolation) from coefficients known
for related species \cite{2001PASP..113.1024F}. These uncertainties
can affect the predicted metal abundances in the absorption
systems. The effect of uncertainty in the dielectronic recombination
rates on commonly measured metal features is discussed by
\textcite{2000ApJ...533..106S}. Some of the methods of approximation
are described in the literature for the rate coefficients (see below).

Sophisticated computer codes have been developed to solve for the
equilibrium ionization states. A widely used publicly-distributed code
is CLOUDY \cite{1998PASP..110..761F}, which solves the radiative
transfer problem for plane-parallel and spherical geometries. The
photoionization of intervening absorption systems has been described
by \textcite{1990Ap&SS.165...27V}, whose formulation is followed here.

Denote by $x_{ij}$ the fraction of a particular element $i$ in the $j^{\rm th}$
ionization state, so that $j$ runs from 1 (neutral) to $Z_i+1$ (fully
ionized) for an element of atomic number $Z_i$. Then $\sum_{j=1}^{Z_i+1}
x_{ij}=1$. The general rate equation for $x_{ij}$ is
\begin{eqnarray}
\frac{dx_{ij}}{dt} &=& -x_{ij}\biggl[n_e(c_{ij}+a_{ij}+r_{ij-1}+
d_{ij-1}) + p_{ij}\nonumber\\
&&+n_{\rm HII}c_{ij}^{\rm H}+n_{\rm HeII}c_{ij}^{\rm He}
+\sum_{k=j+2}^{Z_i+1}\,q_{ijk}\nonumber\\
&&\phantom{{=}-x_{ij}\biggl[}+n_{\rm HI}r_{i,j-1}^{\rm H}+
n_{\rm HeI}r_{i,j-1}^{\rm He}\biggr]\nonumber\\
&&+x_{ij+1}\biggl[n_e(r_{ij}+d_{ij})+n_{\rm HI}r_{ij}^{\rm H}+
n_{\rm HeI}r_{ij}^{\rm He}\biggr]\nonumber\\
&&+x_{ij-1}\biggl[n_e(c_{ij-1}+a_{ij-1})+p_{ij-1}\nonumber\\
&&\phantom{{=}+x_{ij-1}\biggl[}+n_{\rm HII}c_{ij-1}^{\rm H}
+n_{\rm HeII}c_{ij-1}^{\rm He}\biggr]\nonumber\\
&&+\sum_{l=1}^{j-2}x_{il}q_{ilj},
\label{eq:ionizemetal}
\end{eqnarray}
where $c_{ij}$ is the collisional ionization rate
\cite{1997ADNDT..65....1V}, $a_{ij}$ the autoionization rate
\cite{1985A&AS...60..425A}, $p_{ij}$ the photoionization rate without
post-ionization Auger-electron ejection \cite{1996ApJ...465..487V,
1995A&AS..109..125V}, $r_{ij}$ the radiative recombination rate
\cite{VF96, 1982ApJS...48...95S, 1982ApJS...49R.351S, AP73,
1985A&AS...60..425A}, $d_{ij}$ the dielectronic recombination rate
\cite{1983A&A...126...75N, 1986A&AS...64..545N, 1987A&AS...69..123N},
$c_{ij}^{\rm H}$ the rate of charge transfer ionization with $\HII$
\cite{1996ApJS..106..205K}, $c_{ij}^{\rm He}$ the rate of charge
transfer ionization with $\HeII$ \cite{1985A&AS...60..425A},
$r_{ij}^{\rm H}$ the rate of charge transfer recombination with $\HI$
\cite{1996ApJS..106..205K}, and $r_{ij}^{\rm He}$ the rate of charge
transfer recombination with $\HeI$ \cite{1985A&AS...60..425A}.
Photoionization from deep shells followed by Auger-electron ejection,
resulting in a change from an initial ionization state $j$ to a final
state $k\ge j+2$, occurs at the rate $q_{ijk}$
\cite{1993A&AS...97..443K}. The convention that $r_{ij}$ corresponds
to recombination to ionization state $j$ and $c_{ij}$ corresponds to
collisional ionization of state $j$ is used. Because of the low
metallicity of the IGM, the electron density may normally be
well-approximated by the electrons lost only by hydrogen and helium.

Using the above notation for the various species and taking $i=1$ for
hydrogen and $i=2$ for helium, Eqs.~(\ref{eq:phionizeH}) and
(\ref{eq:phionizeHe}) are modified, allowing for collisional
ionization and the charge transfer reactions, to
\begin{eqnarray}
\frac{dx_{11}}{dt} &=& -x_{11}\biggl(p_{11}+n_ec_{11}
+\sum_{i\ne1}\sum_{j=2}^{Z_i+1}r_{ij-1}^{\rm
H}n_{ij}\biggr)\nonumber\\ &&+ x_{12}\biggl(n_er_{11} +
\sum_{i\ne1}\sum_{j=1}^{Z_i} c_{ij}^{\rm
H}n_{ij}\biggr),\nonumber\\ \frac{dx_{12}}{dt} &=&
-\frac{dx_{11}}{dt},\nonumber\\ \frac{dx_{21}}{dt} &=& -x_{21}
\biggl[p_{21} + n_e(c_{21} + a_{21}) +q_{213}\nonumber\\
&&\phantom{{=}-x_{21}}+\sum_{i\ne2}\sum_{j=2}^{Z_i+1}n_{ij}r_{ij-1}^{\rm
He}\biggr]\nonumber\\ && + x_{22}\biggl[n_e(r_{21} + d_{21}) +
\sum_{i\ne2}\sum_{j=1}^{Z_i}n_{ij}c_{ij}^{\rm
He}\biggr],\nonumber\\ \frac{dx_{22}}{dt} &=& -\frac{dx_{21}}{dt}
- \frac{dx_{23}}{dt},\nonumber\\ \frac{dx_{23}}{dt} &=& -x_{23}
\biggl[n_e(r_{22} + d_{22}) +n_{11}r_{22}^{\rm H}\biggr] +
x_{21}q_{213}\nonumber\\ &&+x_{22}\biggl[n_e(c_{22} + a_{22}) + p_{22}
+n_{12}c_{22}^{\rm H}\biggr],
\label{eq:ionizeHHe}
\end{eqnarray}
where the first sum of the double sums is over all elements except the
excluded one indicated.

Noting that the photoionization rates are proportional to the energy
density of the radiation field, it follows from the structure of
Eqs.~(\ref{eq:ionizemetal}) and (\ref{eq:ionizeHHe}) that the
equilibrium fractions are determined by the ratio $U=n_\gamma/n_{\rm
H}$ for a given spectral shape of the radiation field with photon
number density $n_\gamma$, and for a given set of elemental
abundances. The ratio $U$ is know as the {\it ionization parameter}.
Larger values for $U$ generally produce higher ionization stages in
the metals, although these also depend on the spectral shape of the
radiation field. Comparisons with measured ion column densities may
then be used to constrain the natures of the radiation field and the
absorption systems.

Fitting formulas to collisional ionization, radiative recombination
and dielectronic recombination rate coefficients are provided by
\textcite{1982ApJS...49R.351S}. Improved values for many of these
rates have since then become available, and have been referenced
above. Power law fits to collisional and radiative recombination rates
involving hydrogen and helium are provided by
\textcite{1997NewA....2..181A}. Fitting formulas to more recent rates
for the elements hydrogen through nickel are provided by
\textcite{1998A&AS..133..403M}, along with equilibrium ionization
fractions for metals in a hot environment, for which collisional
ionization will normally dominate photoionization. A comprehensive
compilation of the literature is available at D. Verner's {\it Atomic
  Data for Astrophysics} website,
\footnote{http://www.pa.uky.edu/$\sim$verner/atom.html} as described
in \textcite{1998PASP..110..761F}. Since then, improved estimates have
been obtained for some rates, including laboratory measured values of
dielectronic recombination rates (\textcite{2006ApJ...642.1275S} and
references therein). A comprehensive review of the atomic physics
relevant to x-ray lines, as will arise from a warm-hot IGM, is
provided by \textcite{2007RvMP...79...79K}.

\subsection{Thermal equilibrium}
\label{subsec:thermal-eq}

In the presence of heat transfer at the rate $(G-L)/n$ per particle,
where $G$ and $L$ are the thermal gain and loss functions per volume and
$n$ is the particle density, the second law of thermodynamics
requires the entropy per particle $s$ to change at the rate
\begin{equation}
\frac{ds}{dt}=\frac{1}{nT}(G-L),
\label{eq:dsdt}
\end{equation}
where $T$ is the temperature of the system. For an ideal gas, the
entropy per particle is $s=(\gamma-1)^{-1}k_{\rm B}\ln(p/\rho^\gamma)
+ s_0$, where $p$ is the gas pressure, $\rho$ is the mass density,
$\gamma$ is the ratio of specific heats at constant pressure to
constant volume ($\gamma=5/3$ for a monatomic gas), $k_{\rm B}$ is
Boltzmann's constant, and $s_0$ is an arbitrary additive constant. For
numerical computations, it is often useful to express the energy
equation in its entropic form Eq.~(\ref{eq:dsdt}), but through the
entropy parameter
\begin{equation}
\label{eq.Entropy}
S_E \equiv \frac{p}{\rho^\gamma}.
\label{eq:S}
\end{equation}
It follows from Eq.~(\ref{eq:dsdt}) that
\begin{equation}
\frac{dS_E}{dt} = (\gamma-1)\rho^{-\gamma}(G-L).
\end{equation}
The gas temperature is related to $S_E$ by
\begin{equation}
T = \frac{\bar m}{k_{\rm B}} S_E \rho^{\gamma-1} 
\label{eq:TS}
\end{equation}
where $\bar m$ is the mean mass per particle.

Photoionization will provide heat through the excess energy absorbed
by an electron above the threshold energy $h_{\rm P}\nu_T$ required to
ionize a given atom. For a single species $i$ of density $n_i$, the
photoionization heating rate is
\begin{equation}
G_i = n_i c\int_{\nu^i_T}^{\infty}\frac{d\nu}{\nu}u_\nu \sigma_i(\nu)
(\nu-\nu^i_T),
\label{eq:PhotoionizationHeating}
\end{equation}
where $h_{\rm P}\nu^i_T$ is the ionization potential of species $i$.
The total heating rate from all species is
\begin{equation}
G = G_\HIs + G_\HeIs + G_\HeIIs.
\label{eq:netHeating}
\end{equation}
The contributions due to the photoionization of the metals is normally
negligible because of their low abundances.

Cooling is provided by recombinations, collisional excitation of the
excited levels in neutral hydrogen, and inverse Compton scattering off
CMB photons. At high temperatures, free-free losses and collisionally
excited line radiation from atoms and ions other than neutral hydrogen
may become important contributions as well. As for heating, all
species contribute to cooling, giving the total cooling rate
\begin{equation}
L = L_\HIIs + L_\HeIIs + L_\HeIIIs + L_{eH} + L_C + L_{\rm line} + L_{ff}.
\end{equation}

For a single species, recombinations radiate the electron energy $\sim
k_{\rm B}T$ as photons at the rate $L_i = n_e n_i
\beta_i(T)$. Recombination cooling coefficients $\beta_i(T)$ are
provided by \textcite{Seaton59} for recombinations to \HI\ and \HeII.
An expression for the radiative contribution to $\beta_\HeIIs$ for
recombinations to \HeI\ is provided by \textcite{Black81}. The
dielectronic contribution may be approximated as $\beta^{\rm
  diel}_\HeIIs = 3 \ry \alpha^{\rm diel}_\HeIIs$
\cite{1970AnPhy..61..351G} using the second part of the expression for
$\alpha_\HeIIs$ in Table~\ref{table.RecombCoef}, which expresses the
dielectronic component. The total recombination cooling coefficients
are provided in Table~\ref{table.RecombCoef}.

The cooling rate due to the collisional excitation of \HI\ by
electrons is given to an accuracy of 3\% over the temperature range
$4000 < T < 12000$~K by the expression \cite{Spitzer78}:
\begin{equation}
L_{eH} = 7.3\expd{-32} \J \m^3 \s^{-1} n_e  n_\HIs  e^{-118400 / T}.
\end{equation}
For temperatures relevant to the IGM, cooling losses due to
collisional ionization of hydrogen and collisional excitation and
ionization losses from helium are negligible. Similarly, for the
low metallicities encountered in the IGM, radiative losses from
metals are normally negligible, although they may possibly be
important in the WHIM.

Cooling is also produced by the Compton scattering of CMB photons off
the electrons, which is especially important at high redshifts. The
cooling rate is \cite{Weymann65}
\begin{equation}
L_C = 4 n_e c\frac{\sigma_T}{m_e c^2} a T^4_{\rm CMB}(z)
k_{\rm B}\left[T - T_{\rm CMB}(z)\right],
\end{equation}
where $\sigma_T=(8\pi/3)(e^2/[4\pi\epsilon_0]m_ec^2)^2=
6.65\times10^{-29}\,{\rm m^2}$
is the Thomson cross section, $a$ is the radiation density constant,
$m_e$ is the electron mass, $c$ is the speed of light, and $T_{\rm
CMB}$ is the temperature of the cosmic microwave background.

In the WHIM, cooling by thermal bremsstrahlung radiation, or free-free
emission, due to the acceleration of a charge in the Coulomb field of
another charge may be significant. For an ion of number density $n_i$
and charge $Z_i$, the cooling rate is \cite{1979rpa..book.....R}
\begin{equation}
L_{ff} = 1.43\times10^{-40} \J \m^3 \s^{-1} T^{1/2}n_e n_i Z_i^2{\bar g}_B,
\label{eq:tbremss}
\end{equation}
where ${\bar g}_B$ is a frequency average of the velocity-averaged
Gaunt factor, and ranges between 1.1 and 1.5.
\textcite{1987ApJ...318...32S} provide the fitting formula ${\bar
g}_B=0.79464 + 0.1243\log_{10}(T/Z_i^2)$ for $T/Z_i^2<3.2\times10^5$~K
and ${\bar g}_B=2.13164 - 0.1240\log_{10}(T/Z_i^2)$ for
$T/Z_i^2>3.2\times10^5$~K. Cooling rates including collisionally
excited metallic line losses and their individual emissivities are
provided by \textcite{1983ApJS...52..155G} for solar
abundances. Non-solar values may be interpolated with the
zero-metallicity limit for which atomic cooling is determined only by
the hydrogen and helium losses.

Photoelectric heating by x-rays is automatically included in
Eq.~(\ref{eq:PhotoionizationHeating}). X-rays, however, may introduce
two other sources of energy exchange:\ gains through Compton heating
and losses through the effect of secondary electrons.

Sources of high energy radiation, such as a hard x-ray background, may
inject additional energy into the IGM through Compton heating
\cite{1969ApJ...158L..91A, 1969ApL.....4..113R, 1970A&A.....8..410R,
1999ApJ...517L...9M}. For an IGM of electron density $n_e$ and
temperature $T$, an x-ray background with energy density $u_\nu^X$
will heat the gas through Compton scatterings at the rate
\begin{equation}
  G_C=n_ec\frac{\sigma_T}{m_ec^2}\int_0^\infty\,d\nu u_\nu^X(h_{\rm P}\nu-4k_{\rm B}T).
\label{eq:HC}
\end{equation}
For photons with energies exceeding 100~keV, Compton scattering takes
place in the relativistic regime, requiring the Klein-Nishina cross
section instead of the Thomson. The heating rate then
becomes \cite{1974ApJ...188..121B, 1999ApJ...517L...9M}
\begin{eqnarray}
G_C&=&\frac{3}{4} n_ec\sigma_T\int_0^\infty dx u_x^X\frac{1}{x^2}
\biggl[\frac{x^2-2x-3}{2x}\ln(1+2x)\nonumber\\
&&\phantom{{=} n_ec}
+\frac{-10x^4+51x^3+93x^2+51x+9}{3(2+3x)^3}\biggr],
\label{eq:rHC}
\end{eqnarray}
where $x=h_{\rm P}\nu/m_ec^2$. The Compton heating rate $G_C$ should
then be added to Eq.~(\ref{eq:netHeating}).

The secondary electrons ejected by x-rays have only a small effect on
the temperature of highly ionized gas. In partially ionized gas,
however, their contribution to the energy losses may become
appreciable. They play a particularly important role in the cooling of
gas ahead of ionization fronts during the Epoch of Reionization.
Expressions for the energy of a primary electron deposited as heat,
further ionizations, and collisional excitations, for both hydrogen
and helium, in a partially ionized medium are provided by
\textcite{1985ApJ...298..268S}.

\section{The Metagalactic UV Background}
\label{sec:mguvbg}

\subsection{Mean energy density of the UV background}
\label{subsec:unu}

\subsubsection{Origin of the UV background}
\label{subsubsec:uvbg-origin}

Since the ionization structure of the IGM is determined by the
specific radiation energy density $u_\nu$ of the metagalactic
ultraviolet photoionizing background, a complete understanding of the
structure of the IGM requires an estimate for $u_\nu$. In turn,
measured properties of the IGM, like its optical depth to hydrogen and
helium ionizing photons or the ratios of ionization fractions of
metals, place constraints on the shape of the metagalactic UV
background.

In a homogeneous and isotropic expanding universe,
Eq.~(\ref{eq:tdRT-exp-sol}) gives for the integrated specific energy
density produced by sources with a total proper emissivity
$\epsilon_\nu(z)=4\pi j_\nu(z)$
\begin{equation}
u_\nu(z) = \frac{1}{c}\int_z^\infty\, dz^\prime \frac{dl_p}{dz^\prime}
\frac{(1+z)^3}{(1+z^\prime)^3}\epsilon_{\nu^\prime}(z^\prime)
\exp[-\tau_{\rm eff}(\nu,z,z^\prime)],
\label{eq:unu}
\end{equation}
where $\tau_{\rm eff}(\nu,z,z^\prime)$ is an effective optical depth
due to absorption by the IGM and $\nu^\prime=\nu(1+z^\prime)/(1+z)$.
For discrete absorption systems, Eq.~(\ref{eq:taul}) may be used,
adapted to photoelectric absorption by choosing $x=N_{\rm HI}$ and
taking $w=\int d\lambda^{\prime\prime}[1-\exp(-\tau_{\nu^{\prime\prime}})]$,
where $\nu^{\prime\prime}=\nu(1+z^{\prime\prime})/(1+z)$. Then, on
replacing $d\lambda^{\prime\prime}\partial^2 {\cal N}/(\partial N_{\rm HI}
\partial \lambda^{\prime\prime})$ by $dz^{\prime\prime}
\partial^2 {\cal N}/(\partial N_{\rm HI}\partial z^{\prime\prime})$,
\begin{equation}
\tau_{\rm eff}(\nu,z,z^\prime)=\int_z^{z^\prime} dz^{\prime\prime}
\int_0^\infty dN_{\rm HI}\frac{\partial^2 {\cal N}}{\partial N_{\rm HI}
\partial z^{\prime\prime}}[1-\exp(-\tau_{\nu^{\prime\prime}})],
\label{eq:taueff}
\end{equation}
where
\begin{equation}
\tau_{\nu^{\prime\prime}}=N_{\rm HI}a_{\rm HI}(\nu^{\prime\prime})
+ N_{\rm HeI}a_{\rm HeI}(\nu^{\prime\prime})
+ N_{\rm HeII}a_{\rm HeII}(\nu^{\prime\prime}).
\label{eq:taunupp}
\end{equation}

The emissivity $\epsilon_\nu$ consists of two contributions, isolated
radiation sources, such as galaxies and QSOs, and diffuse emission
from the IGM itself. Survey estimates for the contributions from
isolated sources are discussed below. There are four principal
contributions to the diffuse emission capable of photoionizing the IGM
\cite{HM96}:\ 1.\ recombinations to the ground state of \HI\ and
\HeII, 2. \HeII\ \Lya recombination radiation, 3.\ \HeII\
two-photon continuum emission, and 4.\ \HeII\ Balmer continuum
emission. The specific recombination emissivity associated with
each of these processes may be expressed generally in the form \cite{HM96}
\begin{equation}
  \epsilon_r(\nu,z)=h_{\rm P}\nu f(\nu) B(\nu,z) \frac{dz}{dl_p},
\label{eq:epsrec}
\end{equation}
where
\begin{eqnarray}
  B(\nu,z)&=&c\int_0^\infty dN_{\rm HI}\frac{\partial^2{\cal N}}
  {\partial N_{\rm HI}\partial z}p_{\rm rec}
  p_{\rm esc}(\nu)\nonumber\\
  &&\times\int_{\nu_T}^\infty d\nu^\prime\frac{u_{\nu^\prime}(z)}{h_{\rm
      P}\nu^\prime}
  w_{\rm abs}(\nu^\prime),
\label{Bfunc}
\end{eqnarray}
where $\nu_T$ refers to the photoionization threshold for ionizing the
ground state of an ion of atomic number $Z_i$. The function $B(\nu,z)$
expresses the escape rate of recombination photons produced by
discrete absorption systems in terms of the product of the fraction
$w_{\rm abs}(\nu)$ of photons of energy $h_{\rm P}\nu$ absorbed by a
given ion, the probability $p_{\rm rec}$ that a recombination will
result in the particular radiative transition considered, and the
probability $p_{\rm esc}$ that the emitted photon will escape the
absorption system that produced it. These factors may be computed from
detailed radiative transfer computations using a code like CLOUDY
\cite{HM96}.

The capture of electrons into the $n ^2L$ level of a hydrogenic ion
results in the emission of a continuum photon distribution sharply
peaked at the photoelectric edge. The rate of direct recombinations
may be computed from the Milne relation between the free-bound
recombination cross section $\sigma_{fb}(v)$ for an electron moving at
velocity $v$ and the bound-free photoionization cross-section
$a_n(\nu)$ \cite{1979rpa..book.....R}
\begin{equation}
\frac{\sigma_{fb}(v)}{a_n(\nu)}=\frac{h_{\rm P}^2 \nu^2}{m_e^2 c^2 v^2}
\frac{2g_n}{g_i},
\label{eq:Milne}
\end{equation}
where $g_n=2n^2$ is the statistical weight for the atom with electron in
the bound state $n$ and $g_i=2$ is the statistical weight for
the ion. The bound-free cross-section from state $n$ in a hydrogenic atom
of atomic number $Z_i$ may be expressed as
\begin{equation}
a_n(\nu) = \frac{64\pi}{3^{3/2}}\frac{n}{Z_i^2}\alpha a_0^2 \left(\frac{\nu_T}
{\nu}\right)^3G_n,
\label{eq:phion_nZ}
\end{equation}
where $G_n$ is a Gaunt factor and $\alpha=[1/4\pi\epsilon_0]e^2/\hbar c$
is the fine-structure constant. The Gaunt factor for the ground state is
\begin{equation}
G_1 = 8 \pi 3^{1/2} (1+\epsilon^2)^{-1}
\frac{e^{-(4{\rm tan^{-1}}\epsilon)/\epsilon}}
{1-e^{-2\pi/\epsilon}},
\label{eq:G1}
\end{equation}
where $\epsilon=(\nu/\nu_T-1)^{1/2}$ (cf Eq.~[\ref{eq:sigmaHZ}]). The
photon released following recombination will have energy $h_{\rm
  P}\nu=h_{\rm P}\nu_T+(1/2)m_ev^2$, where $m_e$ is the mass of an
electron and $h_{\rm P}$ is Planck's constant. For a Maxwellian
velocity distribution $f_M(v)=4\pi(m_e/2 \pi k_{\rm B} T)^{3/2} v^2
\exp(-m_e v^2/ 2 k_{\rm B} T)$ for the electrons, the free-bound emissivity
due to recombinations to state $n$ in a hydrogenic ion of atomic
number $Z_i$ and number density $n_i$ is then
\begin{eqnarray}
  \epsilon_n(\nu)&=&n_e n_i h_{\rm P}\nu \sigma_{fb}(v) v f_M(v)
  \frac{dv}{d\nu}\nonumber\\
  &=&\frac{4\pi}{c^2} n_e n_i \biggl(\frac{h_{\rm P}^2}{2\pi m_e k_{\rm B}T}\biggr)^{3/2}
  h_{\rm P} \nu^3 a_n(\nu) 2 n^2\nonumber\\
  &&\times\exp\biggl(-\frac{h_{\rm P}\nu-h_{\rm P}\nu_T}{k_{\rm B}T}\biggr).
\label{eq:emiss_fb}
\end{eqnarray}

Recombinations to \HeII\ produce hydrogen-ionizing photons ($h_{\rm
  P}\nu > 13.6\,{\rm eV}$). The recombinations generally occur to high
energy levels, followed by a cascade of bound-bound transitions.
Recombinations that populate the $2 ^2P$ level of \HeII\ result in a
\HeII\ \Lya photon of energy 40.8~eV, which photoionizes neutral
hydrogen in the \Lya forest systems. Recombinations that populate the
$2 ^2S$ level produce \HeII\ $2 ^2S\rightarrow 1 ^2S$ two-photon
emission with a total energy of 40.8~eV and a spectrum peaking at
20.4eV. Direct recombinations to $2 ^2S$ and $2 ^2P$ produce \HeII\
Balmer continuum emission, for which the threshold photoionization
energy is the same as the hydrogen Lyman limit.

\subsubsection{Sources}
\label{subsubsec: sources}

\begin{table*}
\caption{Estimates of the \HI\ mean \Lya transmitted flux (col 2) and
\HI-ionization rate $\Gamma_{\rm HI}^{\rm IGM}$ (in $10^{-12}\,{\rm
s^{-1}}$) required to match the mean \Lya flux according to
\textcite{MW04} (col 3) and to \textcite{2005MNRAS.357.1178B} and
\textcite{2007MNRAS.382..325B} (col 4). Columns 5 and 6 are the
\HI-ionization rates $\Gamma_{\rm HI}^{\rm QSO, PLE}$ and $\Gamma_{\rm
HI}^{\rm QSO, PDE}$ (in $10^{-12}\,{\rm s^{-1}}$) from QSOs using the
Pure Luminosity Evolution (PLE) and Pure Density Evolution (PDE)
models of \textcite{Meiksin05}. Column 7 gives the \HI-ionization rate
$\Gamma_{\rm HI}^{\rm QSO, HRH}$ using the emissivity from
\textcite{2007ApJ...654..731H} and the same attenuation lengths as for
the PLE and PDE models. For the PLE and PDE models, the indicated
redshift entries $z=3.89,\, 4.00,\, 5.00$ and 6.00 correspond to the
ranges $3.6<z<3.9$, $3.9<z<4.4$, $4.4<z<5.0$ and $z>5.7$,
respectively.
}
\begin{ruledtabular}
\begin{tabular}{rllllll}
 $z$ & $\langle \exp(-\tau)\rangle$ & $\Gamma_{\rm HI, -12}^{\rm IGM, MW04}$
& $\Gamma_{\rm HI, -12}^{\rm IGM, B}$
& $\Gamma_{\rm HI, -12}^{\rm QSO, PLE}$ & $\Gamma_{\rm HI, -12}^{\rm QSO, PDE}$
& $\Gamma_{\rm HI, -12}^{\rm QSO, HRH}$\\
 \hline
 2.75   & $0.74\pm0.04$ & $0.86^{+0.36}_{-0.24}$ & $-$ & $-$ & $0.64$ \\
 3.00   & $0.70\pm0.02$ & $0.88^{+0.14}_{-0.12}$ & $0.86^{+0.34}_{-0.26}$ 
& $1.1^{+3.0}_{-0.7}$ & $0.44^{+0.39}_{-0.23}$ & $0.46$ \\
 3.89   & $0.48\pm0.02$ & $0.68^{+0.08}_{-0.07}$ & $-$ & $0.19\pm0.05$ & $0.14^{+0.07}_{-0.05}$ & $0.06$ \\
 4.00   & $0.47\pm0.03$ & $0.76^{+0.12}_{-0.11}$ & $0.97^{+0.48}_{-0.33}$ 
& $0.13^{+0.05}_{-0.03}$ & $0.084^{+0.040}_{-0.030}$ & $0.058$ \\
 5.00   & $0.12^{+0.03}_{-0.04}$ & $0.31^{+0.07}_{-0.09}$ & $0.52^{+0.35}_{-0.21}$ 
& $0.051^{+0.026}_{-0.018}$ & $0.035^{+0.017}_{-0.014}$ & $0.027$ \\
 5.50   & $0.079^{+0.017}_{-0.013}$ & $0.37^{+0.06}_{-0.05}$ & $-$ & $-$ 
& $-$ & $0.019$ \\
 6.00   & $<0.006$ & $<0.14$ & $<0.19^{+0.15}_{-0.10}$ & $0.012^{+0.013}_{-0.005}$ & $0.006^{+0.006}_{-0.002}$ & $<0.014$\\
 \hline
\label{tab:GammaHI}
\end{tabular}
\end{ruledtabular}
\end{table*}

QSOs were early recognized as a prominent source of metagalactic
ionizing photons \cite{1970A&A.....8..410R, 1970ApL.....5..123A,
1970ApL.....5..287A}.  Since then, there have been numerous estimates
of their contribution to the UV metagalactic background
\cite{1979ApJ...230...49S, 1987ApJ...315..180B, MO90,
1992ApJ...389L...1M, 1992ApJ...392...15M, MM93, GS96, HM96,
1998AJ....115.2206F, 1999AJ....118.1450S, 2001cghr.confE..64H,
2003ApJ...584..110S}. The QSO luminosity function is conveniently
parametrized by a double power-law form \cite{1988MNRAS.235..935B}
\begin{equation}
\phi(M,z) = \frac{\phi_*}{10^{0.4(1-\beta_1)(M-M^*)} +
10^{0.4(1-\beta_2)(M-M^*)}},
\label{eq:BoyleLF}
\end{equation}
where $\phi(M,z)dM$ describes the spatial density of QSOs at redshift
$z$ in the absolute magnitude range $M$ to $M + dM$, $\phi^*$ is a
spatial normalization factor (normally expressed in comoving units),
and $M^*$ is a characteristic break magnitude. In a Pure Luminosity
Evolution (PLE) model, $\phi^*$ is held fixed while $M^*$ is allowed
to evolve. In a Pure Density Evolution (PDE) model, $M^*$ is held
fixed while $\phi^*$ is allowed to evolve. Although PLE and PDE models
well fit the data over limited redshift ranges,
\textcite{2001AJ....121...54F} show that the steepness of the bright
end of the luminosity function has evolved between $z<2$ and
$z>3.6$. Estimates of the QSO luminosity function at $z<2.1$ are
provided by \textcite{2004MNRAS.349.1397C} based on the 2dF redshift
survey, and by \textcite{2005MNRAS.360..839R}, combining data sets
from the 2dF and SDSS.  \textcite{2006AJ....131.2766R} use the SDSS to
measure the bright end of the QSO luminosity function and its
evolution over the redshift range $0<z<5$.

A long-standing uncertainty affecting estimates of the QSO
contribution to the UV background at high redshifts ($z>2$) is the
poorly determined number of low luminosity QSOs.
\textcite{2004ApJ...605..625H} have measured the low luminosity end of
the luminosity function at $z\approx3$. An estimate of the QSO
luminosity function over the redshift range $3<z<6$ is provided by
\textcite{Meiksin05} based on fitting PLE and PDE models to the
results of \textcite{2001AJ....121...54F},
\textcite{2004AJ....128..515F} and \textcite{2004ApJ...605..625H}. An
estimate of the bolometric QSO luminosity function over the redshift
range $0<z<6$ is provided by \textcite{2007ApJ...654..731H}.

The QSO emissivity is given by $\epsilon_\nu=\int dM \phi(M)
L_\nu(M)$, where $L_\nu(M)$ is the mean specific luminosity
corresponding to a QSO of magnitude $M$. The spectral shape of QSOs,
however, is uncertain at frequencies above the \HI\ Lyman
edge. Results based on {\it HST} data using the Faint Object
Spectrograph (FOS) give an intrinsic spectral distribution
$f_\nu\propto\nu^{-\alpha_Q}$ with QSO spectral index
$\alpha_Q=1.76\pm0.12$ \cite{2002ApJ...565..773T} over the restframe
energies $0.8-1.8$~Ry. The sample contains QSOs predominantly in the
redshift range $0.3<z<2.3$. Distinguishing between radio-quiet and
radio-loud QSOs gives $\alpha_Q=1.57\pm0.17$ and
$\alpha_Q=1.96\pm0.12$, respectively. Results based on {\it FUSE} data
give instead $\alpha_Q=0.56^{+0.28}_{-0.38}$ (without distinguishing
between radio-quiet and radio-loud QSOs)
\cite{2004ApJ...615..135S}. The origin of the discrepancy is
unclear. The {\it FUSE} sample is at lower redshift ($z<0.67$) and the
sources have lower luminosities, so the difference may reflect a
hardening with decreasing redshift and/or luminosity, for which there
is evidence \cite{2004ApJ...615..135S}. The spectral range used for
the power-law fits to the {\it FUSE} spectra is restricted to
restframe energies $0.8-1.4$~Ry, similar to the range for the {\it
HST} sample. The spectral corrections due to absorption by the
intervening IGM, however, are smaller for the {\it FUSE} sample QSOs
than the higher redshift {\it HST} QSOs, so that the {\it FUSE}
results may be more reliable.

\begin{figure}
\includegraphics[width=8cm, height=8cm]{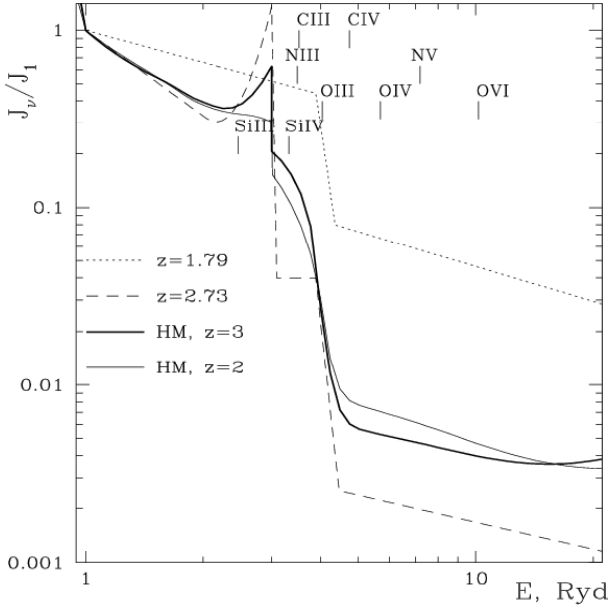}
\caption{The UV metagalactic ionizing background intensity $J_\nu$,
normalized by the intensity $J_1$ at 1~Ry, from \textcite{HM96} at $z
= 3$ (thick line) and $z=2$ (thin line) for a power-law QSO source
spectrum $\nu^{-1.5}$ shortward of the \HI\ Lyman edge. The threshold
energies required to ionize several commonly detected ions are
indicated. Also shown are estimates of the UV background as
constrained by metal absorption systems at $z=2.73$ (dashed line) and
$z=1.79$ (dotted line). From \textcite{2007A&A...461..893A}.
(Figure reproduced by permission of Astronomy \& Astrophysics.)
}
\label{fig:UVBG}
\end{figure}

The angle-averaged intensity $J_\nu=cu_\nu/4\pi$ of the UV background
determined by \textcite{HM96} based on an earlier assessment of the
QSO contribution is shown in Fig.~\ref{fig:UVBG} at redshifts $z=2$
and 3. The steps in the background are produced by photoelectric
absorption of the source radiation by intervening absorbers at the
\HI\ and \HeII\ Lyman edges. The comparison with the purely absorbing
IGM estimate shows the formation of emission ``horns'' due to the
reprocessing of ionizing radiation by the IGM into redshifted \HI\ and
\HeII\ \Lya photons. The redshifted \HeII\ \Lya photons enhance the
flux of \HI-ionizing photons as does the continuum emission resulting
from direct recombinations to the ground levels of \HI\ and \HeII. At
higher redshifts, the UV background more strongly resembles the
spectral shape of the sources because the opacity of the IGM becomes
so large that the UV background is increasingly dominated by local
sources.

Combining Eq.~(\ref{eq:GHI}) with Eq.~(\ref{eq:unu}), the
photoionization rate may be expressed in terms of the emissivity as
\begin{equation}
\Gamma_{\rm HI}=\frac{\epsilon_L(z)a_T}
{h_{\rm P}(3+\alpha_{\rm MG})}r_{\rm eff},
\label{eq:GHI-emiss}
\end{equation}
where $\epsilon_L(z)$ is the emissivity at the Lyman edge,
$\alpha_{\rm MG}$ is the spectral index of the UV background just
shortward of the Lyman edge, $r_{\rm eff}$ is an effective lengthscale
over which the distribution of sources contributes to the radiation
field at any given point, the photoionization cross-section is
approximated as $a_\nu\simeq a_T(\nu/\nu_T)^{-3}$, and $h_{\rm P}$ is
Planck's constant. It is well known that in a uniform infinite
Euclidean universe with point sources of radiation distributed
throughout, the effective lengthscale is infinite and the local
radiation field diverges (Olber's paradox). Cosmological effects, in
particular the finite age of the sources and the effects of
redshifting, will guarantee convergence, and $r_{\rm eff}$ describes
an effective horizon to the sources. Attenuation by a
photoelectrically absorbing IGM will shorten the effective length even
further, and will dominate at high redshifts. These two extremes, a
cosmologically limited lengthscale and an attenuation limited
lengthscale, may be combined into the approximate expression $r_{\rm
eff} \simeq 1/(1/l_{\rm H} + 1/r_0)$, where $l_{\rm H}$ is the
lengthscale limitation due to cosmological expansion and the finite
age of the sources,and $r_0$ is the lengthscale due to attenuation by
the IGM. For sources turning on at $z=z_{\rm on}$ with (comoving)
emissivity parametrized as $\epsilon_\nu(z)=\epsilon_L(1+z)^\gamma
(\nu/\nu_L)^{-\alpha_S}$, the effective (proper) horizon to the
sources is
\begin{eqnarray}
l_{\rm H}&\simeq&\frac{c}{H_0\Omega_m^{1/2}}
\frac{3+\alpha_{\rm MG}}{3+\alpha_S}(1+z)^{-3/2}\nonumber\\
&&\times\frac{1-[(1+z)/(1+z_{\rm on})]^{9/2+\alpha_S-\gamma}}
{9/2+\alpha_S-\gamma},
\label{eq:lH}
\end{eqnarray}
for large $z$ \cite{MW03}. Values for the attenuation length $r_0$ are
provided by \textcite{MW04} based on numerical simulations (see
\S~\ref{subsubsec:Gamma}). The values obtained lie between the medium
and high attenuation models of \textcite{MM93} derived from direct
absorption line counts.

Resulting values for the QSO contribution to the \HI\ ionization rate
are given in Table~\ref{tab:GammaHI} at selected redshifts. The
estimates are derived from the QSO emissivities of
\textcite{Meiksin05} and \textcite{2007ApJ...654..731H} using
Eq.~(\ref{eq:GHI-emiss}) and the attenuation lengths $r_0$ from
\textcite{MW04}, allowing for an extrapolation of the measured Lyman
Limit System distribution to $z>4$ (see \S~\ref{subsubsec:Gamma}
below). Results for both the Pure Luminosity Evolution model and the
Pure Density Evolution models of \textcite{Meiksin05} for the QSO
luminosity function are given. The estimate based on the QSO
luminosity function of \textcite{2007ApJ...654..731H} uses the
emissivity shown in their Figure 11. All the estimates agree within
the errors.

\begin{table*}
\caption{Ratio of $f_\nu({\rm 1500\,\AA})/f_\nu({\rm 912^-\,\AA})$ for
starburst galaxies undergoing continuous star formation with a
Salpeter stellar Initial Mass Function, over the indicated times for
solar ($Z=0.02$) and 1/20 solar ($Z=0.001$) metallicity. Based on the
STARBURST99 models of \textcite{Leitherer99}.
}
\begin{ruledtabular}
\begin{tabular}{rlllll}
$Z$ & 1~Myr & 20~Myr & 100~Myr & 300~Myr & 900~Myr\\
 \hline
 0.02   & 2.7 & 4.9 & 6.2 & 6.7 & 6.8 \\
 0.001  & 2.6 & 4.6 & 6.2 & 6.9 & 7.5 \\
 \hline
\label{tab:flux-convert}
\end{tabular}
\end{ruledtabular}
\end{table*}

Star-forming galaxies are an important potential source of ionizing
radiation. While the spectra are likely too soft to produce
appreciable \HeII-ionizing photons, the hot young stars in
star-forming galaxies could in principle provide even more
hydrogen-ionizing photons than QSOs. An estimate of the contribution
of star-forming galaxies to the UV background requires making a
variety of assumptions regarding the Initial Mass Function,
populations and ages of the stars as well as the numbers of faint
galaxies below the survey limits. Some of these uncertainties are
mitigated by using the results of galaxy surveys that measure the
intrinsic flux from the galaxies close to the Lyman edge (the range
$1300-1800$\AA\ is typical), since the stars that produce the ionizing
radiation are the same as those dominating the detected light of the
galaxies \cite{1999ApJ...514..648M}. Even then, however, uncertainties
in the spectra of the massive stars that dominate the radiation at
early times introduce a comparable level of uncertainty into the
estimate of the combined ionizing emissivity from the galaxy
population \cite{2003MNRAS.344.1000B}. Estimates for the conversion
between the measured flux at 1500\AA\ (restframe) and the predicted
flux shortward of the Lyman discontinuity are provided in
Table~\ref{tab:flux-convert} assuming a Salpeter Initial Mass Function
with $M_{\rm lower}=1\,M_\odot$ and $M_{\rm upper}=100\,M_\odot$ and
either solar ($Z=0.02$) or 1/20 solar ($Z=0.001$) metal abundances,
based on the STARBURST99 models of
\textcite{Leitherer99}.\footnote{The code and model results are
available at http://www.stsci.edu/science/starburst99/.} A similar
value of $f_\nu(1500)/f_\nu(912^-)\simeq6$ is quoted by
\textcite{1999ApJ...514..648M}. The values for the ratio
$f_\nu(1350)/f_\nu(912^-)$ are about 5--10\% larger, and for
$f_\nu(1700)/f_\nu(912^-)$ about 5--10\% smaller, than the values in
Table~\ref{tab:flux-convert}. The lower mass cut-off $M_{\rm
lower}=1\,M_\odot$ appears appropriate for some starbursts; in any
case, choosing a lower value only alters the normalization to the
total star formation rate as the lower mass stars contribute
negligibly to the spectra except at very late times
\cite{1995ApJS...96....9L}.

Another major uncertainty is the escape fraction $f_{\rm esc}(\nu)$ of
ionizing radiation, as this must multiply the emissivity based on
galaxy counts and finally governs the contribution of the star-forming
galaxies to the UV background. There are few direct measurements of
the escape fraction, and virtually all are upper limits.
\textcite{2003ApJ...597..948P} infer a global escape fraction from the
Milky Way of 1--2\% from H$\alpha$ spectroscopic measurements of
nearby high velocity clouds. Based on a detection of Lyman continuum
flux from {\it FUSE} observations of a local starburst galaxy,
\textcite{2006A&A...448..513B} suggest an escape fraction of
$0.04-0.1$, subject to spectral modeling
uncertainties. \textcite{FLC03} place an upper limit of 4\% for
galaxies in the redshift range $1.9<z<3.5$. Similar upper limits based
on {\it FUSE} observations were obtained by
\textcite{2001ApJ...558...56H} and \textcite{2001A&A...375..805D}.
\textcite{2006ApJ...651..688S} report the detection of UV radiation
shortward of the Lyman edge in a sample of Lyman Break Galaxies at
$z\approx3$ with a corresponding escape fraction of 14\%, although the
estimate is sensitive to the assumed calibrating population synthesis
models and corrections for the IGM optical depth, and it may also
apply only to the more luminous galaxies.

In addition to direct observational constraints, several attempts have
been made to estimate the escape fraction from theoretical models
allowing for inhomogeneities in the interstellar medium of galaxies,
the influence of supernovae, and the flow of ionizing radiation
through chimneys produced by large scale outflows
\cite{1994ApJ...430..222D, 2000ApJ...531..846D, 2000ApJ...545...86W,
2002MNRAS.331..463C, 2002MNRAS.337.1299C, 2003ApJ...599...50F}. Escape
fraction values ranging from less than 1\% for quiescent disks to more
than half in an inhomogeneous IGM are obtained. The models suggest
that high star formation rates, such as in Lyman Break Galaxies, may
sufficiently fragment the interstellar medium of the galaxies and/or
create sufficient channels through supernova-driven superbubbles that
escape fractions in excess of 10\% may be achieved, particularly if
star forming regions are distributed throughout the disk of a galaxy.

A composite luminosity function of star-forming galaxies detected by
the Lyman-break method\footnote{The Lyman edge moves into the optical
range at $z\approx3$, permitting the identification of galaxies with
large Lyman breaks, as is expected for galaxies with high star
formation rates \cite{Leitherer99}.} at $z\approx3$ was derived by
\textcite{1998hdf..symp..219D}. \textcite{1999ApJ...514..648M}
combined this with determinations of the Lyman-break population at
$2.75<z<4$ by \textcite{1996MNRAS.283.1388M} and
\textcite{1998ApJ...498..106M} to estimate the emissivity of
star-forming galaxies and their contribution to the metagalactic UV
background.

Establishing the star-formation history of the Universe is currently
an area of intense activity. A revised UV luminosity function of Lyman
Break Galaxies was provided by \textcite{1999ApJ...519....1S}.
\textcite{2004ApJ...600L.103G} measured the luminosity density of
Lyman Break Galaxies in the redshift interval $3.5<z<6.5$, while
\textcite{2004ApJ...604..534S} estimated the star-formation history of
galaxies in the redshift interval
$1.4<z<2.5$. \textcite{2006ApJ...653..988Y} have measured the
luminosity function of Lyman Break Galaxies in the SUBARU Deep Field
at $z\approx4$ and 5. \textcite{2006ApJ...653...53B} provide a
comprehensive analysis of the UV luminosity function of galaxies at
$z\approx6$ and the evolution of the luminosity function to
$z\approx3$.

Despite the heroic efforts to establish the UV luminosity density and
cosmic star formation rates at high redshifts, the contribution of the
galaxies to the metagalactic UV background is still uncertain due to
several factors. In addition to the ionizing radiation escape fraction
and UV-to-Lyman limit conversion factor, the effects of internal
galactic extinction, uncertainty in the amount of intergalactic
extinction by the IGM \cite{2006MNRAS.365..807M} and the minimum
source luminosity on the UV luminosity density and its evolution
create further uncertainty in the net contribution of the galaxies.

The error budget may be illustrated using the results for $3<z<6$
\cite{2004ApJ...600L.103G, 2006ApJ...653...53B}. The uncertainty in
the measured luminosity density of the sources is 20\% or larger.
Another factor of 3 uncertainty results from the uncertain extinction
correction to the number counts \cite{2006ApJ...653...53B}. The
uncertain lower luminosity limit of the sources introduces a factor of
2 uncertainty to the integrated contribution of the sources to the UV
background \cite{2006ApJ...653...53B}. The conversion of the flux from
the measured bands to shortward of the Lyman edge (see
Table~\ref{tab:flux-convert}) introduces at least another 20\%
uncertainty. The attenuation length $r_0$ is uncertain by about 30\%
\cite{MW04}, and the uncertain evolution rate and birth time of the
sources introduces another 10\% uncertainty in the effective length
scale to use in Eq.~(\ref{eq:GHI-emiss}). Since these uncertainties
are primarily systematic, the factors must be multiplied. Altogether,
taking middle values of the expected ranges still leaves a combined
uncertainty of about a factor of 4 for the contribution of the
galaxies to $\Gamma_{\rm HI}$, excluding the uncertainty of the escape
fraction.

\textcite{2004ApJ...600L.103G} obtain for the comoving luminosity of
sources at 1500\AA, $\log_{10}\rho_{1500}\simeq19.25\pm0.05$ (in units
of ${\rm W\,Hz^{-1}\,Mpc^{-3}}$) at $3.4<z<4.1$ and
$\log_{10}\rho_{1500}\simeq19.0\pm0.1$ at $4.6<z<5.2.$ Applying the
conversion factors in Table~\ref{tab:flux-convert} gives for the
corresponding emissivities shortward of the Lyman edge
$\log_{10}\rho_{912^-}\simeq18.4\pm0.06$ at $3.4<z<4.1$ and
$\log_{10}\rho_{912^-}\simeq18.2\pm0.1$ at $4.6<z<5.2$. Using these
values in Eq.~(\ref{eq:GHI-emiss}), combined with the attenuation
length estimates of \textcite{MW04} (see \S~\ref{subsubsec:Gamma}
below), gives $\Gamma_{\rm HI, -12}\simeq2.7f_{\rm esc}$ at
$3.4<z<4.1$ and $\Gamma_{\rm HI, -12}\simeq2.1f_{\rm esc}$ at
$4.6<z<5.2$. Comparing with the required values for $\Gamma_{\rm HI,
  -12}$ in Table~\ref{tab:GammaHI} (see \S~\ref{subsubsec:Gamma}),
allowing for the contribution from QSOs in the PLE model of
\textcite{Meiksin05} (the PDE model estimates are even smaller), and
allowing for a factor 4 uncertainty in the galaxy contribution imposes
the restrictions on the escape fractions of $0.004<f_{\rm esc}<0.9$ at
$3.4<z<4.1$ and $0.02<f_{\rm esc}<0.6$ at $4.6<z<5.2$. Similar escape
fractions are inferred by \textcite{2006MNRAS.371L...1I}. Galaxies
thus plausibly make up the missing ionizing radiation, but the
uncertainties are still too great to preclude additional sources,
which may actually dominate. Since re-radiation by the IGM itself may
contribute some 20--40\% of the total emissivity \cite{MM93, HM96},
these restrictions on $f_{\rm esc}$ may be further eased. The
emissivities are based on source counts uncorrected for the effects of
internal extinction on color. Corrections may increase the
emissivities by as much as a factor of 3--4
\cite{2006ApJ...653...53B}, corresponding to a similar reduction
factor in the required escape fractions.

At $5.5<z<6.5$, \textcite{2006ApJ...653...53B} obtain a comoving
integrated UV emissivity at $\lambda=1350$\AA\ of
$\log_{10}\rho_{1350}\simeq 19.0\pm0.25$ at $5.5<z<6.5$, with the
error range covering the uncertainty in the lowest luminosity of the
contributing sources. Using the same procedure as above, this
translates into the rather large ionization rate of $\Gamma_{\rm HI,
-12}\simeq20f_{\rm esc}$. Correcting the colors for internal
extinction increases this value by about another 50\%. Consistency
with the residual ionization rate at $z\simeq5.5$, after subtracting
the QSO contribution (which is small), gives $0.004<f_{\rm esc}<0.09$
(or somewhat less, allowing for extinction corrections), a very
plausible range, suggesting that these sources may in fact dominate
the UV ionizing background at high redshifts.

\begin{figure}
\includegraphics[width=8cm, height=8cm]{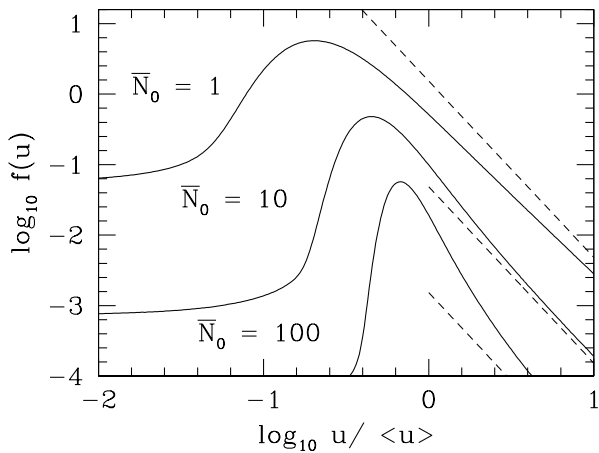}
\caption{The probability density $f(u)$ for the fluctuations in the
energy density $u$ of the UV background for a mean number of sources
within the attenuation volume of ${\bar N}_0 = 1$, 10 and 100 for a
QSO luminosity function (solid lines). The corresponding asymptotic
limits for $u/\langle u\rangle \gg1$ are also shown (dashed
lines). The peak in the distributions shift to lower $u/\langle
u\rangle$ as ${\bar N}_0$ decreases. Adapted from \textcite{MW03}.
(Figure reproduced by permission of the RAS and Blackwell Publishing.)
}
\label{fig:u-dist}
\end{figure}

\subsection{Fluctuations in the UV background}
\label{subsec: uvbg-flucs}

The discreteness of the sources of radiation will produce local
fluctuations in the UV background. Several physical effects contribute
to the fluctuations.  Any single source will dominate the radiation
field sufficiently nearby. The result is a region of enhanced
photoionization around the source and a reduction in the local opacity
of the IGM. Known as the ``proximity effect,'' the diminished strength
of absorption by the \Lya forest near an ionizing source like a QSO
may be used to constrain the intensity of the UV background, as is
discussed in the next section. If the opacity of the IGM to ionizing
photons is sufficiently small that a large number of sources
distributed over large spatial scales contribute to the radiation
field, the fluctuations will be small. If the attenuation length,
which is the characteristic distance over which the radiation from
sources is diminished by IGM absorption, is smaller than the
characteristic clustering scale of the sources, then the radiation
field will be modulated by the large scale spatial distribution of the
sources. In the limit of such a short attenuation length that the
number of sources contributing to the radiation field in any given
region is subject to large Poisson fluctuations, the radiation field
will exhibit substantial fluctuations \cite{1992MNRAS.258...36Z,
  MW03}. This will become a particularly appreciable effect near the
reionization epochs of \HI\ and \HeII, but may be an important effect
well after the reionization epochs as well if the IGM opacity is
sufficiently large at high redshifts.

The magnitude of the fluctuations may be estimated by considering the
effect of Poisson fluctuations in the source numbers on the integrated
intensity. The attenuation length $r_0$ is equivalent to the mean free
path of an ionizing photon, the distance over which the mean number of
absorbers encountered is unity. The probability that a photon survives
a journey of a distance $r$ is then $\exp(-r/r_0)$. In general the
distance will depend on frequency, but most of the absorption takes
place near the photoelectric edge at frequency $\nu_T$. Provided the
distances are sufficiently small that cosmological effects may be
neglected, the generation of the radiation field may be considered in
the Euclidean limit. Any individual source of luminosity $L_\nu$ a
distance $r$ from a given location will then contribute a flux
$L_\nu\exp(-r/r_0)/4\pi r^2$ for $\nu>\nu_T$. Allowing for a
luminosity distribution function $\Phi(L_\nu)$ per unit luminosity per
unit volume with mean spatial density of sources between minimum and
maximum luminosities $L_{\rm min}$ and $L_{\rm max}$ of ${\bar n}=\int
dL_\nu\,\Phi(L_\nu)$, and defining ${\bar N}_0=(4\pi/3)r_0^3{\bar n}$
to be the mean number of sources within an attenuation volume, the
distribution of the dimensionless energy density $u=u_\nu/u^*$, where
$u^*\equiv(L^*/c)/4\pi r_0^2$ for a characteristic luminosity $L^*$,
becomes \cite{MW03}
\begin{eqnarray}
f(u) &=& \frac{1}{\pi}\int_0^\infty ds\, \cos\biggl[s{\bar N}_0
\int_{x_{\rm min}}^{x_{\rm max}}dx\, x\phi(x){\rm Re}G(sx)\nonumber\\
&&\phantom{{=}\frac{1}{\pi}\int_0^\infty ds\cos}-su\biggr]\nonumber\\
&&\times\exp\biggl[-s{\bar N}_0\int_{x_{\rm min}}^{x_{\rm max}}dx\,
x\phi(x){\rm Im}G(sx) \biggr],
\label{eq:u-dist}
\end{eqnarray}
where $x=L_\nu/L^*$ and $\phi(x)$ is the normalized luminosity
function of the sources
\begin{equation}
\phi(x) = \frac{\Phi(xL^*)L^*}{\int_{L_{\rm min}}^{L_{\rm max}} dL_\nu\,
\Phi(L_\nu)},
\label{eq:phix}
\end{equation}
$x_{\rm min}=L_{\rm min}/L^*$ and $x_{\rm max}=L_{\rm max}/L^*$. The
function $G(sx)$ is given by
\begin{equation}
G(sx)\equiv\int_0^\infty d\rho\, e^{isx\rho}\tau^3(\rho),
\label{eq:G}
\end{equation}
with $\tau(\rho)$ defined implicitly through
$\rho(\tau)\equiv\tau^{-2}e^{-\tau}$. For large $u$, the distribution
takes the asymptotic form $f(u)\sim(3/4){\bar N}_0\langle
x^{3/2}\rangle u^{-5/2}$, where $x^{3/2}$ is averaged over
$\phi(x)$. It follows that the variance $\langle u^2\rangle - \langle
u\rangle^2$ of the radiation energy density is undefined, diverging
like $u_{\rm max}^{1/2}$ for $u_{\rm max}\gg1$.

Examples of the resulting distribution function are shown in
Fig.~\ref{fig:u-dist}. A QSO luminosity function of the form
Eq.~(\ref{eq:BoyleLF}) was used with parameters given by the maximum
likelihood Pure Luminosity Evolution fit of \cite{Meiksin05} at $z=3$
(renormalized to $H_0=70\,{\rm km\, s^{-1}\, Mpc^{-1}}$):\
$\beta_1=1.24$, $\beta_2=2.70$, $\phi^*_{\rm
  proper}=2.6\times10^{-5}\, {\rm Mpc^{-3}\, mag^{-1}}$, and
$M^*=-27.2$ at restframe $\lambda=1450$\AA. In anticipation of the
discussion of helium reionization below
(\S~\ref{subsubsec:He-constraints}), the attenuation lengths were
fixed at 7, 15 and 33 (proper) Mpc, corresponding to averages of 1, 10
and 100 sources per attenuation volume. When the average is small, the
distribution becomes quite skewed towards low values, corresponding to
few or no sources within an attenuation volume.

The Poisson fluctuations of sources within an attenuation volume
will also give rise to spatial correlations in the radiation field.
The correlation function is \cite{1992MNRAS.258...45Z}
\begin{equation}
\xi_J(r)=\frac{1}{3\bar N_0}\frac{\langle x^2\rangle}{\langle x\rangle^2}
\frac{r_0}{r}\int_{r/r_0}^\infty\, du\, \frac{1}{u}
\ln\biggl(\frac{u+r/r_0}{u-r/r_0}\biggr)
e^{-u},
\label{eq:xi_J}
\end{equation}
where the averages $\langle\dots\rangle$ are performed over the
dimensionless QSO luminosity function $\phi(x)$ (eq.~[\ref{eq:phix}]).
For the luminosity function above at $z=3$, the coefficient $\langle
x^2\rangle/3\langle x\rangle^2\simeq2.6\times10^4$, so that
substantial correlations are expected on scales $r<r_0$.

\subsection{Observational constraints on the UV background}
\label{subsec: obs-constraints}

\subsubsection{The proximity effect}
\label{subsubsec: proximity}

The number of \Lya forest systems decreases systematically near bright
QSOs \cite{1986ApJ...309...19M, 1987ApJ...321...69T,
1991ApJ...367...19L}. The decrease, known as the ``proximity
effect,'' is expected sufficiently near a QSO where its ionizing flux
dominates the UV background \cite{1987ApJ...319..709C}.
\textcite{1988ApJ...327..570B} quantified the effect in terms of the
statistical properties of the \Lya forest to obtain an estimate of the
UV background.

The factor by which the \HI\ column density in an absorption system
will be reduced is $1+\omega$, where $\omega$ is the ratio of the
ionization rate $\Gamma_{\rm HI}^Q$ due to the QSO at a luminosity
distance $r_L$ between the QSO and the absorber to the metagalactic
ionization rate $\Gamma_{\rm HI}$. The ionization rate due to the
QSO is
\begin{eqnarray}
\Gamma_{\rm HI}^Q &=& \int \frac{d\nu}{h_{\rm P}\nu} \frac{L^Q_\nu}{4\pi r_L^2}
a_{\rm HI}(\nu)\nonumber\\
&=&\frac{1}{1+z_Q}\frac{d_L^2}{r_L^2}
\int d\nu \frac{f^Q_{\nu^\prime}}{h_{\rm P}\nu}a_{\rm HI}(\nu),
\label{eq:GammaQ}
\end{eqnarray}
where $f^Q_{\nu^\prime}$ is the observed flux from the QSO at
frequency $\nu^\prime=\nu/(1+z_Q)$ and $d_L=(1+z_Q)c\int_0^{z_Q}
dz/H(z)$ is the luminosity distance to the QSO at redshift $z_Q$ for a
flat universe ($\Omega_K=0$) \cite{Peebles93}. For a power-law \HI\
column density distribution $dN/dN_{\rm HI}\propto N_{\rm
HI}^{-\beta}$, the increase in intensity will decrease the number of
absorption systems above a given threshold $N_{\rm HI}^{\rm thresh}$
by the factor $(N_{\rm HI}^{\rm thresh})^{1-\beta}/ [N_{\rm HI}^{\rm
thresh}/(1+\omega)]^{1-\beta} =(1+\omega)^{-(\beta-1)}$. A mean line
density $dN/dz = N_0(1+z)^\gamma$ would then become
\begin{equation}
\frac{dN}{dz} = N_0(1+z)^\gamma [1+\omega(z)]^{-(\beta-1)}.
\label{eq:dNdz-omega}
\end{equation}
Measurements of $\omega$ as a function of distance $r_L$ from
the QSO as inferred from the depression in line density $dN/dz$
permits a determination of the metagalactic $\Gamma_{\rm HI}$.

Since \textcite{1988ApJ...327..570B}, several estimates of the
metagalactic ionization rate have been made using the proximity
effect. \textcite{2000ApJS..130...67S} performed a uniform analysis of
the spectra of 99 QSOs at moderate resolution ($\sim1$\AA) over the
redshift range $0<z<5$ to obtain an estimate of $\Gamma_{\rm
HI}=1.9^{+1.2}_{-1.0}\times10^{-12}\,{\rm s^{-1}}$ over
$1.7<z<3.8$. (They also provide references to much of the previous
literature.) The rates have generally exceeded the estimates of the
contribution from quasars. Several systematic effects could account
for some of the discrepancy. Redshifts for the quasars measured from
low ionization permitted lines like the Balmer lines or \MgIIs or from
forbidden lines like [\OIII]~$\lambda\lambda4959,\, 5007$\AA\ are
redshifted compared with \Lya and \CIVs by a few to several hundred
$\kms$ \cite{1992ApJS...80..109B, 1995ApJS...99....1L,
2001AJ....122..549V}. If the higher redshifts reflect the true
distances to the QSOs, then the QSO flux producing the proximity
effect will be overestimated using QSO redshifts based on \Lya or
\CIV, which in turn would result in an overestimate of the
metagalactic ionization rate by as much as a factor of two
\cite{2000ApJS..130...67S}. Numerical simulations suggest that
enhanced large-scale structure in the vicinity of a QSO will partly
compensate for the photoionizing influence of the QSO, again resulting
in an overestimate of the metagalactic rate by as much as a factor of
2--3 \cite{2008ApJ...673...39F}.

Quasar variability will introduce scatter in the estimates for
$\Gamma_{\rm HI}$ and degrade the fit of $\omega$ against $r_L$ for a
model assuming a constant quasar flux \cite{1988ApJ...327..570B}.
Episodic behavior of QSO luminosities could introduce a further
systematic in the estimates of the metagalactic ionization rate. If a
QSO satisfies a survey magnitude threshold as a result of entering a
bright phase, the neutral fraction will lag behind, resulting in an
overestimate of the metagalactic rate. From
Eq.~(\ref{eq:phionizeH-sol}) for $\phi\gg1$, it follows that the
actual metagalactic rate is a factor $1-(1+\omega)/\exp(-\Gamma_{\rm
HI}^{\rm tot}t)$ smaller than the estimate assuming equilibrium, where
$\omega$ is measured from the deficit in \Lya forest absorption lines
near the QSO and $\Gamma_{\rm HI}^{\rm tot}=\Gamma_{\rm HI} +
\Gamma_{\rm HI}^Q$. Time-averaging over a period $t_Q$ during which
the QSO is sufficiently bright to enter the survey produces the
average factor $1-(\Gamma_{\rm HI}t_Q)^{-1}[1-\exp(-\Gamma_{\rm
HI}^{\rm tot}t_Q)]$. For $t_Q\approx10^{12}-10^{13}$~s, this would
result in too low a value for $\omega$, and so too little suppression
of the \Lya forest line density over a distance $ct_Q$ from the QSO,
as is frequently found \cite{2000ApJS..130...67S}. An overestimate of
the metagalactic photoionization rate by several percent could result.

\begin{figure}
\includegraphics[width=8cm, height=6.5cm]{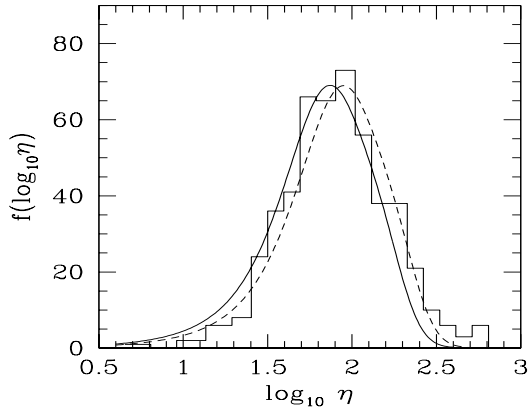}
\caption{Distribution of column density ratio $\eta = N_{\rm HeII}/
N_{\rm HI}$, for $N_{\rm HeII}$ measured from {\it FUSE} observations
and corresponding $N_{\rm HI}$ from Keck observations for absorption
features over the redshift range $2.30<z<2.75$ from
\textcite{2006A&A...455...91F}. Systems detected in spectral bins
affected by metal line absorption have been excluded. The systems are
further restricted to lie in the range $0.01<\tau_{0, \rm HI}<0.1$ to
assure completeness and reliable fits. Without this latter
restriction, the full distribution ranges over
$-0.2<\log_{10}\eta<3$. The solid curve shows the expected $\eta$
distribution allowing for fluctuations in the \HeII-ionizing
background for an average of 10 QSOs per attenuation volume. The
dashed line allows for a small increase in the mean $\eta$ values
within the measurement errors.
}
\label{fig:eta}
\end{figure}

\subsubsection{Constraints from helium ionization}
\label{subsubsec:He-constraints}

The ratio of \HeII\ to \HI\ column densities places a constraint on
the ratio of the metagalactic UV background at the \HeII\ and \HI\
photoelectric edges. From Eq.~(\ref{eq:phionizeHI-eq}) and the
corresponding equation for \HeII, assuming the helium (like the
hydrogen) is highly ionized and so almost entirely in the form of
\HeIII,
\begin{equation}
\eta\equiv\frac{N_{\rm HeII}}{N_{\rm HI}}=\frac{\Gamma_{\rm HI}}
{\Gamma_{\rm HeII}}\frac{n_{\rm He}}{n_{\rm H}}\frac{\alpha_{\rm HeIII}(T)}
{\alpha_{\rm HII}(T)},
\label{eq:eta}
\end{equation}
where $\Gamma_{\rm HI}$ and $\Gamma_{\rm HeII}$ are the photoionization
rates per atom of \HI\ and \HeII, respectively. For an approximate temperature
dependence of $\alpha_{\rm HII}\propto T^{-0.72}$ near $T=10^4$~K (and
similarly for \HeII) and a hydrogen-to-helium number density ratio of
12.9, it follows from Eq.~(\ref{eq:eta}), Eq.(\ref{eq:GHI}) and the
analogous equation for $\Gamma_{\rm HeII}$ that
\begin{equation}
\eta \simeq 1.68\frac{3+\alpha_{\rm L, HeII}}{3+\alpha_{\rm L, HI}}S_L
 = 0.42\frac{\Gamma_{\rm HI}}{\Gamma_{\rm HeII}},
\label{eq:eta_approx}
\end{equation}
where the ``softness parameter'' $S_L=u_{L, {\rm HI}}/u_{L, {\rm
HeII}}$ is the ratio of the metagalactic UV energy density at the \HI\
and \HeII\ Lyman edges \cite{MM94}, and $\alpha_{\rm L, HI}$ and
$\alpha_{\rm L, HeII}$ are the spectral indices just shortward of the
respective edges. (Allowing for the full temperature dependences of
the \HII\ and \HeIII\ radiative recombination rates introduces a weak
temperature dependence of $\eta\propto T^{0.055}$;
\textcite{2004ApJ...600..570S}.)  Because of the steepness of $u_\nu$
at energies just above 4~Ry (Fig.~\ref{fig:UVBG}), instead of the
softness parameter, it is usually more convenient to introduce the
ionization rate ratio $\Psi=\Gamma_{\rm HI}/\Gamma_{\rm HeII}$, as it
is this ratio which governs $\eta$ directly.

It follows from Eq.~(\ref{eq:tau0}) that the ratio of the \HeII\ to
\HI\ optical depths at line center is given by
\begin{equation}
\frac{\tau_{0, {\rm HeII}}}{\tau_{0, {\rm HI}}}=\frac{1}{4}
\frac{b_{\rm HI}}{b_{\rm HeII}}\eta \simeq 0.10\frac{b_{\rm HI}}{b_{\rm HeII}}\Psi.
\label{eq:HeII-to-HI-tauthin}
\end{equation}
The value of $\Psi$ is predicted to be large for a UV background
dominated by QSOs (Fig.~\ref{fig:UVBG}). For an intrinsic spectral
index of $\alpha_Q=1.5$, $\Psi\approx100$ at $z\approx3$, corresponding
to $\eta\approx40$ \cite{MM94, HM96}, while for $\alpha_Q=2$,
$\Psi\approx340$, corresponding to $\eta\approx140$
\cite{MM94}. Interpolating to $\alpha_Q=1.8$ gives $\Psi\approx220$ and
$\eta\approx90$. Additional photoionizing sources, like galaxies, will
increase $\Psi$ and $\eta$ further. The ratio of optical depths is thus
expected to be large, so that a strong \HeII\ \Lya signature is
expected from the IGM. The ratio also depends on the ratio of the
Doppler parameters. For pure thermal broadening, it follows from
Eq.~(\ref{eq:bparam}) that $b_{\rm HI}/b_{\rm HeII}=2$, and $\tau_{0,
{\rm HeII}}/ \tau_{0, {\rm HI}}\approx 0.2\Psi$ will be quite large. But
if velocity-broadening dominates (see Eq.~[\ref{eq:bbparam}]), then
$b_{\rm HI}/b_{\rm HeII}\approx1$ and $\tau_{0, {\rm HeII}}/ \tau_{0,
{\rm HI}}\approx 0.1\Psi$. Systems optically thin in \HI\ \Lya with
$\tau_{0, {\rm HI}}$ as low as 0.1 will form saturated \HeII\ \Lya
lines.

Because of line-blanketing, the measured optical depths will scale
according to Eq.~(\ref{eq:taul}). A large effective optical depth due
to blanketing is expected \cite{1993MNRAS.262..273M, MM94,
1995ApJ...451..477G}. This has been confirmed by observations (see
\S~\ref{subsubsec:helium}). The estimate of \textcite{MM94} requires
$\Psi>200$ for the 95\% lower limit $\tau_{\rm HeII}>1.5$ at $z=3.2$
of \textcite{1997AJ....113.1495H} if the absorbers are thermally
broadened, but $\Psi > 1000$ to reach the 95\% upper limit of
$\tau_{\rm HeII}<3.0$. For velocity-broadened absorbers, however, for
which \textcite{Zheng04} find evidence, the upper limit may be
satisfied by $\Psi<500$. If the lower limit $\tau_{\rm HeII} > 4.8$
found by \textcite{2000ApJ...534...69H} applies, however, then $\Psi >
1000$ is required even for velocity broadened absorbers.

There is additional reason to believe values of $\Psi>1000$ occur at
$z>3$. While the detection of a high \HeII\ \Lya optical depth was
expected, the wide range in fluctuations in $\eta$ came as a
surprise. Some of the systems indicate values as large as $\eta>1000$
at $z\lta3$ based on follow-up optical observations of the \HI\ \Lya
systems corresponding to the \HeII\ systems detected \cite{Zheng04,
  Reimers05, 2006A&A...455...91F}. Combining $N_{\rm HeII}$
measurements from {\it FUSE} observations over the redshift range
$2.30<z<2.75$ with measurements of $N_{\rm HI}$ for the corresponding
\HI\ systems using Keck observation, \textcite{2006A&A...455...91F}
obtain a very broad distribution for $\eta$, as shown in
Fig.~\ref{fig:eta}. Systems detected in spectral bins affected by
metal line absorption have been excluded. The systems are further
restricted to those with line center optical depths in the range
$0.01<\tau_{0, \rm HI}<0.1$ to assure completeness and reliable
fits. Without this latter restriction, the full distribution ranges
over $-0.2<\log_{10}\eta<3$. A trend of decreasing $\eta$ with
decreasing redshift is also found \cite{Zheng04,
  2006A&A...455...91F}. Only fluctuations producing an extremely soft
spectrum with $\Psi>2000$ could account for the largest $\eta$
excursions. Such high values could arise from strong local
\HI-ionizing sources, or from a broad distribution in the
\HeII-ionizing background with occasional low excursions. Higher
values of $\eta$ are found in \HI\ voids, regions with $\tau_{\rm
  HI}<0.05$ \cite{2004ApJ...600..570S, 2006A&A...455...91F}, again
consistent with a local overdensity of hydrogen-ionizing sources
compared with \HeII-ionizing sources producing the \HI\ voids. The
trend is also consistent with an increase in the relative attenuation
length of hydrogen-ionizing photons compared with \HeII-ionizing
photons, so that more distant hydrogen-ionizing (but not
\HeII-ionizing) sources are able to add to the radiation field in the
\HI\ voids.

Monte Carlo realizations of the expected fluctuations in $\tau_{\rm
  HeII}$ were performed by \textcite{1998AJ....115.2206F}, allowing
for the discreteness of the absorption systems and of the quasars. At
$z=3$, they find a range of values $1.3\lta \tau_{\rm HeII}\lta2.3$,
peaking at $\tau_{\rm HeII}\approx1.9$. Since the \HeII\ column
density scales inversely with the local intensity of the UV background
(provided most of the helium is ionized to \HeIII), the contribution
of source discreteness to the distribution in the \HeII\ to \HI\
column density ratio $\eta$ may be predicted from
Eq.~(\ref{eq:u-dist}) as $f(\eta) = f(u)\vert du/d\eta\vert
=f(u)u^2/[\eta(\langle u\rangle)\langle u\rangle]$, where
$\eta=\eta(\langle u\rangle) \langle u\rangle/u$.\footnote{Because the
  \HI-ionizing radiation field will be much more narrowly distributed,
  its fluctuations will contribute much less to the fluctuations in
  $\eta$ and so may be neglected, except for large values due to the
  proximity effect. The joint \HI\ and \HeII\ proximity effect would
  need to be modeled separately, including the effect on changing the
  attenuation length, which may be especially important for
  \HeII-ionizing photons when the attenuation length and range of the
  proximity effect are comparable.}  The flattening of $f(u)$ for
$u\ll1$ shows that the average of $\eta\propto 1/u$ is undefined,
diverging logarithmically at the upper end due to regions of low
ionization.\footnote{The presence of diffuse radiation will limit the
  maximum values of $\eta$, as will the approach to near full
  recombination to \HeII. The maximum value, however, will still be so
  large that no average will be well-determined.}

Using the distribution shown in Fig.~\ref{fig:u-dist} for an average
of $N_0=10$ QSO sources per attenuation volume, the resulting
distribution for $\eta$ is shown in Fig.~\ref{fig:eta}. The
distribution function shown by the solid line is computed as the
average of two distributions matched to the measured distributions
with mean and spread of $\log_{10}\eta=1.70\pm0.32$ at $z\simeq2.33$
and $\log_{10}\eta=1.98\pm0.32$ at $z\simeq2.68$
\cite{2006A&A...455...91F}. The dashed line uses means increased by
$0.25\sigma$ to $\log_{10}\eta=1.78$ and 2.06. The measured values are
well described by the model, except for a small tail at
$\log_{10}\eta\simeq2.6$. These values may be produced by patches of
only partially ionized helium, as would remain during the epoch of
\HeII\ reionization before its completion \cite{Zheng04,
Reimers05}. Alternatively, they may result from local \HI-ionizing
sources reducing the \HI\ column density, but not the \HeII. Effects
of large scale structure and radiative transfer on the fluctuations
were explored using numerical simulations by \textcite{MF05},
\textcite{BHVC06} and \textcite{2007MNRAS.380.1369T}
(see \S~\ref{subsubsec:sim-helium} below).

The measured values for $\eta$ may be compared with the expectation
for a UV background dominated by QSOs. By Eq.~(\ref{eq:eta_approx}),
the mean values for $\eta$ correspond to $140<\Psi<270$. From
Eq.~(\ref{eq:GHI-emiss}), $\Psi\simeq (\epsilon_L^{\rm HI}/
\epsilon_L^{\rm HeII})(a_T^{\rm HI}/ a_T^{\rm HeII})(r_{\rm
eff}^{\rm HI}/ r_{\rm eff}^{\rm HeII})$. Approximating the emissivity
as $\epsilon_\nu\propto \nu^{-\alpha_S}$, taking $r_{\rm eff}^{\rm
HI}\simeq70$~Mpc from Eq.~(\ref{eq:r0}) below and $r_{\rm eff}^{\rm
HeII}\simeq15$~Mpc for 10 QSOs in an attenuation volume at the \HeII\
Lyman edge (\S\ref{subsec: uvbg-flucs}), gives $\Psi\simeq150-300$ for
$1.5<\alpha_S<2.0$, producing a good match to the required average
value of $\Psi$ for a plausible emissivity spectral shape due to QSO
sources.

\begin{figure}
\includegraphics[width=8cm, height=6cm]{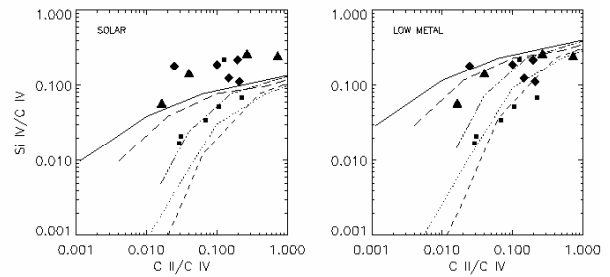}
\caption{Comparison of the observed values of \SiIV/\CIVs vs
\CII/\CIVs with model predictions using CLOUDY with an input
$\nu^{-1.5}$ power law spectrum. Various breaks are allowed across the
\HeII\ Lyman edge:\ dashed line -- no break; dotted line -- factor of
2; dash-dotted line -- factor of 10; long dashed line -- factor of
100; solid line -- factor of 1000. The left panel shows results for a
solar abundance ratio of Si/C; the right panel uses a ratio of Si/C 3
times solar. The differently shaped points indicate different lines of
sight. Small points indicate $z<3.1$ absorbers, while large indicate
$z>3.1$. From \textcite{1996AJ....112..335S}. (Figure reproduced by
permission of the AAS.)
}
\label{fig:metals}
\end{figure}

\subsubsection{Constraints from metal ionization}
\label{subsubsec: metal-constraints}

The presence of metal absorption systems permits further constraints
on the shape of the UV metagalactic background. The ionization edges
for several transitions of measured metal ions, primarily carbon,
nitrogen, oxygen and silicon, lie in the energy range 1--10~Ry, as
shown in Fig.~\ref{fig:UVBG}. The measured column density ratios of
ionized states of these elements have been used to distinguish a UV
background dominated by hard sources like QSOs or Active Galactic
Nuclei (AGN) from a background dominated by galaxies/stars
\cite{1991ApJ...376L..33M}. The inferred ionization fractions may also
be used to convert the measured column densities of the detected ions
compared with the \HI\ column density into physical densities, sizes
and total metallicities \cite{1986A&A...169....1B,
  1991ApJ...383..511D, 1994ApJ...436...33B}. The findings suggest \Lya
forest systems have metallicities less than 0.01 solar, while Lyman
Limit Systems have metallicities from less than 0.01 solar up to 0.1
solar. Some systems show evidence for multiple phases with densities
different by at least a factor of 10 \cite{1994ApJ...436...33B}. Sizes
of typically tens of kiloparsecs are found, with total hydrogen
densities of roughly $10^3-10^5\,{\rm m}^{-3}$.

Based on a comparison of \SiIVs to \CIVs vs \CIIs to \CIV,
\textcite{1996AJ....112..335S} find systematically different behaviors
for systems at $z<3.1$ compared with $z>3.1$, as shown in
Fig.~\ref{fig:metals}. The \SiIVs to \CIVs ratio for a given \CIVs to
\CIIs ratio cannot be satisfied for any softness parameter without
enhancing the Si/C ratio to about 3. In this case, the high redshift
systems require a break of a factor of 10 to 100 at the \HeII\ Lyman
edge, while little or no break is required by the lower redshift
systems. The large break at $z>3$ is similar to the finding based on
the \HeII\ to \HI\ ratios found at $z\simeq3$ discussed above. Refined
estimates of the UV background, including a plausible contribution
from starburst galaxies, suggest the Si/C ratio need not exceed twice
solar \cite{1997AJ....113.1505G}. Based on observations of metal
systems at $z>3$, \textcite{1997A&A...318..347S} and
\textcite{1998A&AS..127..217D} similarly find a much larger ratio of
\SiIVs to \CIVs column densities compared with the expectation for a
QSO or AGN dominated UV background. They find a large softness
parameter between the \HeII\ and \HI\ Lyman edges of $S_L>1000$ is
required to bring the overabundance of silicon over carbon to below a
factor of 10 over solar.

Using 19 high quality QSO spectra over the redshift range $1.5<z<4.5$,
\textcite{2003ApJ...596..768S} and \textcite{2004ApJ...602...38A}
perform a pixel optical depth analysis of \CIV, \SiIVs and \HI\ in
conjunction with simulated spectra. They find evidence for a highly
inhomogeneous distribution of carbon with a cosmic mean of ${\rm
[C/H]}=-2.80\pm0.13$ at $z\simeq3$ and no evidence for evolution over
the redshift range $1.8<z<4.1$, adopting the galaxy-QSO metagalactic
spectrum of \textcite{2001cghr.confE..64H}, rescaled to reproduce the
measured mean \Lya\ transmitted flux. Introducing a large break in the
metagalactic background at energies above 4~Ry, the \HeII\
photoelectric threshold, for all redshifts indicates a decrease in
metallicity with time, which could only be physical if large
quantities of metal-poor gas were accreting onto the metal-enriched
absorption systems or replacing them in the surveys as they accreted
onto galaxy halos, possibilities that perhaps should not be
discounted. Introducing the large break only at $z>3.2$, as would
occur if \HeII\ were ionized late, however, yields a near constant (at
most a weakly increasing with time) carbon abundance. They find no
redshift evolution in the ratio $\tau_{\rm SiIV}/\tau_{\rm CIV}$, and
the silicon and carbon data require no sharp spectral change in the UV
background at $z\simeq3$. They find a strong correlation between
$\tau_{\rm SiIV}/\tau_{\rm CIV}$ and $\tau_{\rm CIV}$, and argue the
redshift effect reported by \textcite{1996AJ....112..335S} may result
from a different range in $\tau_{\rm CIV}$ probed as a function of
redshift, possibly due to different quality spectra for samples at
$z<3$ vs $z>3$. They also infer a Si/C ratio of ${\rm
[Si/C]}=0.77\pm0.05$, with no indication for evolution in the ratio.

In a study of metal absorption systems in the spectra of $z\approx2.5$
QSOs with intervening \CIVs and \OVIs detections,
\textcite{2004ApJ...606...92S} find roughly 30\% of the \Lya forest is
enriched to abundances below $[{\rm C,\, O/H}]\lta-3.5$. The systems
are adequately modeled using a UV background dominated by QSOs or AGN
having a power-law source spectrum $\nu^{-1.8}$, assuming a solar
abundance ratio of carbon to oxygen. If instead the absorption systems
are overabundant in oxygen by $[{\rm C/O}]=-0.5$, as found for
metal-poor Galactic halo stars, then it is necessary to include a
galactic radiation component.

Modeling metal absorption systems at $z\simeq2$ using CLOUDY and an
assumed UV metagalactic background with contributions from both QSO
sources and galaxies, \textcite{2006ApJ...637..648S} infer a trend of
decreasing sizes and increasing densities from high to low ionization
states. The largest, lowest density systems, with sizes of 0.1--10~kpc
and densities of $2.5\lta \log_{10}n_{\rm H}\lta3.5$ (in units of
${\rm m^{-3}}$), are seen exclusively in \CIVs or \SiIVs. These sizes
are comparable to the minimum inferred sizes based on coincident
absorption systems along neighboring lines of sight
(\S~\ref{subsubsec:metals}).  The lower ionization species \CIIs and
\SiIIs have sizes of 0.1--1~kpc and densities of
$n_{\rm}\simeq10^4\,{\rm m^{-3}}$, while \MgIIs and \AlIIs systems
have inferred sizes of about 100~pc and densities of $n_{\rm
H}\simeq10^5\,{\rm m^{-3}}$. The smallest and densest structures are
\FeIIs absorbers, with sizes of about 1~pc and densities of $n_{\rm
H}\simeq10^5\,{\rm m^{-3}}$.

Comprehensive analyses of metal absorption systems between $1.4<z<3$
that are optically thin at the \HI\ Lyman edge were carried out by
\textcite{2005A&A...441....9A}, \textcite{2006A&A...449....9R} and
\textcite{2007A&A...461..893A}. They employ a Monte Carlo Inversion
procedure which allows for a random distribution of gas density and
velocity along the line of sight. The procedure assumes a constant
metallicity for each system, that the gas is optically thin to the
ionizing UV radiation, assumed uniform, and that the gas is in thermal
and ionization equilibrium. The ionizing UV background is expressed in
parametric form. All parameters of the model are then varied to obtain
the best fit to the metal absorption features, using CLOUDY to compute
the ionization states and temperatures. The method is described in
detail by \textcite{2000A&A...360..833L},
\textcite{2005A&A...441....9A} and references therein. Applied to
individual systems, a range in \HeII\ to \HI\ column densities is
inferred from the data \citep{2005A&A...441....9A}, suggestive of a
variable UV radiation field. Since the variability is generally based
on systems optically thick to \HeII-ionizing radiation, the method
does not distinguish between a UV radiation field that is variable on
the metagalactic scale external to the absorbers and a radiation field
that differs between the absorbers as a consequence of internal
absorption.

These authors find a range of hydrogen densities from the analyses,
with values as low as $n_{\rm H}<10^2\,{\rm m}^{-3}$ to as high as
$n_{\rm H}>10^4\,{\rm m}^{-3}$, similar to the findings of
\textcite{2006ApJ...637..648S}. The line-of-sight sizes of the metal
absorbers range from a few parsecs to over 100~kpc. Systems at
$z\lta3$ are best fit by imposing a strong reduction in flux between
3~Ry and 4~Ry, as shown in Fig.~\ref{fig:UVBG}. Such a reduction could
be accounted for by strong \HeII\ \Lya absorption. At $z<1.8$, a much
shallower break at 4~Ry is required compared with the model of
\textcite{HM96}, suggesting there may be fewer high column density
systems filtering the \HeII\ ionizing radiation than assumed by the
model. The radiation field at $z<3$ is dominated by sources with
QSO/AGN spectra. Only a very small soft spectrum component from
galaxies may contribute, suggesting an escape fraction of $f_{\rm
  esc}<0.05$. Large fluctuations in the UV background are also
required. At $z\lta3$, these are likely produced predominantly by the
effects of radiative transfer through an IGM optically thick to
\HeII-ionizing photons. The optical depth of \HeII-ionizing photons is
dominated by systems with $15<\log_{10}N_{\rm HI}<16$, while much
rarer systems with $\log_{10}N_{\rm HI}>17.2$ dominate the
\HI-ionizing photon optical depth. As a result, the mean free path of
\HeII-ionizing photons is much shorter than for \HI-ionizing photons,
and relatively fewer sources contribute to the \HeII-ionizing
radiation background in any given region. By $z\lta2$, however,
$\log_{10}N_{\rm HI}<16$ systems are sufficiently sparse that the IGM
should become optically thin to \HeII-ionizing radiation to large
distances, and the fluctuations are instead likely produced
predominantly by the spread in the spectral indices of
quasars. Fluctuations in $\eta=N_{\rm HeII}/ N_{\rm HI}$ and an
anti-correlation between $\eta$ and $N_{\rm HI}$ may be accounted for
by the filtering through the IGM of radiation from QSO sources with a
variety of spectral shapes and by differences in the mean densities of
the absorbing clouds. The progressive hardening of the UV background
towards lower $z$ would also account for the decrease in the median
values for $\eta$ as $z$ decreases \cite{Zheng04,
  2006A&A...455...91F}.

The modeling has some limitations, especially for systems with large
values of $\eta$. A system with $\log_{10} N_{\rm HI}<17.2$ is
optically thin to \HI-ionizing radiation but thick to \HeII-ionizing
radiation, so that a break at 3--4~Ry must be introduced to account
for the radiative transfer internal to the system. The assumption of
thermal equilibrium assumes adiabatic cooling or heating are
negligible. The timescales are also long for the more rarefied
systems. For a temperature of $T\approx20000$~K, the cooling time due
to radiative recombination to \HI\ exceeds $10^9$~yr for $n_{\rm
HI}<200\, {\rm m}^{-3}$, which is becoming long, particularly if flows
occur on shorter timescales so that adiabatic gains or losses are not
negligible. \textcite{2004ApJ...606...92S} also point out that the
\OVIs abundances are sensitive to the poorly determined soft x-ray
background. Still, the studies have lent considerable insight to the
physical nature of the metal absorption systems and the background UV
radiation field that ionizes them.

\section{Early Models}
\label{sec:models}

Several models have been suggested for describing the physical
structure of the \Lya forest absorption systems. All of the early
models have been supplanted by numerical cosmological simulations,
which provide a largely successful description of the \Lya forest and
associated absorption systems within the broader context of
cosmological structure formation. Vestiges of the early models,
however, survive. In this section, elements of those models are
discussed that continue to provide physical insight into the findings
of the simulations, or which may still offer viable alternative
descriptions of some absorption systems.

\subsection{Pressure-confined clouds}
\label{subsec:PC-clouds}

The earliest physical model for the absorbers was one in which
the absorption features arise from distinct clouds pressure-confined
by a hot IGM \cite{1980ApJS...42...41S}. Denoting the cloud temperature
and hydrogen number density by $T_c$ and $n_c$, and those of the IGM
by $T_{\rm IGM}$ and $n_{\rm IGM}$, the condition of pressure
equilibrium is $n_cT_c=n_{\rm IGM}T_{\rm IGM}$. Allowing for
clouds photoionized by a QSO/AGN-dominated UV background,
small enough to be Jeans stable and able to withstand thermal evaporation
by the hot confining medium, while requiring the IGM to meet the
Gunn-Peterson constraint, \textcite{1980ApJS...42...41S} found for
the cloud parameters at $z\approx2.4$, $n_c=100\,{\rm m^{-3}}$,
$T_c=3\times10^4$~K, $n_{\rm IGM}=10\,{\rm m^{-3}}$ and
$T_{\rm IGM}=3\times10^5$~K. These values were somewhat modified
by \textcite{1983ApJ...268L..63O} and others.

The pressure-confined models suffered from a serious shortcoming:\ a
wide range of sizes was required to reproduce the range in observed
\HI\ column densities. The higher column density systems would need to
be so large that they would become unstable under their own
gravity. Jeans stability limits the size of a self-gravitating cloud
of density $n_c$ and temperature $T_c$ to be sufficiently small that a
sound wave can cross it in less than the time for it to
gravitationally collapse. This requirement is necessary to ensure the
internal gas pressure gradient of the cloud is able to balance its
weight. For an isothermal gas, this corresponds to a maximum size of
\begin{eqnarray}
\lambda_J &=& \biggl(\frac{\pi k_{\rm B}T}{\bar m}\biggr)^{1/2}
\frac{1}{(G\rho_0)^{1/2}} \nonumber\\
&\simeq& 5.2\times10^{21}
\biggl(\frac{n_{\rm IGM}T_{\rm IGM}}{10^6\,{\rm K\,m^{-3}}}\biggr)^{1/2}
\biggl(\frac{100\,{\rm m^{-3}}}{n_c}\biggr)\,{\rm m}.
\label{eq:Jeanslength}
\end{eqnarray}
In ionization equilibrium with an ionization rate $\Gamma_{\rm
HI}\approx10^{-12}\, {\rm s}^{-1}$, Eq.~(\ref{eq:phionize-xiHI-eq})
gives a neutral hydrogen density for a cloud at $T_c=3\times10^4$~K of
$n_{\rm HI}\simeq0.002\,{\rm m^{-3}}$. This gives a maximum column
density for a Jeans stable cloud of $N_{\rm HI,
max}\simeq10^{15}\,{\rm cm}^{-2}$. The observed distribution extends
to values two orders of magnitude larger before the absorbers become
Lyman Limit Systems. A further failure of the model is that a system
with a column density of $N_{\rm HI}=10^{14}\,{\rm cm}^{-2}$ would
have a typical size of 17~kpc. This conflicts with the sizes inferred
from coincident absorbers along neighboring lines of sight to lensed
QSO images, which give cloud sizes in excess of 70~kpc for systems
with $N_{\rm HI}<10^{14}\,{\rm cm}^{-2}$ (\S~\ref{subsubsec:sizes}).

Elements of the pressure-confined model, however, may still describe
some absorption systems. The clouds could, for example, be confined by
a wide range of pressures in different environments, which are
themselves gravitationally confined \cite{1989ApJ...337..609B}. The
association of metal absorption systems with galaxies suggests they
may be pressure-confined systems within galactic halos, as may be
Lyman Limit Systems \cite{1996ApJ...469..589M}.

\subsection{Dark matter minihalos}
\label{subsec:minihalos}

A more viable physical model for the absorbers is gravitational
confinement by a dark matter halo. \textcite{1986MNRAS.218P..25R}
recognized that the \Lya forest is a necessary consequence of the Cold
Dark Matter (CDM) scenario for structure formation. The same
normalization of the CDM primordial matter power spectrum that
recovers the clustering of galaxies predicts an abundance of low mass
minihalos with virial velocities comparable to the sound speed of
photoionized gas. The gas trapped in these halos will produce
absorption features with \HI\ column densities and line widths
comparable to those observed \cite{1986Ap&SS.118..509I,
1986MNRAS.218P..25R}.

To fix parameters, consider the formation of a minihalo through
gravitational collapse. The formation of a dark matter halo may be
approximated by the spherical collapse model. The evolution of the
radius $r(t)$ of a spherical shell of mass $M$ in an Einstein-deSitter
universe, which provides a good description at high redshifts, is
given by the parametric solution \cite{1980lssu.book.....P}
\begin{equation}
r = r_V(1-\cos\theta), \qquad t = \frac{r_V}{v_{\rm circ}}(\theta-\sin\theta),
\label{eq:tophat}
\end{equation}
where $r_V$ is the virial radius of the shell and $v_{\rm
circ}=(GM/r_V)^{1/2}$ is the circular velocity of the shell at the
time of virialization. For a top hat initial spherical perturbation,
the central dark matter density at the time of virialization is
$\rho_{\rm DM}=(9/8)(3\pi+2)^2\rho_{\rm crit}(z_V)$, where $\rho_{\rm
crit}(z_V)$ is the critical Einstein-deSitter density at the
virialization epoch at redshift $z_V$. Since the gas is dissipative,
its core radius will be smaller than the virial radius of the dark
matter halo. Approximating the gas as being in hydrostatic equilibrium
within a uniform density dark matter halo, the gas will have a
gaussian profile $\rho(r)=\rho(0)\exp(-r^2/r_c^2)$ with core radius
\begin{equation}
r_c=\biggl(\frac{k_{\rm B}T/{\bar m}}{2\pi G\rho_{\rm DM}/3}\biggr)^{1/2}
\simeq13(1+z_V)^{-3/2}T_4^{1/2}h^{-1}\,{\rm kpc},
\label{eq:rcore}
\end{equation}
where $T_4$ is the gas temperature in units of $10^4$~K. A minihalo
virialized at $z_V=3$ will be only a few kiloparsecs in size. A
minihalo of mass $10^9\,M_\odot$ would have an adequate circular
velocity to confine the gas. The \HI\ column densities produced
will range roughly over $13<\log_{10}N_{\rm HI}<16$. Higher
column density systems become self-gravitating and Jeans
unstable \cite{Meiksin94}, and will likely produce stars.

\textcite{1988ApJ...324..627B} point out that the gas will in general
be in a dynamical state, either collapsing with an increasing core
density or expanding after photoionization. The observational
signatures and thermal properties of minihalos were explored by
\textcite{Meiksin94}. Gas associated with the minihalos gives rise to
absorption features with $N_{\rm HI}\lta10^{13}\,{\rm cm}^{-2}$ on
scales of 100~kpc at $z=3$. Even though the gas is in motion, the line
profiles are found to remain nearly Doppler in shape, although
broadened by the gas motion. In the idealized case of an isothermal
gas in free expansion, so that the gas profile is gaussian with a
linear velocity profile $v=v_c(r/r_c)$, the Voigt profile is exactly
preserved for lines of sight passing through the cloud, but with the
Doppler parameter broadened to $b=(2k_{\rm B}T/m_{\rm
H}+v_c^2)^{1/2}$. Typical temperatures in the inner region of
minihalos with circular velocities of $v_c\simeq40\kms$, which give
rise to systems with $N_{\rm HI}>10^{15}\,{\rm cm^{-2}}$, are
$30-40\times10^4$~K, consistent with the measurements of
\textcite{1996ApJ...467L...5R} for metal absorption systems expected
to be associated with comparable \HI\ column densities (see
\S~\ref{subsubsec:metals} above). In the inner 10--30~kpc,
temperatures of $T\simeq30-60\times10^4$~K are predicted, with
substantial deviations from thermal equilibrium due to adiabatic
heating of the infalling gas. In the outer 10--100~kpc, the gas is
significantly adiabatically cooled by Hubble expansion. Temperatures
half these values are expected in smaller halos of $v_c\simeq20\kms$.

Further consequences of the minihalo model, including number counts,
effects of photoionization, and ablation due to ram pressure on
minihalos moving relative to their environments, have been explored by
\textcite{1989MNRAS.236P..21I, 1989PASJ...41.1095I},
\textcite{1990PASJ...42L..11M}, \textcite{1993MNRAS.260..617M}, and
\textcite{1993ApJ...409...42M, 1994ApJ...420...68M}.

\subsection{Caustics and sheets}
\label{subsec:sheets}

A gravitational perturbation will generically form sheets since the
enhanced density along a given direction will accelerate the collapse
along that direction. Once a growing density fluctuation becomes
non-linear, it will produce a caustic. The caustic formation is
well-described by the Zel'dovich approximation
\cite{1970A&A.....5...84Z}
\begin{equation}
{\bf x}({\bf q}, t) = {\bf q} - D(t){\bf \nabla}\Psi({\bf q}),
\quad {\bf v}({\bf q}, t) = -a{\dot D}(t){\bf \nabla}\Psi({\bf q}),
\label{eq:Zeld}
\end{equation}
where ${\bf x}$ is the comoving position of a particle with initial
comoving position ${\bf q}$, ${\bf v}=a\dot{\bf x}$ is the peculiar
velocity, $\Psi({\bf q})$ describes the initial deformation of the
density field, the gradient is with respect to ${\bf q}$, and $D(t)$
is the growth factor for linear perturbations $D(t) = ({\dot
a}/a)\int^a da/ {\dot a}^3$ \cite{Peebles93}. The physical radius is
related to the comoving radius through ${\bf r}=a(t){\bf x}$, where
$a(t)=1/(1+z)$ is the expansion factor of the Universe at the epoch
corresponding to redshift $z$. Conservation of mass gives for the
density, up until the time of caustic formation, $\rho({\bf x}, t) =
\rho({\bf q})/\vert d^3{\bf x}/d^3{\bf q}\vert$, where the denominator
is the determinant of the Jacobian of the coordinate
transformation. For 1D collapse, the density prior to caustic
formation grows as $\rho(x, t)=\rho_0 a(t)^{-3}/
[1-D(t)d^2\Psi/dq^2]$, which is exact for the 1D collapse of a slab.
Gravitational instability ensures that a uniform density ellipsoidal
density perturbation will collapse most rapidly along its shortest
axis, forming a ``Zel'dovich pancake.''

Baryons infalling along with the dark matter will produce \Lya forest
absorption features. The line profiles will become distorted at the
time the caustic forms before the baryons come into dynamical
equilibrium \cite{1990MNRAS.242..544M, 1993MNRAS.260..617M,
Meiksin94}. The \HI\ column densities produced will range over about
$13<\log_{10}N_{\rm HI}<15$, with postshock temperatures peaking at
$50-60\times10^4$~K \cite{Meiksin94}.

\subsection{Galactic models}
\label{subsec:cheshire}

Ever since the earliest papers on the \Lya forest, it was recognized
that galaxies should give rise to intervening absorption
systems. Their characteristics and frequency, however, are
unclear. The interstellar medium of disk galaxies is expected to
produce damped \Lya forest absorbers, although relatively few DLAs
have been shown to arise from disks (see \S~\ref{subsubsec:DLA-galaxy}
below). Other suggestions include condensates in cooling outflows from
dwarf galaxies \cite{1995ApJ...444L..17W}, the gas-rich dwarf galaxies
themselves \cite{1986ApJ...311..610Y}, debris from satellite galaxies
lost through tidal stripping by a central galaxy or expelled by
supernovae \cite{1993ApJ...415..174W, 1994ApJ...427..696M}, or
pressure-confined clouds in a two-phase gaseous galactic halo medium
\cite{1996ApJ...469..589M}. The outflow model in particular may give
rise to distorted absorption line features if a line of sight
intercepts the outflowing diffuse gas.

Another suggestion is that the absorption lines arise from gas in
extended disks with too low surface density to produce many stars, the
``Vanishing Cheshire Cat'' model of \textcite{1993AJ....106.1265S}.
In this model, an inner disk gravitationally-confined region of radius
about 15~kpc is surrounded by an extended pressure-confined disk of
radius about 250~kpc. The inner region has a low gas surface density
compared with most disk galaxies, but sufficiently large to produce a
DLA feature in any line of sight passing through it. The outer region
has such a low surface density that the hydrogen is photoionized by
the metagalactic UV background. The outer region will give rise to
\Lya forest features. A transition region will produce Lyman Limit
Systems. The relative abundances of the various \HI\ column densities
is accounted for by the progressively diminishing geometric cross
section for increasing column densities. The transition between
gravitational confinement and pressure confinement may be adjusted to
account for deviations from a perfect power law in the column density
distribution. The ``Cheshire Cat'' nature of the galaxy arises once
star formation begins, producing a starburst galaxy. Metal-enriched
gas ejected by the stars will accelerate the cooling and strengthen
the burst, resulting in the expulsion of the gas in the central region
and quenching further star formation. The result will be an outer
``smiling'' gaseous disk, possibly enriched in gas and metals by a
galactic fountain if material from the starburst falls onto the disk,
and a very low surface brightness ``invisible'' galaxy in the center.

Aspects of the model, and the balance of gravitational and pressure
forces in slab models generally, have been explored by
\textcite{1993ApJ...402..493C}, \textcite{1994ApJ...430L..29C} and
\textcite{1995ApJ...441...51S}.

\section{Numerical Simulations}
\label{sec:simuls}

\subsection{Cosmological structure formation}
\label{subsec:struct-form}

\subsubsection{Dynamical evolution of the dark matter and baryons}
\label{subsubsec:DMbflucs}

The successful description of the large scale distribution of galaxies
by the Cold Dark Matter dominated scenario of cosmological structure
formation \cite{1985ApJ...292..371D} led to attempts to account for
the \Lya forest within the same scenario. Using a combined
hydrodynamics and gravity code, \textcite{Cen94} demonstrated that the
baryons will trace similar small scale structure as does the dark
matter due to gravitational instability in the Cold Dark Matter
scenario. Several other groups soon followed with similar results
\cite{ZAN95, 1997ApJ...485..496Z, 1996ApJ...457L..51H,
1996ApJ...471..582M, 1997ASPC..123..332W}.

In the gravitational instability scenario of structure formation,
cosmological structures grow from smaller structures through gravity.
The growth of structures is governed by the coupled gravity-fluid
equations describing both the collisionless gravitating matter
component and the collisional baryonic component. Because of the
enormous change in spatial scales as the Universe expands, it is most
convenient to solve the equations in a frame comoving with the
expansion \cite{Peebles93}. The physical position ${\bf r}$ and
velocity ${\bf u}$ are then expressed in terms of the comoving
position ${\bf x}$ and peculiar velocity ${\bf v}$ according to
\begin{equation}
{\bf r} = a(t){\bf x}, \quad {\bf u} = {\dot {\bf r}} = {\dot a}{\bf x}
+{\bf v},
\label{eq:comcoords}
\end{equation}
where ${\bf v}=a{\dot{\bf x}}$ and $a(t)=1/(1+z)$ is the cosmological
expansion factor corresponding to the cosmological epoch at the
redshift $z$.

The dynamical equations for the dark matter component are
\begin{eqnarray}
\frac{d{\bf x}_d}{dt} &=& {\bf v}_d({\bf x}, t),
\label{eq:dxddt}\\
\frac{d{\bf v}_d({\bf x}, t)}{dt} + \frac{\dot a}{a}{\bf v}_d &=&
-\frac{1}{a}{\bf \nabla}\phi({\bf x}, t),
\label{eq:dvddt}
\end{eqnarray}
where $\phi({\bf x}, t)$ is the gravitational potential, and the
gradient is with respect to the comoving coordinate system. The fluid
equations expressing mass and momentum conservation are
\begin{eqnarray}
\frac{\partial\rho({\bf x},t)}{\partial t} + 3\frac{\dot a}{a}\rho +
\frac{1}{a}{\bf \nabla}\cdot [\rho({\bf x}, t){\bf v}({\bf x, t})] &=&0,
\label{eq:drhodt}\\
\frac{\partial{\bf v}({\bf x, t})}{\partial t} +
\frac{\dot a}{a}{\bf v}({\bf x}, t)
+\frac{1}{a}[{\bf v}({\bf x, t})\cdot{\bf \nabla}]{\bf v}({\bf x}, t)&=&
\nonumber\\
-\frac{1}{\rho a}{\bf\nabla}p({\bf x}, t)
-\frac{1}{a}{\bf \nabla}\phi({\bf x}, t),
\label{eq:dvdt}
\end{eqnarray}
where $\rho({\bf x}, t)$ and $p({\bf x}, t)$ are the gas density and
pressure at ${\bf x}$ at cosmological time $t$. Because the evolution
equation for the entropy, Eq.~(\ref{eq:dsdt}), is already in the
comoving frame, it remains the same with the understanding that $d/dt$
is replaced by $\partial/\partial t + (1/a){\bf v}({\bf
x},t)\cdot{\bf\nabla}$ \cite{Peebles93}. The heating and cooling terms
include the physical processes discussed above in
\S~\ref{subsec:thermal-eq}. The evolution of the ionization state of a
mixture of hydrogen and helium is given by Eqs.~(\ref{eq:phionizeHe}).
The inclusion of ionization assumes a given set of \HI, \HeI\ and
\HeII\ photoionization rates and photoheating rates. These must be
computed from the ambient radiation field. The initial ionization
state of the gas is set during the passage of ionization fronts
produced by the radiation sources that photoionized the gas during the
epochs of hydrogen and helium reionization. This is discussed in
\S~\ref{sec:reionization} below.

The dark matter and fluid are coupled through Poisson's equation
for the gravitational potential
\begin{equation}
{\nabla^2}\phi = 4\pi Ga^2[\rho_{\rm tot}({\bf x}, t) -
\langle\rho_{\rm tot}\rangle],
\label{eq:Poisson}
\end{equation}
where $\rho_{\rm tot}=\rho_d+\rho$ is the total mass density of the
dark matter and fluid combined, $\langle\rho_{\rm tot}\rangle$ is the
spatial average of $\rho_{\rm tot}$ over the Universe, and
$p\ll(\rho_{\rm tot}-\langle\rho_{\rm tot}\rangle)c^2$ has been
assumed. In terms of the present day matter closure parameter
$\Omega_m$, $\langle\rho_{\rm tot}\rangle=3H_0^2\Omega_m/(8\pi
Ga^3)$. Eqs.~(\ref{eq:dxddt}) -- (\ref{eq:Poisson}) are valid for any
Friedmann-Robertson-Walker universe, including those with non-zero
vacuum energy (such as an effective cosmological constant).

These are the same equations that describe the mix of dark matter,
baryons and radiation that give rise to the Cosmic Microwave
Background, except for the addition of the radiation energy density
and pressure. The principal difference is that while the density
fluctuations are linear for the CMB up to the time the baryons begin
to recombine and decouple from the radiation, the description of the
growth of cosmological structures extends the computation of the
fluctuations into the nonlinear regime. Numerical solutions to
Eqs.~(\ref{eq:dxddt}) -- (\ref{eq:Poisson}) generally begin at a
sufficiently high redshift that the primordial density fluctuations
are still linear on the scales of interest, but after the matter and
radiation have decoupled. The most convenient set of boundary
conditions for solving the PDEs for the coupled dark matter and gas on
cosmological scales is to assume all the variables are periodic over
the simulation volume. It is therefore important that the simulation
volume be sufficiently large to represent a fair sample of the
Universe.

No structures will form unless the matter is initially inhomogeneous.
The inhomogeneity is usually specified in terms of a primordial power
spectrum of matter fluctuations, which is the ensemble average of the
squared modulus of the Fourier modes of the density field. The mass
density expanded in Fourier modes within a periodic Cartesian volume
$V_{\rm box}$ may be expressed as
\begin{equation}
\rho({\bf x}, t) = \langle\rho\rangle\biggl[1+
\sum_{k\neq 0}\delta_{\bf k}(t)\exp(-i{\bf k}\cdot{\bf x})\biggr],
\label{eq:rhoFE}
\end{equation}
where the Fourier amplitudes $\delta_{\bf k}(t)$ are assumed to have
random phases and to be distributed according to a gaussian random
process. The power spectrum, which quantifies the amplitude of the
density fluctuations, may be defined as $P(k)=V_{\rm box}\langle
\vert\delta_{\rm k}\vert^2\rangle$, where the average is over
independent statistical realizations of the Fourier amplitudes.

The real space analogue of the power spectrum is the spatial correlation
function $\xi(r)$ of the density field
\begin{equation}
\langle\rho({\bf x}^\prime + {\bf x})\rho({\bf x}^\prime)\rangle
=\langle\rho\rangle^2[1+\xi({\bf x})],
\label{eq:xi}
\end{equation}
where the average is carried out over all spatial locations ${\bf
x}^\prime$. The correlation function is related to the power spectrum
through a Fourier transform
\begin{equation}
\xi({\bf x}) = \frac{1}{(2\pi)^3}\int\, d^3k\biggl[P(k) -
\frac{1}{n_p}\biggr]\exp(-i{\bf k}\cdot{\bf x}),
\label{eq:xiP}
\end{equation}
where $n_p$ is the mean spatial density of the particles, and the
Poisson noise contribution due to an average number of $n_pV_{\rm
box}$ particles per simulation volume has been subtracted.

It is convenient to define the amplitude of density fluctuations
\begin{equation}
\Delta^2(k) = \frac{k^3}{2\pi^2}P(k).
\label{eq:Deltak}
\end{equation}
The amplitude of mass fluctuations $(\delta M/ M)^2$ in a sphere of
radius $r_f$ is then given by
\begin{equation}
\sigma^2_{r_f} = \int_0^\infty\, \frac{dk}{k}\Delta^2(k)
\biggl[\frac{3j_1(kr_f)}{kr_f}\biggr]^2,
\label{eq:sigmaf}
\end{equation}
where $j_1(x)=(\sin x)/x^2 - (\cos x)/x$ is a spherical Bessel
function. A common fiducial normalization measure is the mass
fluctuation $\sigma_8$ filtered on a scale of $r_f=8h^{-1}$~Mpc, where
$h=H_0/ 100\,{\rm km\,s^{-1}\,Mpc^{-1}}$. The spectrum of matter
fluctuations may be expressed in the form
\begin{equation}
\Delta^2(k, z) = \delta^2_H\biggl(\frac{ck}{H_0}\biggr)^{3+n}
\frac{D^2(z)}{D^2(0)}T^2(k),
\label{eq:Pk-form}
\end{equation}
where $T(k)$ is the transfer function describing the evolution of
density perturbations as they cross the present-day horizon scale with
amplitude $\delta_H$, $n$ is the spectral index, and $D(z)$ is the
growth factor for the density perturbations (see
\S~\ref{subsec:sheets}). The transfer function for a Cold Dark Matter
dominated universe was first computed by
\textcite{1981ApJ...248..885P}. A fast numerical code for computing
the transfer function to high accuracy is provided by
\textcite{1999ApJ...511....5E}.

\subsubsection{Methods of numerical simulations}
\label{subsubsec:simmethods}

Several numerical methods have been developed for solving the combined
gravity-fluid equations. The most common methods for solving the dark
matter evolution equations sample the distribution function of the
dark matter using massive pseudo-particles and solve for their
evolution either on a mesh or using a hierarchical tree. The most
common mesh codes use either the Particle-Mesh (PM) or
Particle-Particle Particle-Mesh (${\rm P^3M}$) techniques. Tree
algorithms instead arrange the particles into a hierarchy of groups
and approximate the forces exerted by distant groups on a given
particle by their lowest multiple moments. The fluid equations are
frequently solved using a finite-difference representation of the
fluid equations, handling shocks either with artificial viscosity or
through shock capturing. An alternative scheme, Smoothed Particle
Hydrodynamics (SPH), smooths the fluid equations and samples the
smoothed fluid variables using pseudo-particles. The algorithm is
intrinsically Lagrangian in nature. A comparison of two popular
publicly available codes using these methods, ENZO and GADGET, is
provided by \textcite{2005ApJS..160....1O}. The paper also serves as a
good entry point to the vast literature on the subject. A more general
overview of $N$-body methods is provided by
\textcite{1998ARA&A..36..599B}.

All of the techniques share a common method for initiating the density
fluctuations and associated peculiar velocity field based on the
Zel'dovich approximation \cite{1985ApJS...57..241E}. Substituting
${\bf v}_d=a{\dot{\bf x}_d} = -a{\dot D(t)}{\bf \nabla}\Psi({\bf
q}_d)$, using Eq.~(\ref{eq:Zeld}) for the dark matter, into the equation
for peculiar velocity from Eq.~(\ref{eq:dvddt}) gives
\begin{equation}
\Psi({\bf q}_d) = \frac{1}{4\pi G\langle\rho_{\rm tot}\rangle a^2 D}
\phi({\bf q}_d, t) + {\rm constant},
\label{eq:Psi_phi}
\end{equation}
after noting that $a{\ddot D}+2{\dot a}{\dot D}=4\pi G
\langle\rho_{\rm tot}\rangle a D$, which follows from
Eq.~(\ref{eq:drhodt}) applied to the combined dark matter and gas
density. The initial deformation is therefore proportional to the
gravitational potential $\phi$ (up to an arbitrary integration
constant). This offers a straightforward prescription for specifying
the initial conditions. First, choose the $\delta_{\bf k}$'s according
to a gaussian random process:\ $\delta_{\bf
k}=\sqrt{P(k)}g\exp(i\theta)$, where $g$ is a gaussian random deviate
between $-\infty<g<\infty$ with zero mean and unit variance, and
$\theta$ is a random phase uniformly distributed over
$0<\theta<2\pi$. In Fourier space, the gravitational potential is
$\phi_{\bf k}= -4\pi G a^2\langle\rho_{\rm tot}\rangle\delta_{\bf
k}k^{-2}$ since the Fourier transform of $\nabla^2\phi$ is
$-k^2{\phi}_{\bf k}$, where ${\phi}_{\bf k}(t)$ is the Fourier
transform of $\phi({\bf q}, t)$. The gravitational potential may then
be constructed from
\begin{eqnarray}
\phi({\bf q}, t) &=& \sum\phi_{\bf k}\exp(-i{\bf k}\cdot{\bf q})\nonumber\\
&=&-4\pi G a^2\langle\rho_{\rm tot}\rangle\sum\frac{\delta_{\bf k}}{k^2}
\exp(-i{\bf k}\cdot{\bf q}).
\label{eq:phixk}
\end{eqnarray}
The deformation function $\Psi({\bf q}_d)$ then follows from
Eq.~(\ref{eq:Psi_phi}). The displacements and velocities
of the particles then follow from Eqs.~(\ref{eq:Zeld}),
\begin{eqnarray}
{\bf x} &=& {\bf q} - \sum\frac{i{\bf k}}{k^2}\delta_{\bf k}
\exp(-i{\bf k}\cdot{\bf q}),\\
{\bf v} &=& -a\frac{\dot D}{D}\sum\frac{i{\bf k}}{k^2}\delta_{\bf k}
\exp(-i{\bf k}\cdot{\bf q}).
\label{eq:Zeldk}
\end{eqnarray}
The optimal form of these equations allows for convolution with the
differencing operator used to evaluate the gradient of the
gravitational potential \cite{1985ApJS...57..241E}. Once initiated,
any of the various numerical methods described above may be used to
evolve forward the dark matter and baryon fluctuations into the
nonlinear regime.

The resolution requirements of the simulations depend on the
statistical quantity for which convergence is sought. While the
demands for convergence on the \HI\ column density distribution over
the range $13<N_{\rm HI}<15$ are not very severe, the early
simulations tended to over-estimate the width of the absorption
features. It is now recognized that good convergence on the Doppler
parameter distribution and of the pixel flux distribution requires at
the minimum good resolution of the Jeans length \cite{Theuns98,
1999ApJ...517...13B}. A resolution of $9h^{-1}$~kpc (comoving) is
about the right level at $z=2$, but still inadequate by $z\ge3$
\cite{1999ApJ...517...13B}, although the rate of convergence is
sensitive to the gas temperature and assumed cosmology. Slightly
coarser resolution is adequate for converging on the pixel flux
distribution. A large simulation volume is required to ensure that the
full contribution of long wavelength modes is included
\cite{1997ASPC..123..332W, Theuns98, 1999ApJ...517...13B}, requiring a
box size of at least $5h^{-1}$~Mpc (comoving). Similar demands are
found using pure dark matter pseudo-hydrodynamics simulations, for
which sequences of large high resolution simulations have been
performed to test convergence \cite{MW01}. Matching both the
resolution and box size requirements demands simulations of minimally
$512^3$ zones, and preferably $1024^3$ zones for grid based codes to
achieve good convergence on the properties of low \HI\ column density
systems. Similar numbers of particles are required for SPH
computations. Even with these parameters, converging on the extremes
in the temperature distribution proves difficult. Such large
hydrodynamics simulations are only now becoming feasible.

The simulations show that the baryons closely trace the distribution
of dark matter down to the Jeans length scale, when pressure forces
prevent further collapse of the baryons. This has prompted an
alternative approach to IGM simulations based solely on the evolution
of the dark matter \cite{1995A&A...295L...9P, 1998ApJ...495...44C,
MW01}. The gas overdensity is either set equal to the dark matter
overdensity or by allowing for a pressure by treating it as a
modification to the gravitational potential in the Hydrodynamics
Particle Mesh (HPM) approach \cite{1998MNRAS.296...44G}. The gas
temperature is either solved for allowing for photoelectric heating
and radiative cooling \cite{1995A&A...295L...9P} or through a
polytropic equation of state
$T=T_0(\rho/\langle\rho\rangle)^\gamma$. This alternative is based on
the recognition by \textcite{1997MNRAS.292...27H} that the
temperatures found in numerical hydrodynamics simulations are
determined largely through a balance between photoelectric heating and
adiabatic cooling losses for structures with overdensities
$\rho/\langle\rho\rangle<5$. Comparisons with full hydrodynamical
computations show that using a PM code or HPM code with a polytropic
equation of state gives a reasonably good description of the
properties of the IGM, including its absorption statistics \cite{MW01,
2002MNRAS.336..685V, 2006MNRAS.367.1655V}, although full hydrodynamics
simulations should still be used for detailed comparisons to the data.

Yet another approach altogether is to generate the density
fluctuations stochastically from a lognormal distribution
\cite{1992A&A...266....1B, 1993ApJ...405..479B,
1997ApJ...479..523B}. This approach captures many of the qualitative
features of the structure of the IGM, and provided an early argument
that the \Lya forest itself may be accounted for by density
fluctuations in a diffuse IGM.

\subsection{Physical properties of the intergalactic medium}
\label{subsec:IGMphysprops}

\begin{figure*}
\includegraphics[width=8cm]{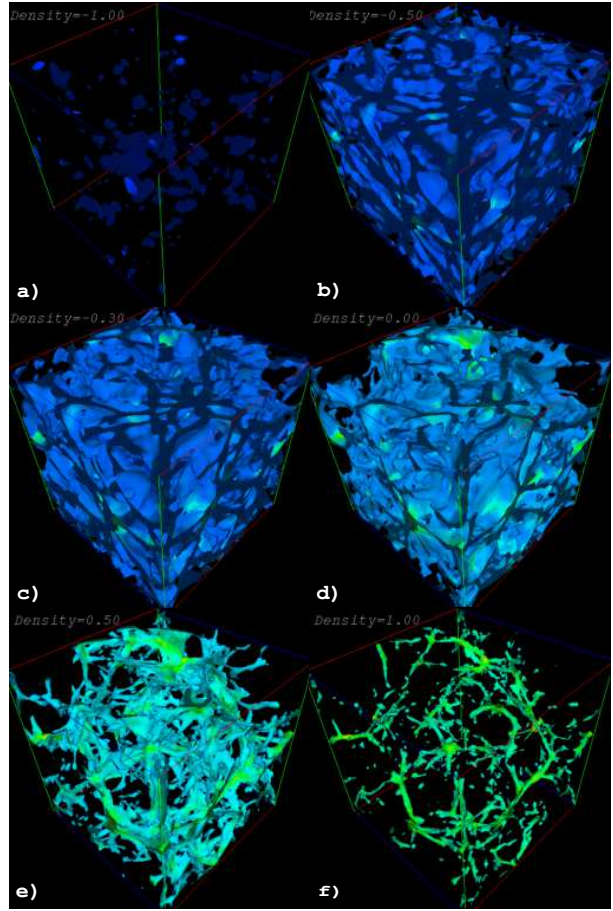}
\caption{Isodensity contour surfaces of log baryon overdensity at
$z=3$ for a flat Cold Dark Matter dominated universe. The contour
levels are $\log_{10}(\rho/\langle\rho\rangle)=-1.0, -0.5, -0.3,
0.0, 0.5$ and 1.0. Low density regions are amorphous structures
filling sheets at the mean density which intersect at overdense
filaments. The filaments intersect at highly overdense collapsed
halos. From \textcite{ZMAN98}. (Figure reproduced by permission
of the AAS.)
}
\label{fig:IGM-structures}
\end{figure*}

\begin{figure*}
\includegraphics[width=8cm]{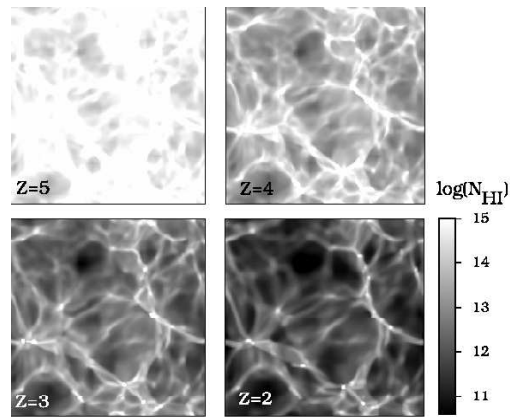}
\caption{Evolution of the \HI\ column density. The filamentary
structure of the IGM is nearly stationary in the comoving frame. By
contrast, the decline in the physical gas density results in rapid
evolution of the \HI\ column density, and the IGM becomes increasingly
transparent with time. From \textcite{ZMAN98}. (Figure reproduced by
permission of the AAS.)
}
\label{fig:HI-structures}
\end{figure*}

The results of structure formation simulations show that the IGM
fragments into an interconnected network of sheets, filaments and
halos \cite{Cen94, ZAN95, 1996ApJ...457L..57K, 1996ApJ...471..582M,
1997ASPC..123..332W}. The various morphologies correspond to different
physical densities and different \HI\ column density absorption
systems. These dependences are illustrated in
Figures~\ref{fig:IGM-structures} and \ref{fig:HI-structures}, taken
from the simulation of \textcite{ZMAN98} of a Cold Dark Matter
dominated Einstein-deSitter universe with the radiation field from
\textcite{HM96} based on QSO sources. The simulation volume was
9.6~Mpc (comoving) on a side, and the power spectrum was normalized to
$\sigma_8=0.7$ with $h=0.5$ used. At $z=3$, systems with $N_{\rm
HI}\lta10^{13}\,{\rm cm}^{-2}$ arise from the amorphous underdense
regions enclosed by the sheets. The underdense regions have
characteristic diameters of a few comoving Mpc, and occupy most of the
volume of the Universe. The sheets have densities comparable to the
mean baryon density, and give rise to absorption systems with
$10^{13}\,{\rm cm}^{-2}\lta N_{\rm HI}\lta 10^{14}\,{\rm
cm}^{-2}$. Systems with $10^{14}\,{\rm cm}^{-2}\lta N_{\rm HI}\lta
10^{15}\,{\rm cm}^{-2}$ originate in the filaments, which are
moderately overdense structures ($1\lta
\rho/\langle\rho\rangle\lta5$). Higher column density systems, with
$N_{\rm HI}\gta10^{15}\,{\rm cm}^{-2}$, arise in spheroidal nodes with
overdensities of several to over a hundred embedded within the
filaments.

The mean temperatures of overdense systems range over $10^4\,{\rm K}$
to $3\times10^4\,{\rm K}$ \cite{Cen94, ZMAN98, Theuns98}, while
\textcite{1996ApJ...471..582M} find mean temperatures up to
$6\times10^4\,{\rm K}$. These larger values may be due to the hard UV
background assumed. Much higher temperatures are also found, extending
up to $10^6-10^7$~K, for gas associated with collapsed halos
\cite{Theuns98}. There is less agreement on the temperatures in the
underdense regions. The time to achieve thermal equilibrium exceeds a
Hubble time for underdense gas \cite{Meiksin94, MR94}. The gas
temperature will therefore reflect the thermal history of the gas
following reionization, and so is sensitive to the particular
reionization model. The \Lya forest simulations also did not take into
account the heating effects of reionization itself
\cite{2007MNRAS.380.1369T}.

The thicknesses of the filaments are approximately 100~kpc (proper),
with coherence lengths of about a Mpc. The scales are consistent with
the observational constraints based on neighboring line-of-sight
statistics discussed in \S~\ref{subsubsec:sizes} above. This is
comparable to the isothermal Jeans length. For a characteristic
hydrogen number density of $55\,{\rm m}^{-3}$, corresponding to 5
times the mean baryon density at $z=3$, and a characteristic
temperature of $2\times10^4$~K, the pressure is $10^6\,{\rm
K\,m^{-3}}$ and the Jeans length from Eq.~(\ref{eq:Jeanslength}) is
320~kpc.

The baryonic overdensity at $z=3$ is found to scale roughly as
$\rho/\langle\rho\rangle \simeq 3(N_{\rm HI}/10^{14}\,{\rm
cm^{-2}})^{1/2}$, with the neutral fraction scaling as $x_{\rm
HI}\simeq 1.3\times10^{-5}(N_{\rm HI}/10^{14}\,{\rm
cm^{-2}})^{1/2}T^{-0.75}_4\Gamma^{-1}_{\rm HI, -12}$
\cite{ZMAN98}. The neutral fraction dependency is comparable to the
estimate in \S~\ref{subsubsec:sizes} above, and results in virtually
all the baryons' residing in the \Lya forest at $z>1$.

The structure of the IGM at $z\ll1$ differs somewhat from its
structure at $z>2$. \textcite{1999ApJ...511..521D} obtain the relation
between overdensity and \HI\ column density of
$\rho/\langle\rho\rangle \simeq 20(N_{\rm HI}/10^{14}\,{\rm
cm^{-2}})^{0.7} e^{-0.4z}$, similar to the finding of
\textcite{ZMAN98} at $z=3$. A given overdensity results in \HI\ column
densities smaller by a factor of 10--50 at $z\ll1$ than at $z>2$. Most
absorption systems at low redshift arise in systems with hydrogen
densities of $n_{\rm H}\simeq0.1-10\,{\rm m^{-3}}$ for \HI\ column
densities of $N_{\rm H}\simeq10^{13}-10^{15}\,{\rm
cm^{-2}}$. Structures retain coherence over scales of several hundred
kiloparsecs. The gas temperatures for structures with
$\rho/\langle\rho\rangle<10$ are about a factor of 2 smaller at
$z\ll1$ than at $z>2$. A hot, intermediate-density gas phase also
arises from shocks in filaments and in galaxy, galaxy group and galaxy
cluster halos, reaching temperatures of over $10^7$~K.

\subsection{Statistical properties of the absorption systems}
\label{subsec:IGMstatprops}

\subsubsection{Mean metagalactic ionization rates}
\label{subsubsec:Gamma}

\begin{figure}
\includegraphics[width=8cm]{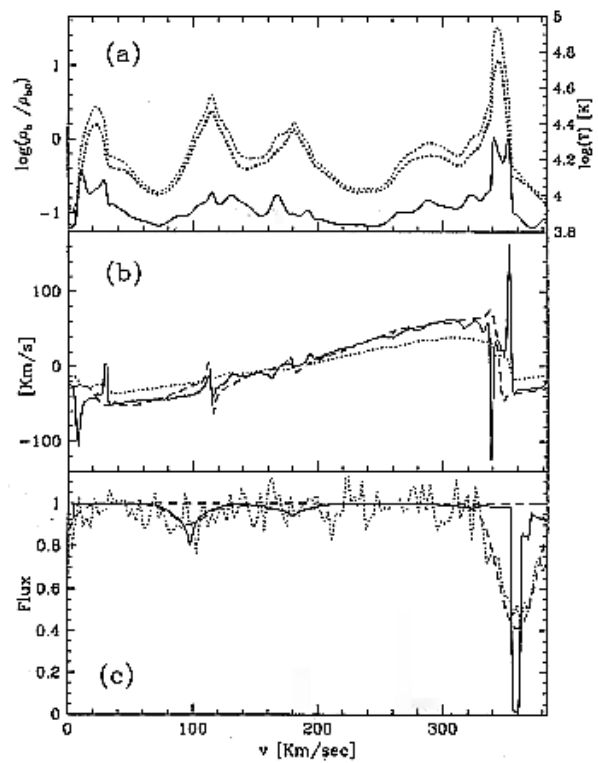}
\caption{A line-of-sight cut through a $\Lambda$CDM simulation of the
\Lya forest at $z=3$. The top panel shows the gas temperature (solid
line), gas density (thick dotted line) and gas pressure, in arbitrary
units (thin dotted line). All are shown as a function of the spatial
coordinate $r$ expressed in velocity units through $v=Hr$, for
$H=512h\,{\rm km\,s^{-1}\,Mpc^{-1}}$. The gas temperature and density
are closely correlated. The middle panel shows the peculiar velocity
(dotted line), the gravitational acceleration divided by the Hubble
constant (dashed line), and the total hydrostatic and gravitational
acceleration divided by the Hubble constant (solid line). An infalling
peculiar velocity pattern forms around a collapsing structure near
$v\simeq360\kms$. The bottom panel shows the transmission factor
$\exp(-\tau_\nu)$ before convolving with a Doppler profile (solid
line), after convolution with a Doppler profile (dashed line), and
after adding the effects of instrumental resolution and noise (dotted
line). From \textcite{Cen94}. (Figure reproduced by permission of the
AAS.)
}
\label{fig:sim-spectrum}
\end{figure}

Almost all that is directly known about the structure of the IGM is
derived from spectra. The simulation results must therefore be
translated into spectra for comparison against observational data to
test the validity of the models. The spectra, normalized to a unit
continuum level, may be expressed as $\exp(-\tau_\nu)$, where
$\tau_\nu$, the IGM optical depth as a function of frequency, is given
by Eq.~(\ref{eq:taunu}).  Allowing for a peculiar velocity $v$ varying
along the line of sight,
\begin{equation}
\tau_\nu=\int\,dz^\prime\frac{dl_p}{dz^\prime}n(z^\prime)
\sigma\biggl[\nu(1+z^\prime)-\nu_{lu}\biggl(1-\frac{v}{c}\biggr)
\biggr].
\label{eq:taunuspec}
\end{equation}
Setting $n = \langle n\rangle(1+\delta)$, where $\delta$ is the local
density contrast, Eq.~(\ref{eq:taunuspec}) is identical to
Eq.~(\ref{eq:tauGP}) for the Gunn-Peterson effect in the absence of
perturbations ($v=\delta=0$). For this reason, the varying optical
depth of the IGM is sometimes referred to as giving rise to a
``fluctuating Gunn-Peterson effect'' \cite{1993MNRAS.260..617M, Cen94}.

An example of a spectrum generated from the $\Lambda$CDM simulation of
\textcite{Cen94} with $\Omega_v=0.6$, $\Omega_m=0.4$, $h=0.65$, and
normalization $\sigma_8=0.79$ is shown in
Figure~\ref{fig:sim-spectrum}. The temperature variations closely
trace the baryon density fluctuations. The absorption features are
slightly offset from the density features due to the peculiar motion
of the gas. Mild overdensities produce weak features while large
overdensities produce stronger features. The apparently isolated
absorption features arise not from distinct clouds but from
modulations in the density field produced by the gravitational
instability of primordial density fluctuations.

Any comparison between the simulated and measured spectra requires
normalization by the uncertain ionization rates for \HI\ and \HeII.
The uncertainty in the normalization rate for \HeII\ is compounded by
the unknown epoch of \HeII\ reionization, which may have been as late
as $z\simeq3$. Late \HeII\ reionization will also affect the gas
temperatures, particularly for underdense gas. This in turn affects
the optical depth through the gas. Much of the uncertainty on global
flux properties may be removed by normalizing to the measured mean
\Lya transmission $\langle \exp(-\tau_\nu)\rangle$ of the
IGM. Although an error in the temperature of the gas will result in a
small offset in the inferred ionization rate (see
Eq.~[\ref{eq:phionizeHI-eq}]), once normalized, the probability
distribution of the emissivity is only slightly affected. Altering the
temperature of the gas as required by the line-widths, either directly
or by boosting the heating rate, can further improve the fits.

Various groups have provided estimates for the \HI-ionization rate
$\Gamma_{\rm HI}$ based on matching their simulation results to
measurements of the mean \Lya transmissivity. Results from
\textcite{MW04}, \textcite{2005MNRAS.357.1178B} and
\textcite{2007MNRAS.382..325B} are shown in
Table~\ref{tab:GammaHI}. All the predicted rates are in reasonably
good agreement. \textcite{2007MNRAS.382..325B} provide an additional
estimate at $z=2$ of $\Gamma_{\rm HI, -12}=1.29^{+0.80}_{-0.46}$,
which agrees well with the estimate from
\textcite{2004ApJ...617....1T} of $\Gamma_{\rm HI,
-12}=1.44\pm0.11$. An additional uncertainty of a few to several
percent may result from fluctuations in the ionizing background
\cite{2002MNRAS.334..107G, 2004ApJ...610..642C, MW04}. A comparison of
the required ionization rates with the predictions from QSO sources
shows that unless there is a large population of low luminosity AGN so
far undetected at high redshifts, QSOs alone are unable to provide the
required ionization rates at $z\gta3$.

Specifying $\Gamma_{\rm HI}$ fixes the entire equilibrium ionization
structure of the gas (except in isolated optically thick
regions). This permits the contribution of the IGM to the attenuation
length of ionizing photons to be determined directly from a
simulation. In principle, a self-consistent solution for the
equilibrium structure of the IGM may then be obtained, including the
magnitude of the required emissivity for any given spectral shape
using Eq.~(\ref{eq:GHI-emiss}). \textcite{MW04} find for a
$\Lambda$CDM cosmology the (proper) attenuation length is approximated
to 10\% accuracy by
\begin{equation}
r_0\simeq1.7\times10^4 (1+z)^{-4.2}\, h^{-1}\, {\rm Mpc}
\label{eq:r0}
\end{equation}
for $2.75<z<5.5$. Because the simulation underpredicts the number of
Lyman Limit Systems by a large factor, these are added using the
results of \textcite{1995ApJ...444...64S}. It is found that the
diffuse gas comprising the \Lya forest and the Lyman Limit Systems
contribute about equally to the attenuation length, consistent with
models based on direct line counts \cite{MM93}. The implied (comoving)
metagalactic emissivity at the Lyman edge required to recover the \HI\
ionization rates is \cite{Meiksin05}
\begin{equation}
\epsilon_L\simeq A_{\rm MG}\biggl(\frac{3+\alpha_{\rm MG}}{3}\biggr)
(1+z)^{\gamma}\,h\,{\rm W\,Mpc^{-3}},
\label{eq:epsL}
\end{equation}
where $A_{\rm MG} = 8.4^{+9.7, 31}_{-4.5, 6.5}\times10^{18}$ and
$\gamma=-1.6-0.6\ln(A_{\rm MG}/8.4\times10^{18})$
and $\alpha_{\rm MG}$ is the spectral index of the metagalactic
emissivity, $\epsilon_\nu=\epsilon_L(\nu/\nu_L)^{-\alpha_{\rm MG}}$.
The errors on $A_{\rm MG}$ are the $1\sigma$ and $2\sigma$ limits.

At $z>4$, Eq.~(\ref{eq:epsL}) is well above the predictions from QSO
counts. As discussed in \S~\ref{subsubsec: sources} above,
Eq.~(\ref{eq:epsL}) may be matched by the emissivity of galaxies at
the Lyman edge for plausible escape fractions of the ionizing
radiation.

\subsubsection{Spectral properties of the IGM}
\label{subsubsec:specprops}

\begin{figure}
\includegraphics[width=8cm]{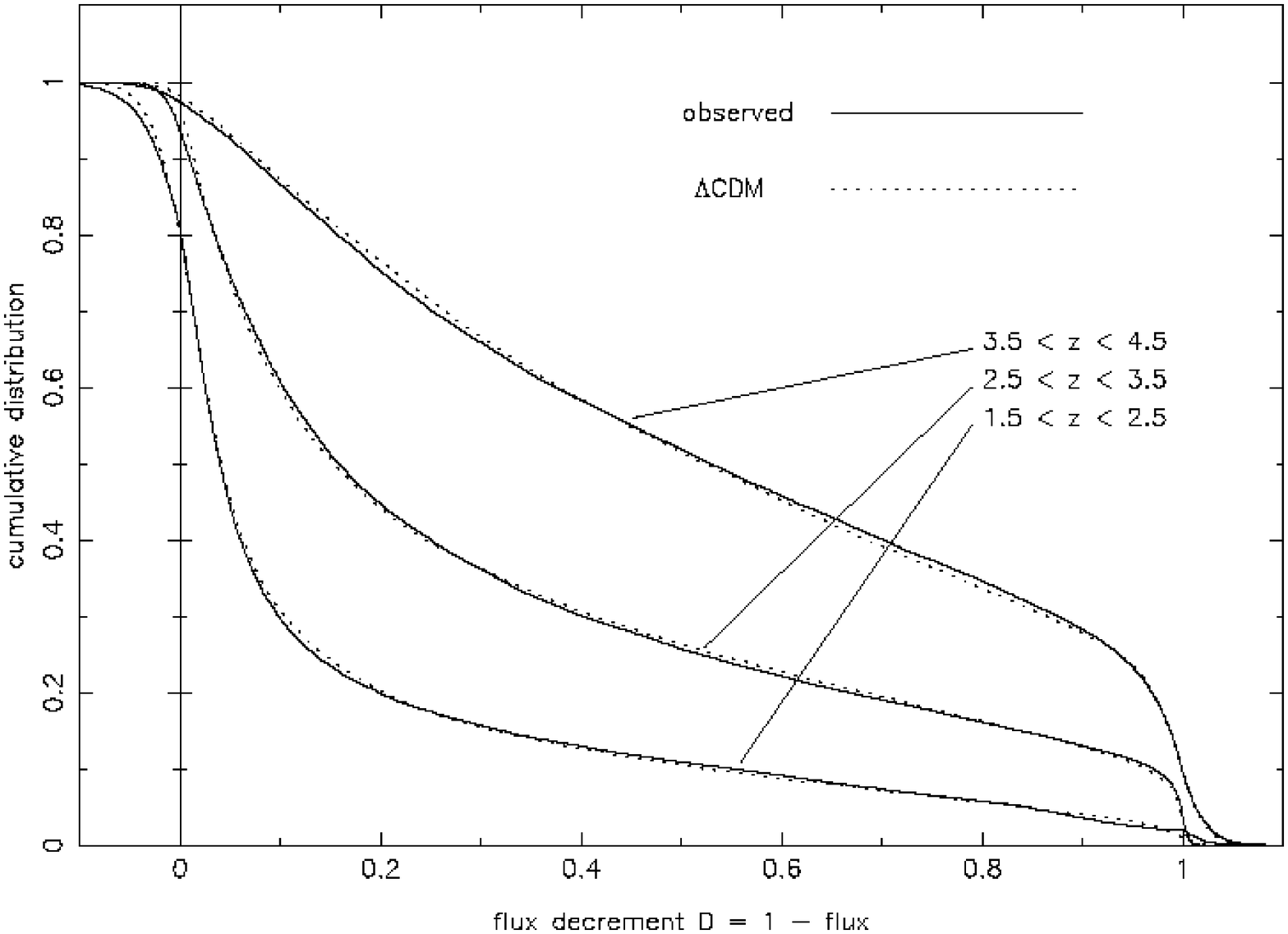}
\caption{Comparison of the cumulative flux per pixel distribution from
a $\Lambda$CDM simulation with data based on high resolution Keck
spectra of 7 QSOs. The data have been grouped into three redshift
intervals and the observations corrected to take into account
evolution in the mean transmission within each interval. Noise has
been added to the simulated data to match the noise properties of the
measured spectra. From \textcite{1997ApJ...489....7R}. (Figure
reproduced by permission of the AAS.)
}
\label{fig:sim-FPPD}
\end{figure}

The predicted spectral properties of the \Lya forest are sensitive to
the cosmological parameters used \cite{1997ApJ...489....7R,
1997ASPC..123..332W, 1999MNRAS.303L..58T, 2000ApJ...532..118M,
2005MNRAS.361...70J}. Comparisons with the spectral properties of the
\Lya forest have shown impressive agreement for plausible cosmological
models. \textcite{2000ApJ...532..118M} show that the distribution of
optical depth per pixel in the models is well-fit by a lognormal
distribution for values between $0.02<\tau<4$, roughly the measurable
range. This prediction is confirmed by the observations of
\textcite{2007ApJ...662...72B}, who find the flux distributions in 55
high resolution QSOs are well-reproduced by lognormal distributions
for the underlying optical depths. Detailed comparisons show that the
predicted cumulative distributions of pixel fluxes agree to a
precision of a few percent with measurements in individual QSO spectra
\cite{MBM01}.

These results suggest the \Lya forest may be used as an effective tool
for selecting between rival cosmological models of structure
formation. An example of a comparison between the $\Lambda$CDM
simulation above and high signal-to-noise ratio high resolution Keck
spectra of 7 QSOs is shown in Figure~\ref{fig:sim-FPPD}
\cite{1997ApJ...489....7R}. \textcite{2000ApJ...532..118M} and
\textcite{MBM01} find that the shape of the pixel flux distribution
and the magnitude of the Doppler parameters vary systematically with
the normalization on small scales, such as $\sigma_J$ filtered on the
Jeans scale, as defined by Eq.~(\ref{eq:sigmaf}) with filter scale
$r_f=\lambda_J$ from Eq~(\ref{eq:Jeanslength}). A detailed comparison
between the predicted flux distributions for several cosmological
models and measurements from several high resolution Keck spectra
constrains $\sigma_J$, defined for an isothermal gas at
$T=2\times10^4$~K, to the range $1.3\lta\sigma_J\lta1.7$ at $z=3$
\cite{MBM01}, although this is somewhat subject to the uncertain
thermal properties of the gas. The closest agreement is found for the
larger values of $\sigma_J$. A range of $\Lambda$CDM models, however,
shows that agreement with $\sigma_8$ inferred from {\it WMAP}
measurements of CMB fluctuations prefers lower values \cite{MW04}, so
that there is some tension between the higher normalization preferred
by the \Lya forest simulations and the lower normalization preferred
by the CMB measurements.

Voigt profile fitting to the spectra permits comparison between the
predicted and measured \HI\ column density and Doppler parameter
distributions. Here the models fair somewhat less well compared with
the pixel flux distribution. The models successfully reproduce the
\HI\ column density distributions to good accuracy over a wide range
of values of $14\lta\log_{10}N_{\rm HI}\lta16$ at high redshifts
($2<z<4$) \cite{Cen94, ZAN95, 1996ApJ...457L..51H,
1997ApJ...485..496Z, 1997ASPC..123..332W, MBM01} and at moderate to
low redshifts ($0<z<2$) \cite{1999ApJ...511..521D,
2008arXiv0802.3730P}. A discrepancy is found at higher column
densities, with simulations generally finding fewer systems with
$\log_{10}N_{\rm HI}>16$ than measured. This is particularly true for
Lyman Limit Systems \cite{1996ApJ...457L..57K, MW04}, although
extrapolating the simulation results to smaller mass halos suggests
agreement is possible, depending on the appropriate minimum circular
velocity of halos able to produce high \HI\ column density systems in
the metagalactic photoionizing background
\cite{2001ApJ...559..131G}. A smaller discrepancy is found for
DLAs. The simulations of \textcite{2007ApJ...660..945N} suggest the
incidence of DLAs may be met if they are associated with halos having
masses as low as $10^{8.5}h^{-1}\,M_\odot$.

It is less clear that all Lyman Limit Systems arise from collapsed
halos. Allowing for radiative transfer effects,
\textcite{2007ApJ...655..685K} find their simulations produce Lyman
Limit Systems from material bound to and falling onto galaxies rather
than from minihalos. It is possible the systems have more than a
single origin.

\begin{figure}
\includegraphics[width=8cm]{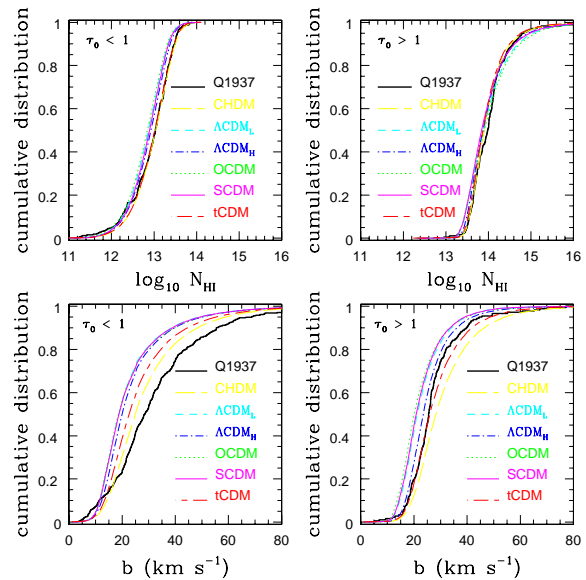}
\caption{Comparison of the cumulative \HI\ column density (upper
panels) and Doppler parameter (lower panels) distributions for the
\Lya forest measured along the line of sight to the QSO Q1937--1009
with predictions for six cosmological models:\ a standard CDM model
(SCDM), two $\Lambda$CDM models with high ($\Lambda$CDM$_{\rm H}$) and
low ($\Lambda$CDM$_{\rm L}$) baryon densities, an open CDM model
(OCDM), a tilted CDM model (tCDM), and a mixed hot dark matter and
cold dark matter model (CHDM). Much better agreement is found for
absorption line systems with line center optical depths $\tau_0>1$
(right panels) than for $\tau_0<1$ (left panels). From
\textcite{MBM01}. (Figure reproduced by permission of the RAS and
Blackwell Publishing.)
}
\label{fig:sim-lineparams}
\end{figure}

Because the detailed distribution of the absorption line parameters is
sensitive to the method of line finding and fitting (see
\S~\ref{subsubsec:physprops}), a definitive comparison between the
simulations and the measured absorption line properties requires
identical line analysis methods. Automating the line analysis is
necessary for processing the large number of lines of sight from
simulations to build up reliable statistics. A few codes are
publicly available. An industry standard is
VPFIT,\footnote{http://www.ast.cam.ac.uk/$\sim$rfc/vpfit.html} a sophisticated
code that handles in addition to \HI\ features a wide range of metal
lines. A simpler alternative well-adapted to automated fitting of
\HI\ features is AutoVP \cite{1997ApJ...477...21D}. Based on fits with
AutoVP, \textcite{1997ApJ...477...21D} find that the simulations
predict a greater fraction of lower Doppler parameter systems than
measured. Similar discrepancies were found by
\textcite{1999ApJ...517...13B} and \textcite{1999MNRAS.303L..58T}.

A detailed comparison between several high resolution QSO spectra and
numerical simulations for a variety of models, analysed using AutoVP
and a similar code, SPECFIT, by \textcite{MBM01} shows that it is not
possible to match simultaneously the pixel flux distributions, column
density distributions, and Doppler parameter distributions using
simulations with optically thin reionization. While adjusting
$\sigma_J$ may help to match the Doppler parameters, it destroys
agreement with the pixel flux distribution. In
Figure~\ref{fig:sim-lineparams}, a comparison is shown between the
predicted and measured line parameter distributions for the \Lya
forest detected in the spectrum of the $z=3.8$ QSO Q1937--1009. The
systems have been isolated by their line center optical depths. The
comparisons show that the discrepancies are due primarily to the low
optical depth systems, with line center optical depths $\tau_0<1$.

The origin of the discrepancies is most likely the temperature of the
gas. Artificially broadening the absorption features gives excellent
agreement between the measurements and predictions for the
$\Lambda$CDM$_{\rm L}$ model for the pixel flux and column density
distributions, and good agreement between the Doppler parameter
distributions. Since the optically thin absorbers arise in underdense
regions, the temperature of the gas will retain a memory of its
heating history. Since the underdense gas occupies most of the volume
of the Universe, whatever heated the gas must have been
ubiquitous. The best candidate is late reionization of \HeII, in the
redshift interval $3<z<4$. This would boost the gas temperature,
possibly by the required amount. The effects of heating by late \HeII\
reionization may be mimicked to some extent by boosting the \HeII\
heating rate \cite{2000ApJ...534...57B, 2004ApJ...617....1T}, but the
only fully satisfactory solution requires incorporating radiative
transfer \cite{2007MNRAS.380.1369T}.

With the goal of developing a physical description for the origin of
the Doppler parameter distribution, \textcite{1999ApJ...517..541H}
introduce a semi-analytic approach based on the growth of the density
perturbations to derive a Doppler parameter distribution of the form
\begin{equation}
\frac{d{\cal N}}{db}=\frac{4b_\sigma^4}{b^5}
\exp\left(-\frac{b_\sigma^4}{b^4}\right),
\label{eq:dNdbHR}
\end{equation}
where $b_\sigma$ is a fit parameter. Restricting their analysis to
systems with column densities $12.5\le\log_{10}N_{\rm HI}\le14$,
corresponding to the physical conditions of the IGM for which their
analysis best applies, \textcite{1999ApJ...517..541H} find the
distribution provides statistically acceptable matches to the Doppler
parameter distributions of the \Lya forest along the lines of sight to
four QSOs with high resolution Keck spectra
\cite{1995AJ....110.1526H}. Adequate matches, however, were not found
for Q0000--26 \cite{1996ApJ...472..509L} or HS~1946$+$7658
\cite{1997ApJ...484..672K}. The Doppler parameter distributions
measured in the spectra of all six of the QSOs, without any
restriction on the column density range applied, are well-fit by
lognormal distributions (see Table~\ref{tab:bfit}).
\textcite{2000ApJ...534...41R} generalize Eq.~(\ref{eq:dNdbHR}) to
\begin{equation}
\frac{\partial^2{\cal N}}{\partial{\cal N}\partial b}\propto
{\cal N}^{-\alpha}\left(\frac{b_*}{b}\right)^\varphi\exp\left[
-\left(\frac{b_*}{b}\right)^\Phi\right],
\label{eq:d2NdNdbRGS}
\end{equation}
where $\alpha$ is a fit constant, and $\varphi$, $\Phi$ and $b_*$ are
allowed mild dependences on the column density. The best fits
obtained for absorbers in several lines of sight over the redshift
range $2.5\lta z\lta 4$ have $1.6<\alpha<2.1$, $2.6<\varphi<5.9$ and
$2.0<\Phi<7.2$, with values almost independent of column density.

Several attempts have been made to infer the temperature of the IGM
directly from the Doppler parameter distribution. One approach is to
normalize the distributions using simulations, invoking additional
sources of heat input to allow for a range in temperatures
\cite{2000ApJ...534...57B}. Another exploits a polytropic relation
between the gas density and temperature \cite{1997MNRAS.292...27H,
1999MNRAS.310...57S}, calibrated using numerical simulations. Both
methods rely on numerical simulations that neglect the effect of
reionization on the temperature of the underdense gas, which may be
substantial \cite{1999ApJ...520L..13A, BMW04,
2007MNRAS.380.1369T}. Modeling the Doppler parameter distributions
according to Eqs.~(\ref{eq:dNdbHR}) and (\ref{eq:d2NdNdbRGS}) to
obtain lower cutoff values, \textcite{2000ApJ...534...41R} and
\textcite{Schaye00} find evidence for a temperature jump and reduction
in the polytropic index towards isothermality near $z=3.0$. Both
groups attribute their results to late \HeII\ reionization. \HeII\
reionization simulations predict a more gradual change
\cite{2008arXiv0807.2799M}. No significant change in the temperature
or polytropic index over the redshift range $2.4<z<3.9$ is found by
\textcite{2001ApJ...562...52M, 2003ApJ...598..712M}. Although any
conclusion using these methods is subject to the accuracy of the
modeling assumptions made, the possibility of measuring an unambiguous
temperature jump in this redshift range merits further investigation.

\subsubsection{Intergalactic helium absorption}
\label{subsubsec:sim-helium}

As for the \HI\ absorption features, the numerical simulations predict
the properties of \HeII\ absorption as well. The systems that dominate
the measured \HeII\ optical depths arise from the diffuse structures
in underdense regions \cite{ZAN95, 1997ApJ...488..532C}. This makes
predictions for the \HeII\ optical depths difficult for two reasons:\
1. high spatial resolution is required to accurately compute the
structures in the underdense regions, and 2.\ the gas temperature is
sensitive to the history of reionization since the time for the
temperature to reach equilibrium is comparable to or longer than a
Hubble time. The latter affects both the ionization fraction of \HeII\
because of the temperature dependence of the radiative recombination
rate and the Doppler widths of the absorption features. Another
uncertainty is the epoch of full helium reionization. The results of
\textcite{ZAN95} and \textcite{ZMAN98} suggest matching to both the
\HI\ and \HeII\ \Lya mean optical depth measurements at $2.5\lta z\lta
3.3$ requires an ionization rate ratio of $\Psi\simeq250-400$,
comparable to the inferred values from line statistics
(\S~\ref{subsubsec:He-constraints}) and corresponding to an intrinsic
QSO spectral index of 1.8--2.0, allowing for filtering through the IGM
\cite{HM96}. A similar estimate is obtained from high resolution
numerical simulations by \textcite{BHVC06}, who find values for $\Psi$
rising from $139^{+99}_{-67}$ at $z=2.1$ to $301^{+576}_{-151}$ at
$z=2.8$, assuming a spatially uniform UV background. Most of the
uncertainty stems from the uncertainties in the measurements of the
\HeII\ \Lya mean optical depths. Allowing for spatial fluctuations in
the QSO contribution lowers the inferred values for $\Psi$ by about a
factor of two. Allowing for fluctuations also reproduces much of the
observed scatter in the \HeII\ to \HI\ column density ratio $\eta$
(\S~\ref{subsubsec:He-constraints}), except for the extreme excursions
at $\eta<10$ and $\eta>200$.

\subsubsection{Metal absorption systems}
\label{subsubsec:sim-metals}

By providing the physical scales of density structures and their
associated temperatures, the simulations permit estimates of the
ionization states of metals for any assumed UV radiation field.
Several observed features are reproduced by the models. The metal
absorption features arise within the substructure of systems that give
rise to the \HI\ \Lya features, and may result in multiple features
corresponding to a single \HI\ system \cite{1996ApJ...465L..95H,
1997ApJ...485..496Z, 1997ApJ...487..482H,1997ApJ...481..601R}. The
frequencies of \CII, \CIV, \SiIVs and \NVs systems are comparable to
those observed at $z\simeq3$ assuming a QSO/AGN-dominated UV
background and metallicities of $[{\rm C/H}]\simeq-2.5$
\cite{1997ApJ...487..482H}. The \CIVs to \HI\ column density ratio
decreases with increasing \HI\ column density, since higher column
densities correspond to systems with higher densities and so lower
ionization parameters \cite{1997ApJ...487..482H}.

\textcite{1998ApJ...499..172H} rank a variety of metal ions by their
degree of detectability. The strongest absorption lines are the
\OVI$\lambda\lambda$1032,1038\AA\ doublet for low column density
systems, with $N_{\rm HI}<10^{15}\,{\rm cm^{-2}}$. These lines
unfortunately are very difficult to extricate from the \Lya
forest. The measured levels of \OVIs associated with \CIVs systems at
$z\simeq3$ requires an overabundance of $[{\rm O/C}]\simeq0.5$,
assuming a QSO/AGN-dominated UV background, suggestive of enrichment
by Type II supernovae \cite{1998ApJ...509..661D}. If helium was not
fully ionized by $z\simeq3$, the resulting break between the \HI\ and
\HeII\ Lyman edges requires a much higher overabundance of oxygen. At
least half the IGM must have helium fully reionized to keep $[{\rm
O/C}] < 10$.

Longward of the \HI\ \Lya\ line, the strongest lines are the
\CIV$\lambda\lambda$1548, 1551\AA\ doublet for all systems with
$N_{\rm HI}<10^{17}\,{\rm cm^{-2}}$. In high column density systems,
the regions giving rise to \CIIs absorption sometimes do not coincide
with those giving the strongest \CIVs absorption, resulting in
velocity offsets between the two.

At low redshift, the formation of galaxy groups and clusters will
produce an outer region of high density
$\rho/\langle\rho\rangle\simeq100$ and temperatures of
$T\simeq10^5-10^7$~K, which will give rise to strong absorption lines
of \OVIs (for temperatures $5.3\lta \log_{10}T\lta6.3$), \OVIIs
($5.3\lta \log_{10}T\lta6.7$) and \OVIIIs ($6\lta \log_{10}T\lta7.3$)
in a warm-hot intergalactic medium (WHIM), detectable in the extreme
ultraviolet and x-ray \cite{1998ApJ...509...56H, 1998ApJ...503L.135P,
1999ApJ...514....1C, 2001ApJ...552..473D, 2003ApJ...594...42C,
2003MNRAS.341..792V, 2005ApJ...618L..91Y}. Especially of interest is
the \OVI$\lambda\lambda$1032,1038\AA\ doublet, which has already been
detected in numerous systems at low redshifts
\cite{2005ApJ...624..555D, 2008ApJ...679..194D,
2008ApJS..177...39T}. For relatively high metallicities (at least 0.1
of solar), a few higher ionization species features may be detected by
{\it Chandra} and {\it XMM-Newton} \cite{2003ApJ...594...42C}. These
along with lines of carbon, nitrogen and possibly neon and iron may be
detectable by the proposed x-ray satellites {\it
XEUS}\footnote{http://www.rssd.esa.int/index.php?project=XEUS}
\cite{2003ApJ...594...42C, 2003MNRAS.341..792V} and {\it
Constellation-X}\footnote{http://constellationx.nasa.gov}
\cite{2007SPIE.6686E..11P}. The interpretation of the abundances,
however, may be complicated by imperfect equilibration between the
electron and ion temperatures in hot regions with $T\simeq10^7$~K
\cite{2005ApJ...618L..91Y}, unless collective plasma processes, such
as scattering off Alv{\`e}n waves, are adequate for driving the
electrons and ions into thermal equilibrium.

Predictions for low ionization species are much more difficult.
Observations suggest sizes for \CII, \SiII, \MgIIs and \AlIIs systems
of typically below a kiloparsec, down to sizes of only about a parsec
\cite{2006ApJ...637..648S}. Such small, dense clumps are a challenge
for any cosmological numerical simulation to produce.

\subsubsection{The flux power spectrum}
\label{subsubsec:PFk}

In addition to the direct flux and line parameter measurements, the
models may also be tested by appealing to higher order moments of the
flux distribution. The one which has received the most attention is
the flux power spectrum, as a vehicle for using the \Lya forest to
constrain cosmological models. Two broad approaches have been pursued.
In one, measurements of the flux power spectrum are inverted to obtain
a measurement of the underlying dark matter power spectrum
\cite{1998ApJ...495...44C, 1999ApJ...520....1C, 1999ApJ...516..519H,
1999MNRAS.303..179N}. In the second, the flux power spectrum itself is
used directly as a statistic to be matched by models
\cite{2000ApJ...543....1M, 2003MNRAS.344..776M, 2005ApJ...635..761M}.

The basis of the inversion methods is the relation between the 1D and
3D matter power spectra \cite{1989MNRAS.238..293L}
\begin{equation}
\Delta^2_{\rm 1D}(k)=k\int_k^\infty dy \frac{\Delta^2(y)}{y^2}
\label{eq:D3D1}
\end{equation}
and
\begin{equation}
\Delta^2(k) = -k^2\frac{d}{dk}\biggl[\frac{\Delta^2_{\rm 1D}(k)}{k}\biggr],
\label{eq:D1D3}
\end{equation}
where $\Delta^2_{\rm 1D}(k)$ is the 1D mass fluctuation spectrum, related
to the matter spatial correlation function $\xi(x)$ by
\begin{equation}
\xi(x) = \int_0^\infty\frac{dk_x}{k_x}\Delta^2_{\rm 1D}(k_x)\cos(k_x x).
\label{eq:xiD1}
\end{equation}
Eq.~(\ref{eq:D3D1}) shows that $\Delta^2_{\rm 1D}$ depends on all
small scales, well into the nonlinear range of density
fluctuations. Because the flux along the line of sight to a QSO probes
the 1D density field of the IGM, it might appear that any predictions
require a computation of the full nonlinear extent of the power
spectrum to all small scales. Even the highest resolution simulations
are unable to achieve convergence on $\Delta^2_{\rm 1D}$ even at small
$k$ \cite{MW04}. The statistics of the 1D power spectrum are also
notoriously non-gaussian.

Such stringent demands for interpreting the flux power spectrum are
avoided for two reasons:\ 1.\ the width of the absorption features
acts to filter the 3D density field, erasing the smallest scale,
nonlinear contributions to the 1D flux power spectrum, and 2.\ the
inversion Eq.~(\ref{eq:D1D3}) at low $k$ depends only on the behavior
of the 1D flux power spectrum at low $k$. Numerical convergence on the
1D flux power spectrum $\Delta^2_F(k)$ at small $k$ is achievable with
sufficiently fine spatial resolution and a sufficiently large box
size. A comoving box size of $25h^{-1}$~Mpc with a force resolution of
$30-60h^{-1}$~kpc (comoving) gives a flux power spectrum convergent to
10\% on scales $k<0.01\,{\rm km^{-1}\,s}$ at $z>3$, with more severe
requirements at lower redshifts \cite{MW04}. The greater ease of
convergence for the flux power spectrum, however, pays the price of
insensitivity to the underlying matter power spectrum. As a result, a
large number of spectra are required to measure the matter power
spectrum to high accuracy. A precision of 5\% requires some $10^4$
lines of sight \cite{2005MNRAS.363.1145Z}.

Inversion methods have been further developed and applied by
\textcite{2000ApJ...543....1M} and \textcite{2002ApJ...581...20C}, who
assume the 1D flux power spectrum is proportional to the 1D matter
spectrum up to a scale dependent bias factor,
$\Delta^2_F(k)=b^2(k)\Delta^2_{1D}(k)$, and by
\textcite{1999MNRAS.303..179N, 2000MNRAS.313..364N} and
\textcite{2006MNRAS.369..734Z}, who convert local optical depth
measurements to local density estimates, assuming a power-law relation
between the two.

In principle, the flux autocorrelation function may also be used to
test models. Simulations, however, are found to converge even more
slowly on this statistic, with convergence to better than 10\% for
separations exceeding 3\% of the box size problematic \cite{MW04}.

Methods that rely on direct comparisons between the measured and
predicted flux power spectra require a large grid of simulations to
cover the range in expected cosmological model parameters. Because of
the considerable computational expense involved in running full
hydrodynamical simulations, \textcite{2005ApJ...635..761M} rely
instead on a large grid of HPM pseudo-hydrodynamics simulations (see
\S~\ref{subsubsec:simmethods}), calibrated by a few full hydrodynamics
runs. These are computed using $256^3$ cells in box sizes of
$10h^{-1}$~Mpc (comoving) or smaller. This is well below the box size
recommended by \textcite{MW04}, who also find that the prediction for
$\Delta^2_F(k)$ is sensitive to the method of smoothing at the several
percent level near the peak and beyond (towards smaller values of
$k$). In a comparison of full hydrodynamics simulations with HPM
simulations in $30h^{-1}$~Mpc (comoving) boxes,
\textcite{2006MNRAS.367.1655V} find HPM simulations with many more
grid zones ($1200^3$) than particles ($2\times200^3$ for gas and dark
matter) are required to converge to within 5\% of the full
hydrodynamics derived flux power spectrum on scales $k<0.03\,{\rm
km^{-1}\,s}$ at $z=3$ and 4. For $400^3$ grid zones, the discrepancy
is several percent and scale-dependent. At $z=2$, convergence is even
poorer, with disagreement as large as 20--30\%, as shocks play an
increasingly important role in heating the gas, an effect which an HPM
simulation cannot account for.

A large number of physical systematic effects may also affect the flux
power spectrum. The impact of galactic winds
\cite{2002ApJ...580..634C, 2005MNRAS.360.1471M} and ionization
radiation fluctuations \cite{1999ApJ...520....1C, 2004ApJ...610..642C,
MW04, 2005MNRAS.360.1471M} are generally found not to have much effect
on the inferred matter power spectrum, except possibly at $z=5-6$ when
the attenuation length to ionizing photons becomes so short few
sources contribute to the local UV background.

More of a concern are known systematics that affect measurements of
the \Lya forest power spectrum. The effect of the wings of Damped \Lya
Absorbers and absorption by intervening metal absorption systems
\cite{1999ApJ...520....1C, 2002ApJ...581...20C} must be corrected for
to obtain high precision measurements of the flux power spectrum
\cite{2005MNRAS.360.1471M}. \textcite{2006ApJS..163...80M} provide an
estimate for the influence of the damping wings from DLAs. They also
correct for metal absorption, uncertainties in sky subtraction, and
calibration errors. They find clear evidence for contamination by
\SiIII, and devise a method for removing its influence.

\begin{figure}
\includegraphics[width=8cm]{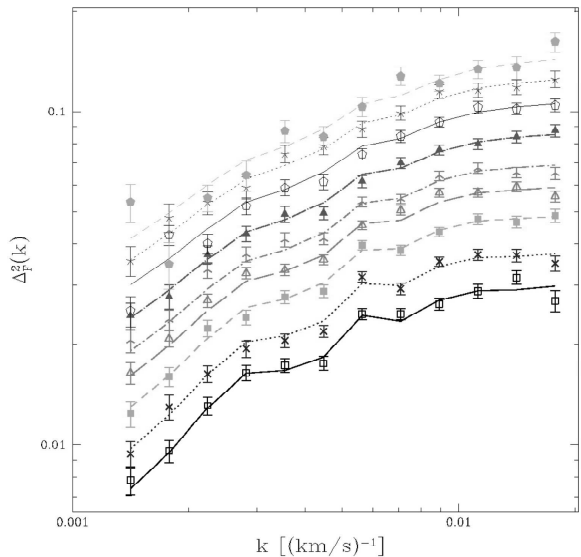}
\caption{Comparison of the best fitting model (lines) and the mean
power spectrum measured from the \Lya forest in 3035 SDSS QSO spectra
(points with error bars). The model includes the effect of \SiIIIs
absorption. The curves show a sequence in redshifts, increasing from
$z=2.2$ at the bottom to $z=3.8$ at the top, in steps of $\Delta z =
0.2$. From \textcite{2006ApJS..163...80M}. (Figure reproduced by
permission of the AAS.)
}
\label{fig:PFk}
\end{figure}

Any measurement of the flux power spectrum and comparison with models
also require a careful estimate of the QSO continuum. An optimal
method for estimating the QSO continuum in the context of flux power
spectra measurements is formulated by
\textcite{2001ApJ...552...15H}. The flux power spectrum was measured
from high resolution spectra of 27 QSOs by
\textcite{2004MNRAS.347..355K, 2004MNRAS.351.1471K}. The largest and
most precise set of measurements by far is that made using a sample of
3035 QSO spectra from the Sloan Digital Sky Survey
\cite{2006ApJS..163...80M}. The evolution of the measured flux power
spectrum is shown in Figure~\ref{fig:PFk}.

Despite their differences, the various approaches have yielded
reasonably good agreement on the inferred constraints of cosmological
parameters. Using the inversion method,
\textcite{2006MNRAS.369..734Z}, assuming $n=1$ for the matter power
spectrum, obtain $\sigma_8=0.92(\Omega_m/0.3)^{-0.3}$. A joint
analysis with the {\it WMAP} CMB data and a prior on the Hubble
constant yields $\sigma_8=0.92\pm0.04$ and
$\Omega_m=0.30\pm0.05$. Combining the \Lya flux power spectrum
measurements from the SDSS \cite{2005ApJ...635..761M,
2006ApJS..163...80M} with CMB, supernovae and galaxy clustering
constraints, \textcite{2006JCAP...10..014S} find the \Lya flux power
spectrum increases $\sigma_8=0.80\pm0.03$ to
$\sigma_8=0.85\pm0.02$. They obtain an index of $n=0.965\pm0.012$ with
no evidence for a running spectral index, $dn/d\ln k=-0.020\pm0.012$.
In a separate analysis of the measured SDSS flux power spectra,
\textcite{2006MNRAS.365..231V} obtain $\sigma_8=0.91\pm0.07$,
$\Omega_m=0.28\pm0.03$ and $n=0.95\pm0.04$. Using pure dark matter
simulations, \textcite{2005MNRAS.361.1257D} obtain $\sigma_8=0.7-0.9$.

The flux power spectra measurements tend to pull the normalization of
the matter power spectrum to larger values compared with estimates
based on other methods. This is consistent with the larger values of
$\sigma_8=0.9-1$ preferred by cosmological models that match
measurements of the pixel flux distribution function and {\it WMAP}
constraints \cite{MBM01, MW04}. On the other hand, the model
comparisons all neglect radiative transfer effects. Allowing for an
inverted equation of state over the redshift range $2<z<3$, in which
underdense regions are warmer than overdense ones, lowers the allowed
range for $\sigma_8$ \cite{2008MNRAS.386.1131B}. There is growing
evidence that the equation of state is inverted, or at least not as
monotonic as the power spectra and pixel flux distribution analyses
have assumed \cite{2008MNRAS.386.1131B, 2007ApJ...662...72B}.
Simulations including radiative transfer show a broad spectrum of
temperatures at a given density, including hot underdense regions, in
scenarios with late \HeII\ reionization \cite{2007MNRAS.380.1369T}.

\section{Reionization}
\label{sec:reionization}

\subsection{The epoch of reionization}
\label{subsec:EoR}

The reionization of the IGM is one of the principal unsolved problems
of cosmological structure formation. Three epochs of reionization of
the IGM may be identified, one for hydrogen and two for
helium. Because the same radiation sources capable of photoionizing
hydrogen would likely also photoionize \HeI\ to \HeII, only the epochs
of hydrogen reionization and full helium reionization to \HeIII\ are
likely to be distinct. As discussed in \S\S~\ref{subsubsec:helium} and
\ref{subsubsec:He-constraints} above, there is strong evidence full
helium reionization occurred at $z\simeq3.5$. The most likely \HeII\
reionization sources are QSOs. The epoch of hydrogen reionization is
much less certain. The topic attracted considerable attention
following measurements of the IGM \Lya optical depth in high redshift
QSOs and CMB measurements of intergalactic Thomson scattering. Unlike
full helium reionization, the sources of hydrogen reionization may
well be the first sources of radiation to turn on in the Universe,
ending the Dark Ages that followed the recombination of the primordial
hydrogen and helium plasma. Because of the numerous recent
reionization reviews (see \S~\ref{sec:intro}), I concentrate here only
on some of the most salient aspects.

The \Lya optical depth of the IGM measured in the spectra of several
$z>5.8$ QSOs discovered by the Sloan Digital Sky Survey has shown that
the IGM was reionized by $z\simeq5.7$ \cite{2000AJ....120.1167F,
2001AJ....122.2833F}, but the \Lya optical depth rises rapidly towards
higher redshifts \cite{2001AJ....122.2850B, 2001ApJ...560L...5D,
2002AJ....123.2151P, 2004AJ....128..515F, 2006AJ....132..117F,
2007ApJ...662...72B}. By $z\simeq6$, the spectra become so black
shortward of \Lya, usually only lower limits to the optical depth may
be obtained, even using the higher order Lyman resonance
transitions. The rapid rise is
suggestive of an approach to the EoR.  This interpretation is not
unique, as only a small neutral fraction is adequate for providing a
large \Lya optical depth by Eq.~(\ref{eq:tauGP}). A QSO at
$z\simeq6.4$ moreover shows transmitted flux at $z\simeq6.1$
\cite{2003AJ....126....1W, 2005AJ....129.2102W}. Gaps, however, are
expected as \HII\ regions develop and evolve prior to their complete
overlap \cite{2004ApJ...613....1F, 2006ApJ...646..696W}.

The interpretation of the IGM transmission measurements at $z>6$ as
evidence for the approach to the EoR is further confounded by {\it
WMAP} measurements of the intergalactic Thomson optical depth
\begin{eqnarray}
\tau_e &=& \int_{z_l}^{z_u} dz n_e
c\sigma_T\biggl\vert\frac{dt}{dz}\biggr\vert\\
&\simeq&0.004\biggl\{\biggl[\Omega_m(1+z_u)^3+\Omega_v\biggr]^{1/2}\nonumber\\
&&\phantom{{=}\frac{2}{3}\frac{n_e(0)c\sigma_T}{H_0\Omega_m}}
-\biggl[\Omega_m(1+z_l)^3+\Omega_v\biggr]^{1/2}\biggr\}\nonumber,
\label{eq:taue}
\end{eqnarray}
where $\sigma_T$ is the Thomson cross section, $n_e(0)$ is the
comoving electron density (assumed constant and without full helium
reionization), and a flat universe is assumed. The 5-year {\it WMAP}
polarization data yield a marginalized optical depth of $\tau_e =
0.087\pm0.017$ \cite{2008arXiv0803.0586D} up to the EoR, corresponding
to an EoR redshift of $z_u=11.0\pm1.4$, assuming reionization was a
single, instantaneous event throughout the Universe. The $2\sigma$ and
$3\sigma$ lower limits are $z_u>8.2$ and $z_u>6.7$, respectively
\cite{2008arXiv0803.0586D}. These values are lowered slightly to
$\tau_e=0.084\pm0.016$ and $z_u=10.8\pm1.4$, allowing for additional
astrophysical constraints on the preferred cosmological model from
Type Ia supernovae and Baryonic Acoustic Oscillations
\cite{2008arXiv0803.0732H}.

It is worth noting that a non-negligible contribution to the total
Thomson optical depth will arise from the IGM {\it prior} to the EoR
because of the residual electron density frozen out after the
Recombination Era. A residual fraction of as little as
$(2-4)\times10^{-4}$ over the redshift range $7<z<500$
\cite{2007ApJS..170..335P} will produce an additional optical depth of
$\Delta\tau_e=0.005-0.01$, using Eq.~(112). Careful modeling
of the contribution to the polarization signal from residual electrons
between the Recombination Era and the Epoch of Reionization is
required to estimate the contribution due to reionization sources
alone.

\subsection{The growth of cosmological \HII\ regions}
\label{subsec:HII-regions}

The reionization of the Universe occurs as a consequence of the
overlap of \HII\ regions driven into the IGM by discrete radiation
sources.\footnote{Excepted are scenarios invoking reionization by
decaying or annihilating dark matter particles
\cite{1982MNRAS.198P...1S, 1991A&A...250..295S, 1998A&A...335...12S,
2006MNRAS.369.1719M}. \textcite{2006MNRAS.369.1719M} find that while
some popular dark matter candidates may cause partial reionization,
none are able to fully reionize the IGM prior to $z=6$.} Unlike the
\HII\ region around a star in a galaxy, the growth of cosmological
\HII\ regions will generally not terminate in a Str\"omgren sphere,
defined as the surface within which the rate of radiative
recombinations balances the rate of production of ionizing photons by
the source. As the ionization front (I-front) of the \HII\ region
advances, the expansion of the Universe reduces the density of the gas
into which the I-front moves. Unless the I-front encounters a large
density inhomogeneity, it will travel unimpeded, lagging further and
further behind the Str\"omgren radius \cite{1986PASP...98.1014S},
until either the source dies, or the I-front overlaps with another
I-front produced by another source or sources.

The advance of the I-front of an isotropic cosmological \HII\ region
into a multicomponent IGM is described by
\cite{1986PASP...98.1014S, SG87, 1987ApJ...323L..13D, MM93}
\begin{equation}
4\pi r_I^2 n_{\rm H} \biggl(\frac{dr_I}{dt} - H r_I\biggr) = S(0)
-\frac{4}{3} \pi r_I^3 n_{\rm H}^2 \sum_i c_i f_{e{\rm H},i} \alpha_B(T_i),
\label{eq:drIdt}
\end{equation}
where $r_I$ is the proper radius of the I-front, $n_{\rm H}=\sum_i f_i
n_{{\rm H}, i}$ is the mean hydrogen density of the IGM averaged over
the various components with volume filling factors $f_i$, $H$ is the
Hubble parameter, $S(0)$ is the emission rate of ionizing photons from
the central sources, $c_i=f_i n^2_{{\rm H}, i}/n^2_{\rm H}$ is the
partial recombination clumping factor of the gas from component $i$,
$f_{e{\rm H}, i}$ is the number of electrons per hydrogen atom in
component $i$ (which will contain helium and possibly metals), and
$\alpha_B(T_i)$ is the Case B radiative recombination rate to the
$n=2$ level of hydrogen, allowing for ionizations from the ground
state by the diffuse radiation emitted following recombinations to the
ground state (see \S~\ref{subsubsec:uvbg-origin} above). If the
diffuse radiation is treated separately, then Case A recombination
should be used, and a second set of partial ionization clumping
factors must be introduced to represent the volume-averaged product of
the local ionization rate and \HI\ density in each subcomponent $i$
\cite{2007ApJ...657...15K}. If the electron fractions $f_{e{\rm H},
i}$ and temperatures of the various components are approximately the
same, then $f_{e{\rm H}, i}\alpha_B(T_i)$ may be approximated as
$f_{e{\rm H}}\alpha_B(T)$ and it is convenient to introduce a total
recombination clumping factor over all subcomponents, ${\cal
C}(z)=\sum_i c_i(z)$. The recombination term in Eq.~(\ref{eq:drIdt})
is then $(4\pi/3) r_I^3 n_{\rm H}^2 {\cal C} f_{e{\rm H}}
\alpha_B(T)$. (If the diffuse radiation is treated separately, then a
similar total ionization clumping factor may be similarly defined.)

Eq.~(\ref{eq:drIdt}) is an approximation for which it is assumed the
I-front, within which the gas transforms from neutral to highly
ionized, is narrow compared with the radius of the \HII\ region. Is
also assumes the I-front propagates rapidly compared with the sound
speed, so that the effects of gas motion induced by the pressure
within the \HII\ region may be neglected. It assumes as well that the
I-front propagates slowly compared with the speed of light, which
formally breaks down in the early stages. Near luminal expansion may
be approximately described by replacing $4\pi r_I^2n_{\rm H}$ on the
LHS of Eq.~(\ref{eq:drIdt}) by $4\pi r_I^2n_{\rm H} + S^\prime/c$,
where $S^\prime$ is the RHS of Eq.~(\ref{eq:drIdt})
\cite{2003AJ....126....1W, 2004Natur.427..815W, 2006ApJ...648..922S}.
This will ensure $dr_I/dt<c$.

A key uncertainty in the rate at which an \HII\ region expands is the
recombination clumping factor. Because the clumping factor may be
dominated by rare very dense structures, establishing its value from
numerical simulations is not straightforward. In particular, systems
that produce ionizing photons should be excluded. The dynamical
response of clumped structures to an impinging I-front must also be
computed, which requires radiative transfer and high spatial
resolution. The degree to which a dense cold clump bound by a minihalo
impedes the propagation of an I-front in general depends on the mass
of the clump and the intensity and spectral shape of the incident
radiation field \cite{2004MNRAS.348..753S, 2005MNRAS.361..405I}. A
wide range of values for the clumping factor are available in the
literature. On the basis of high resolution simulations,
\textcite{2007MNRAS.376..534I} find the recombination clumping factor
is well-fit over the redshift range $6<z<30$ by
\begin{equation}
{\cal C}(z) \simeq 26.2917 \exp[-0.1822z+0.003505z^2],
\label{eq:cfac-evol}
\end{equation}
for models using the preferred {\it WMAP} 3-year cosmological
parameters \cite{2007ApJS..170..377S}.

The evolution in the filling factor of ionized gas as the \HII\
regions grow may be quantified by the porosity parameter $Q_{\rm
HII}$, which is the product of the volume ionized per source of a
given ionizing photon luminosity, integrated over the luminosity
function of the sources \cite{1987ApJ...323L..13D, MM93}. For $Q_{\rm
HII}>1$, the IGM becomes completely ionized. Its growth is governed by
\cite{1999ApJ...514..648M, Meiksin05}
\begin{equation}
\frac{dQ_{\rm HII}}{dz}=\biggl[\frac{\dot n_S(z)}{n_{\rm H}(0)}
-\frac{{\cal C}(z)Q_{\rm HII}}{t_{\rm rec}(z)}\biggr]
\biggl[-\frac{1}{H(z)(1+z)}\biggr],
\label{eq:dQdz}
\end{equation}
where $n_{\rm H}(0)$ is the comoving number density of hydrogen, $\dot
n_S(z)$ is the production rate of ionizing photons per comoving volume,
and $t_{\rm rec}(z) = 1/[n_{\rm H}(z)f_{e{\rm H}} \alpha_B(T)]$ is the
recombination time at the mean electron density at redshift $z$.
If the comoving emissivity of the sources may be approximated as
$\epsilon^S_\nu(z)=\epsilon^S_L(\nu/\nu_L)^{-\alpha_S}(1+z)^{\gamma_S}$,
the ionizing photon production rate is
\begin{equation}
{\dot n}_S=\int_{\nu_L}^\infty d\nu\frac{\epsilon^S_\nu (z)}{h_{\rm P}\nu}
=\frac{\epsilon^S_L}{h_{\rm P}\alpha_S} (1+z)^{\gamma_S}.
\label{eq:nSdot}
\end{equation}

To solve Eq.~(\ref{eq:dQdz}), it is convenient to define $\tau_{\rm H,
HI}(z)={\cal C}(z) t_{\rm H}(z)/ t_{\rm rec} \simeq 0.02 {\cal C}(z)
(1+z)^{3/2}$, which is the Hubble time, $t_{\rm H}=2/3H(z)$, measured
in units of the recombination time, normalized here at $T=10^4$~K and
including the recombination clumping factor. It is equivalent to the
mean number of recombinations per hydrogen atom over a Hubble
time. Similar definitions may be introduced for the reionization of
\HeI, $\tau_{\rm H, HeI}\simeq0.03{\cal C}(z) (1+z)^{3/2}$ (assuming
all the ionized \HeI\ is in the form of \HeII), and of \HeII,
$\tau_{\rm H, HeII}\simeq0.2{\cal C}(z) (1+z)^{3/2}$. Adopting the
clumping factor from Eq.~(\ref{eq:cfac-evol}) gives nearly constant
values for the various $\tau_{\rm H}$ over $6<z<8$ of $\tau_{\rm H,
HI}\simeq4.7$, $\tau_{\rm H, HeI}\simeq5.3$, and $\tau_{\rm H,
HeII}\simeq32$. For constant $\tau_{\rm H, HI}$, Eq.~(\ref{eq:dQdz})
may be directly integrated, yielding
\begin{eqnarray}
Q_{\rm HII}(z) &\simeq& \frac{1}{1+\tau_{\rm H, HI}-2\gamma/3}
\biggl[\frac{{\dot n}_S(z)t_{\rm H}(z)}{n_{\rm H}(0)}\nonumber\\
\phantom{{\simeq}}
&&-\frac{{\dot n}_S(z_i)t_{\rm H}(z_i)}{n_{\rm H}(0)}
\biggl(\frac{1+z}{1+z_i}\biggr)^{3\tau_{\rm H, HI}/2}\biggr]\nonumber\\
&\simeq&\frac{1}{1+\tau_{\rm H, HI}-2\gamma/3}
\frac{{\dot n}_S(z)t_{\rm H}(z)}{n_{\rm H}(0)},
\label{eq:QHII}
\end{eqnarray}
where $z_i$ is the turn-on redshift of the ionizing sources, and the
assumption $z_i\gg z$ has been made to obtain the final expression.
Similar expressions apply for \HeI\ and \HeII\ reionization, on
replacing $\tau_{\rm H, HI}$ by the appropriate form for \HeI\ or
\HeII, using the production rate of ionizing photons for \HeI\ or
\HeII\ for $\dot n_S$, and replacing $n_{\rm H}(0)$ by $n_{\rm
He}(0)$. In principle, a recombination clumping factor appropriate to
each species must by used since the depth of self-shielding in clumps
will differ for different species. To date, simulations have focused
on the clumping of ionized hydrogen gas only.

The (comoving) production rate of ionizing photons from the UV background
emissivity estimate of Eq.~(\ref{eq:epsL}) follows from Eq.~(\ref{eq:nSdot}),
\begin{equation}
{\dot n}_S\simeq A_S\biggl(\frac{3+\alpha_{\rm MG}}{3\alpha_S}\biggr)
(1+z)^\gamma h\,{\rm ph\,s^{-1}\,Mpc^{-3}},
\label{eq:dotns}
\end{equation}
where $A_S=1.3^{+(1.5, 4.6)}_{-(0.7, 1.0)}\times10^{52}$,
$\gamma=-1.6-0.6\ln(A_S/1.3\times10^{52})$, and $\alpha_S$ is the
spectral index at frequencies above the Lyman edge
\cite{Meiksin05}. (This neglects any contribution to the required
emissivity from the diffuse recombination radiation from the IGM.  If
the IGM contributes 30\% of the emissivity, $A_S$ should be reduced by
the factor 1.3.) Substitution into Eq.~(\ref{eq:QHII}) and
extrapolating Eq.~(\ref{eq:dotns}) to high redshifts shows that the
Universe would be reionized by $z_{\rm ri}\simeq5.5$ for $\alpha_{\rm
MG}=\alpha_S=0.5$, but as late as $z_{\rm ri}\simeq3.8$ for
$\alpha_{\rm MG}=\alpha_S=1.8$. Reionization by $z_{\rm ri}\gta6$, as
suggested by observations, requires either a hard spectrum for the
sources or an increase over the evolution rate of Eq.~(\ref{eq:epsL})
in the comoving emissivity at $z>6$ \cite{2003ApJ...597...66M,
Meiksin05, 2007MNRAS.382..325B}.

\subsection{Sources of reionization}
\label{subsec:reion-sources}

\subsubsection{Galaxies}
\label{subsubsec:reion-galaxies}

High redshift star-forming galaxies are generally considered the most
likely candidates for the reionization of the Universe. The discovery
of high redshift galaxies in deep surveys has sparked much lively
debate over whether the population of galaxies that reionized the
Universe has now been identified \cite{2004ApJ...616L..79B,
2004MNRAS.355..374B, 2004ApJ...610L...1S, 2004ApJ...600L...1Y,
2006NewAR..50...94B, 2006ApJ...653...53B}. The uncertainties involved
preclude any definite decision.

Consider the (comoving) luminosity density at $5.5<z<6.5$ from
\textcite{2006ApJ...653...53B}, corrected upward by 50\% to allow for
internal extinction, of $\rho_{1350}\simeq10^{19.4}\,{\rm
W\,Hz^{-1}\,Mpc^{-3}}$. The conversion rates between
$L_\nu(1350\,{\rm\AA})$ (in ${\rm W\,Hz^{-1}}$), and the
production rate of ionizing photons for a Salpeter Initial Mass
Function with $M_{\rm lower}=1~M_\odot$ and $M_{\rm
upper}=100~M_\odot$ and solar metallicity are
\begin{eqnarray}
{\dot N}_{\rm S, HI} = 10^{32.0}\,{\rm ph\,s^{-1}}L_\nu(1350\,{\rm \AA});
\quad{\rm HI}\nonumber\\
{\dot N}_{\rm S, HeI} = 10^{31.3}\,{\rm ph\,s^{-1}}L_\nu(1350\,{\rm \AA});
\quad{\rm HeI}\nonumber\\
{\dot N}_{\rm S, HeII} = 10^{29.6}\,{\rm ph\,s^{-1}}L_\nu(1350\,{\rm \AA});
\quad{\rm HeII}
\label{eq:QLnu}
\end{eqnarray}
\cite{Leitherer99}. For a metallicity of $1/20$ solar, the rates are
about 0.1~dex higher for \HI, 0.2~dex higher for \HeI, and 0.4~dex
smaller for \HeII. A similar conversion for \HI\ is inferred from
\textcite{1998ApJ...498..106M} and \textcite{1999ApJ...514..648M} for
an alternative Salpeter model with $M_{\rm lower}=0.1\,M_\odot$ and
$M_{\rm upper}=125\,M_\odot$. Assuming that all the ionizing photons
escape, applying these rates to the luminosity density at $5.5<z<6.5$,
Eq.~(\ref{eq:QHII}) (and its variants for \HeI\ and \HeII\
reionization, assuming the same recombination clumping factors, which
may in fact differ) shows that the galaxies are able to reionize the
\HI\ by $z_{ri}=12$ and the \HeI\ by $z_{ri}=21$, while they are
unable to reionize the \HeII\ even by the present epoch. The results
for \HI\ and \HeI, however, are very sensitive to the uncertainty in
the counts and the escape fractions. If only 10\% of the ionizing
radiation escapes, then the reionization epochs for \HI\ and \HeI\ are
reduced dramatically to $z_{ri}=1.8$ and 3.7,
respectively. Reionization of \HI\ by $z=6$ requires $f_{\rm
esc}>0.6$, while for \HeI, $f_{\rm esc}>0.3$ is needed. Allowing for
the upper limits on the counts eases these constraints, as would lower
metallicities for the stars \cite{2000ApJ...528L..65T} or an Initial
Mass Function heavily tilted to very massive ($\sim300\,M_\odot$)
stars \cite{2001ApJ...552..464B}. On the other hand, a decline in the
comoving luminosity density of the sources at $z>6$ would make the
case that these sources reionized the Universe less secure. While the
population of sources that reionized the Universe may have been
discovered in high redshift surveys, a definite conclusion is still
premature. There is some evidence for a population of low-luminosity
star-forming galaxies at redshifts as high as $z\simeq8-10$ that may
contribute substantially to the total budget of reionizing photons
\cite{2007ApJ...663...10S}.

Several semi-analytic estimates for the expected contribution of
galaxies to the reionization of both hydrogen and helium have been
made \cite{2003ApJ...584..621V, 2003ApJ...595....1H,
2006ApJ...650....7H, 2008arXiv0806.0392S}. The possible halo masses of
the objects that dominate reionization cover a broad range, including
systems as small as minihalos with masses $M_{\rm halo}<10^6\,{\rm
M_\odot}$ and virial temperatures $T_{\rm vir}<1000$~K which would
have so far evaded detection. Definitive predictions, however, are
hampered by the need to incorporate several poorly determined
quantities, including the fraction of baryons in a collapsed halo that
are converted into stars, the number of ionizing photons produced by
the stars, and the escape fraction of the ionizing radiation from the
galaxies. Recent estimates suggest that the efficiency of ionization
per galaxy cannot be too high, or a Thomson optical depth would be
produced that is too large compared with CMB polarization measurements
\cite{2006ApJ...650....7H, 2008arXiv0806.0392S}.

\subsubsection{QSOs}
\label{subsubsec:reion-QSOs}

The emissivity of QSOs lies well below the requirements for
reionization, unless there is a large population of very low
luminosity Active Galactic Nuclei (AGN) \cite{Meiksin05,
2007MNRAS.374..627S, 2007MNRAS.382..325B}. The estimated production
rate of ionizing photons from accretion onto massive black holes in
the Universe is comparable to the rate from nucleosynthesis in stars
\cite{2006MNRAS.365..833M}:\ low luminosity AGN may still await
discovery at high redshifts. By contrast, QSOs likely fully ionized
helium by $z\simeq3$, and possibly as early as $z\simeq5$ if they have
hard spectra extending down to the \HeII\ photoelectric edge
\cite{Meiksin05}.

An interesting alternative class of reionization sources is a possible
population of miniquasars, systems produced by intermediate mass black
holes with masses $200-1000\,M_\odot$ growing by accretion onto seed
black holes produced by the collapse of very massive Population III
stars \cite{Madau04}. Such sources could plausibly reionize the
Universe by $z\simeq15$.

\subsubsection{Other sources}
\label{subsubsec:reion-other-sources}

Various alternative sources have been proposed for the reionization of
the IGM, including globular star clusters \cite{2004ASPC..322..509R},
massive Population III stars more generally, with their formation
possibly triggered early by primordial magnetic fields
\cite{2006MNRAS.368..965T}, and an early generation of black holes
\cite{2001ApJ...563....1V, 2004MNRAS.352..547R,
2005MNRAS.357..207R}. While these mechanisms are intriguing and have
received a variety of theoretical investigations, observational
evidence for any of them is as yet wanting. The possibility that
sources other than galaxies reionized the IGM, however, should not be
discounted:\ the Universe can surprise.

\subsection{21cm signature of reionization}
\label{subsec:reion-21cm}

An altogether novel method for discovering the Epoch of Reionization
is through the detection of the 21cm signature from the neutral IGM before
and during the reionization process.
The emission or absorption of 21cm radiation from a neutral IGM is
governed by the spin temperature $T_S$ of the hydrogen, defined by
\begin{equation}
\frac{n_1}{n_0}=3\exp\left(-\frac{T_*}{T_S}\right),
\end{equation}
where $n_0$ and $n_1$ are the singlet and triplet $n=1$ hyperfine
levels and $T_*\equiv h_{\rm P}\nu_{10}/k_{\rm B}\simeq0.07\,$K, where
$\nu_{10}$ is the frequency of the 21cm transition. In the presence of
only the CMB radiation, the spin temperature will come into
equilibrium with the CMB on a timescale of $T_*/ (T_{\rm CMB}A_{10})
\simeq3\times10^5(1+z)^{-1}$~yr, where
$A_{10}=2.85\times10^{-15}\,{\rm s^{-1}}$ is the spontaneous decay
rate of the transition. In this case, the spin temperature will be the
same as the CMB temperature, and no emission or absorption relative to
the CMB will be detectable. A mechanism is needed to decouple the two
temperatures. Two mechanisms may couple the spin temperature to the
kinetic temperature of the gas. One is the collisions between hydrogen
atoms \cite{1956ApJ...124..542P}. The collision-induced coupling
between the spin and kinetic temperatures is dominated by
spin-exchange between the colliding hydrogen atoms, with possible
significant contributions from electron-hydrogen collisions and
proton-hydrogen collisions, depending on the ionization fraction and
temperature \cite{1966P&SS...14..929S, 1967RvMP...39..850D,
1969ApJ...158..423A, 2005ApJ...622.1356Z, 2007MNRAS.374..547F,
2007MNRAS.375.1241H}. The second, less obvious mechanism, is
scattering by \Lya photons \cite{1952AJ.....57R..31W,
1958PrIRE...46..240F}. The Wouthuysen-Field effect mixes the hyperfine
levels of neutral hydrogen in its ground state via an intermediate
transition to the $2p$ state. An atom initially in the $n=1$ singlet
state may absorb a \Lya photon that puts it in an $n=2$ state,
allowing it to return to the triplet $n=1$ state by a spontaneous
decay. The combined atomic collision rates are generally too small at
the average IGM densities at the redshifts of interest, although
collisions may be important in overdense regions,
$\delta\rho/\rho\gta30[(1+z)/10]^{-2}$ \cite{MMR97}. Instead the
dominant mechanism at detectable redshifts for the near future is
likely to be \Lya scattering.

A patchwork of either 21cm emission, or absorption against the Cosmic
Microwave Background, will result \cite{1979MNRAS.188..791H}. Large
radio telescopes like LOFAR, the MWA, PAPER, or the SKA offer the
prospect of measuring this signature, and so detecting the
transitional epoch from a dark universe to one with light. The spin
temperature of the neutral hydrogen is
\begin{equation}
T_S=\frac{T_{\rm CMB}+y_\alpha T_\alpha + y_c T_K}{1+y_\alpha + y_c},
\label{eq:Tspin}
\end{equation}
where
\begin{equation}
y_\alpha\equiv\frac{P_{10}}{A_{10}}\frac{T_*}{T_\alpha} \quad{\rm and}
\quad y_c\equiv\frac{C_{10}}{A_{10}}\frac{T_*}{T_K}
\end{equation}
are the \Lya and collisional pumping efficiencies, respectively. Here
$P_{10}$ is the indirect de-excitation rate of the triplet state via
absorption of a \Lya photon to the $n=2$ level, and $T_S\gg T_*$ is
assumed. Consideration of the net transition rates between the various
hyperfine $n=1$ and $n=2$ levels shows that the $1\rightarrow0$
transition rate via \Lya scattering is related to the total rate
\begin{equation}
P_\alpha = c\int_0^\infty\,d\nu\,\sigma_\nu\frac{u_\nu}{h_{\rm P}\nu},
\label{eq:Palpha}
\end{equation}
where $\sigma_\nu$ is the resonance line cross section given by
Eq.~(\ref{eq:sigmaV}) and $u_\nu$ is the radiation energy density, by
$P_{10}=4P_\alpha/27$ \cite{1958PrIRE...46..240F}. The quantity
$T_\alpha$ is the color temperature of the radiation field, given by
the harmonic mean $T_\alpha=1/\langle T_u^{-1}(\nu)\rangle$ of the
frequency-specific temperature $T_u(\nu)=-(h_{\rm P}/k_{\rm B}) (d\ln
u_\nu/d\nu)^{-1}$, weighted by $u_\nu\varphi_V(a,\nu)$
\cite{Meiksin06}. The factor $C_{10}$ is the rate of collisional
de-excitation of the triplet level.  In the absence of collisions,
there exists a critical value of $P_\alpha$ which, if greatly
exceeded, would drive $T_S\rightarrow T_K$. This thermalization rate
is \cite{MMR97}
\begin{equation}
P_{\rm th}\equiv {27A_{10}T_{\rm CMB}\over 4T_*}\approx 
(5.3\times 10^{-12} ~{\rm s}^{-1})~ \left({1+z\over 7}\right).
\label{eq:ptherm}
\end{equation}
It corresponds to a \Lya intensity of
$J_\alpha\simeq0.6S_{\alpha}^{-1}\times10^{-24}\,{\rm W\,
m^{-2}\,Hz^{-1} \,sr^{-1}}$, where $S_{\alpha}$ is a suppression
factor of order unity that takes into account the effect of recoils in
an expanding medium on the shape of the intensity near the center of
the line profile \cite{2004ApJ...602....1C, 2006PhR...433..181F}. The
value of the intensity corresponds to a comparable number of \Lya
photons as hydrogen atoms at $z=8$, and so is comparable to the
intensity expected from sources able to reionize the Universe
\cite{MMR97}. It is this coincidence that promises that the
Wouthuysen-Field mechanism may reveal the IGM prior to its complete
reionization.

To illustrate the basic principle of the observations, consider a
patch of neutral material with spin temperature $T_S\neq T_{\rm CMB}$,
having an angular size on the sky which is large compared with a
beamwidth, and radial velocity extent due to the Hubble expansion
which is larger than the bandwidth. Following an analysis similar to
the derivation of the Gunn-Peterson optical depth
(Eq.~[\ref{eq:tauGP}]), and taking into account the reduction of the
absorption coefficient due to stimulated emission by the factor
$1-e^{-T_*/T_S}\simeq T_*/T_S$ for $T_*\ll T_S$, the intergalactic
optical depth at $21(1+z)\,{\rm cm}$ along the line of sight is
\begin{eqnarray}
\tau(z) &=& \frac{3}{32\pi}x_{\rm HI}[1+\delta(z)]{\bar n_{\rm
H}}\lambda_{10}^3 \frac{A_{10}}{H(z)}\frac{T_*}{T_S} \nonumber\\
&\simeq& 0.0032h^{-1}[1+\delta(z)]\biggl[\frac{x_{\rm
HI}(z)}{T_S}\biggr]\nonumber\\
&&\times\Omega_m^{-1/2}(1+z)^{3/2}\biggl[1+\frac{1-\Omega_m}{\Omega_m(1+z)^3}
\biggr]^{-1/2},
\label{eq:tau21cm}
\end{eqnarray}
which will typically be much less than unity. Here, $x_{\rm HI}$ is
the neutral fraction, $\delta$ accounts for fluctuations about the
mean hydrogen density, $n_{\rm HI}=\bar n_{\rm HI}(1+\delta)$, and a
factor of $1/4$ is introduced since only the hydrogen atoms in the
singlet state will take part in the absorption. A flat universe is
assumed for the second form in Eq.~(\ref{eq:tau21cm}) and
$\Omega_bh^2=0.022$. \textcite{1959ApJ...129..525F} applied
Eq.~(\ref{eq:tau21cm}) to a search for intergalactic emission or
absorption against a background radio galaxy. No clear emission or
absorption signature was detected, with an upper limit on the 21cm
optical depth of $\tau<0.0075$. For $\Omega_m=0.3$, $h=0.7$, and
$T_S=T_{\rm CMB}$, this corresponds to $x_{\rm HI}(1+\delta)<1.5$
along the line of sight. Adopting $\Omega_m=1$ and $\delta=0$ would
correspond to an upper limit on the baryon density of $\Omega_b<0.2$
assuming the hydrogen was all neutral, showing that early measurements
could already have ruled out a baryon-dominated universe closed by a
neutral IGM.

In the context of radio measurements of the IGM planned today, an
alternative observational strategy consists of cross-correlating
measurements of patches separated in either angle or
frequency. Possible detections include large isolated \HII\ regions,
as would be produced by the first QSOS \cite{MMR97} or, possibly more
straightforwardly, a global statistically averaged fluctuation signal
from the IGM \cite{2000ApJ...528..597T}. Foreground contamination,
particularly fluctuations in the ionosphere and discrete sources in
the Galaxy, pose formidable challenges to detecting the 21cm signature
from the Epoch of Reionization. But the development of novel detection
strategies suggests the challenges may not be
insurmountable. Discussions of the feasibility of the measurements are
provided by \textcite{2005ApJ...634..715W},
\textcite{2006ApJ...638...20B}, \textcite{2007MNRAS.382..809D}, and in
the reviews by \textcite{2006ARA&A..44..415F},
\textcite{2006PhR...433..181F}, and \textcite{2007RPPh...70..627B},
which also discuss specific reionization scenarios.

\section{The Absorber--Galaxy Connection}
\label{sec:abs-gal}

Ever since their discovery, intervening absorption systems have held
the promise of opening a new window on the evolution of galaxies by
probing the gas in galaxies as it was converted into stars, and
tracing the feedback of the stars on the gas as supernovae and stellar
winds pollute their environment with metals. Despite long and hard
searches, direct evidence relating intervening absorbers to galaxies
has been scant until only the past few years. Today, the
absorber--galaxy connection is fast becoming a boom area in IGM
studies.

\subsection{Galaxy-associated IGM absorption}
\label{subsec:gal-assoc-abs}

\subsubsection{Association of DLAs with galaxies}
\label{subsubsec:DLA-galaxy}

The motivation for the original systematic Damped \Lya Absorber survey
was the search for the gaseous disks of galaxies at high redshifts
\cite{1986ApJS...61..249W}. A large number of surveys have been
conducted since, summarized in the extensive review by
\textcite{2005ARA&A..43..861W}. There are several properties of DLAs
suggestive of galactic disk material:\ 1.\ They are the dominant
reservoir of neutral hydrogen at high redshifts ($z>3$)
\cite{2005ApJ...635..123P, 2006ApJ...646..730J}, 2.\ with the cosmic
gas mass density $\Omega_g^{\rm DLA}$ decreasing and metallicity
increasing with cosmic time \cite{2003ApJ...595L...9P}. 3.\ There is
evidence for depletion of Fe and Si onto dust grains similar in
strength to that in cold gas in the Galaxy
\cite{2002MNRAS.332..383P}. 4.\ There is photometric evidence for
reddening by dust, with the amount of extinction increasing with iron
column density similar to the trend found for interstellar clouds
within the Galaxy \cite{2006A&A...454..151V}. 5.\ Molecular hydrogen
is present in about 15--20\% of the systems
\cite{2006ApJ...640L..25L}, with the metallicity and depletion factor
increasing with the amount of $H_2$ \cite{2003MNRAS.346..209L}. 6.\
The metallicity of DLAs increases with the velocity width of the line
profiles of the low-ionization metals, which presumably trace the
gravitational potentials of the systems, and so is suggestive of a
mass-metallicity correlation \cite{2006A&A...457...71L}. 7.\ In
addition to a cold low ionization phase, a warm phase is indicated by
\CIV\ and \SiIV\ absorption systems, and a hot phase is indicated by
\OVI\ absorption detections \cite{2007A&A...465..171F}, demonstrating
a complex interstellar medium in DLAs much as is found in disk
galaxies. 8.\ Using a method that deduces the star formation rate from
the strength of \CII$^*\lambda\,1335.7$ as a measure of the amount of
gas heating by the photoelectric dust mechanism due to UV heating by
massive stars, \textcite{2003ApJ...593..215W} infer star formation
rates in a presumed cold neutral medium in DLAs similar to that in the
interstellar medium of the Galaxy. 9.\ The inferred global cosmic star
formation rate in DLAs at $z\simeq3$ is comparable to that of Lyman
Break Galaxies at the same redshift \cite{2003ApJ...593..235W}. 10.\ A
search for Lyman Break Galaxies at $z>1$ associated with DLAs shows
evidence for a significant spatial correlation between the two, with a
cross-correlation length comparable to the autocorrelation length of
Lyman Break Galaxies \cite{2006ApJ...652..994C}. 11.\ Optical
counterparts to DLAs at $z<1$ suggest the DLAs arise as part of large
(radii of $\sim30\,h^{-1}$~kpc), rotating \HI\ disks
\cite{2005ApJ...620..703C}. 12.\ The incidence rate and column density
distribution of DLAs at $z<1$ is comparable to those of low redshift
galaxies directly measured in \HI\ \cite{2005MNRAS.364.1467Z}.

Despite the circumstantial evidence favoring a galactic disk
connection for DLAs, there is strong contrary evidence as well:\ 1.\
While they are the dominant reservoir of neutral gas, DLAs contain
less than half the mass density of present day stars. 2.\ The amount
of neutral gas in DLAs exceeds the stellar mass in present day disks
\cite{2005ARA&A..43..861W}, which suggests they trace the disk
component of galaxies, but their metal depletion patterns most often
resemble that of warm disk clouds in the Galaxy or halo clouds
\cite{2002A&A...385..802L}. The difference in the UV metagalactic
backgrounds at low and high redshifts may partially account for the
apparent discrepancy. 3.\ The \CII$^*$ inferred star formation rates
suggest that a mass in metals should have been produced a factor 30
larger than measured in DLAs \cite{2003ApJ...593..235W}. 4.\ The
\CII$^*$ inferred star formation rates somewhat exceed the prediction
of the Kennicutt-Schmidt \cite{1959ApJ...129..243S,
1998ApJ...498..541K} relation between the star formation rate and gas
mass in galaxies \cite{2005ApJ...630..108H}. Possibly the principal
source of heat powering the measured \CII$^*$ cooling rates is star
formation in centrally located, compact Lyman Break Galaxies rather
than in the bulk of a spatially extended DLA system
\cite{2006ApJ...652..981W}. 5.\ DLAs are sufficiently abundant that if
they obeyed the Kennicutt-Schmidt relation, then 3\% of the sky should
be lit up with star-forming DLAs at high redshifts, contrary to
observations which find a comoving star formation density a factor
30--100 smaller than predicted \cite{2006ApJ...652..981W}. 6.\ Most
puzzling is the rarity of successful follow-up optical
identifications. While there is mounting circumstantial evidence that
DLAs and Lyman Break Galaxies are associated with the same systems at
$z>1.6$, only a handful of DLAs show \Lya in emission
\cite{2004A&A...422L..33M}. Galactic counterparts have been confirmed
in the optical and near-infrared for about a dozen DLAs at $z<1.6$
\cite{2003ApJ...597..706C, 2003ApJ...595...94R, 2004ApJ...600..613S,
2005ApJ...620..703C}. The numbers are too few, however, to establish
whether or not the galaxies represent a random sampling of the main
population of galaxies \cite{2005ARA&A..43..861W}. The nature of DLAs
and their relation to present day galaxies remain an enigma.

Simulations set in a $\Lambda$CDM cosmology reasonably well reproduce
the number of DLAs and their evolution at high redshifts when
dissipation due to cooling, star formation and subsequent supernovae
feedback and galactic winds are included
\cite{2004MNRAS.348..421N}. The DLAs are found to arise in halos no
smaller than those with total masses of about $10^8 h^{-1}M_\odot$ at
high redshifts \cite{2004MNRAS.348..421N}, and almost all in dark
matter halos less than $10^{13} h^{-1}M_\odot$
\cite{2007ApJ...660..945N}. Detailed comparisons between the simulated
DLAs and observations, however, reveal some discrepancies. The
predictions for the number of systems along a line of sight lie
somewhat below the measured number. This may be due to overly compact
DLAs in the simulations, which also show narrower velocity widths than
measured. High resolution simulations with radiative transfer suggest
that gas in tidal tails and quasi-filamentary structures also produce
DLAs \cite{2006ApJ...645...55R}, which may help to reconcile the
predicted and measured numbers. The star formation rate in the
simulations is consistent with the observational estimates in DLAs,
however an order of magnitude more metals are produced in the
simulations than measured \cite{2004MNRAS.348..435N}. Simulations with
radiative transfer suggest that the UV radiation from stars produced
in DLAs and the ambient UV metagalactic background may regulate the
rate of star formation within DLAs
\cite{2006MNRAS.368.1885I}. Detailed kinematic comparisons between the
simulations and the observations have yet to be made. While the
simulations have gone a long way toward accounting for the DLAs within
the framework of the $\Lambda$CDM model for structure formation, much
work still remains to demonstrate that this model correctly describes
the origin of most DLAs.

\subsubsection{Association of metal absorbers with galaxies}
\label{subsubsec:metal-abs-galaxy}

Optical and infrared searches for counterparts to metal absorption
systems, such as \MgII\ or \CIV, have proven an effective means of
discovering galaxies \cite{1991A&A...243..344B, 1994ApJ...437L..75S,
1995AJ....110.2519S, 1997ApJ...480..568S, 2007ApJ...669L...5B}. On the
whole, the galaxies appear to sample the general population of
galaxies in the field, although there occur fewer dim blue galaxies in
the absorber-selected samples. The morphological and kinematic
properties of galaxies associated with \MgII\ systems are reviewed by
\textcite{2005ASPC..331..387C}.

The spectra of QSOs proximate to galaxies on the sky probe galactic
environments. \textcite{2005ApJ...629..636A} find galaxies at $2\lta
z\lta 3$ generally show very strong \CIV\ absorption, $N_{\rm
CIV}\gg10^{14}\,{\rm cm^{-2}}$, within 40~kpc (proper) of the galaxy,
and \CIV\ absorbers with typical column densities of $N_{\rm
CIV}\simeq10^{14}\,{\rm cm^{-2}}$ are found as far out as 80~kpc. The
key question is why the metals are near the galaxies. Galaxies are
known to be associated with large scale baryonic overdensities, at
least as probed by the \Lya forest. In the nearby Universe, the
stronger absorbers ($13.2<\log_{10}N_{\rm HI}<15.4$) are spatially
correlated with galaxies, although with mixed views over the strength
of the cross-correlation compared with the galaxy autocorrelation
function \cite{2002ApJ...565..720P, 2006MNRAS.367.1251R,
2007MNRAS.375..735W}. At higher redshifts,
\textcite{2005ApJ...629..636A} similarly find galaxies are associated
with enhanced amounts of \Lya absorption. If the IGM is polluted by
metals, then the proximity of metal absorption systems with galaxies
may merely reflect the correlation of \HI\ with
galaxies. Alternatively, the metals associated with galaxies may have
been ejected by the galaxies in a wind following an episode of intense
star formation. Either alternative is intriguing. The possibility that
galaxies polluted their environment with metals suggests these metals
may serve as tracers of the past star formation history of the stars
in the galaxy, and so provide independent constraints on their
formation. On the other hand, if metal pollution is widespread
throughout the IGM, it raises the question of whether the pollution
was {\it in situ}, produced by some ubiquitous undetected class of
stellar systems like a population of high redshift globular star
clusters, or are the collective ejecta of galaxies that underwent a
superwind phase. In the latter case, it may mean that the evolution of
the stellar population of galaxies was not isolated, but that galaxies
``talked'' with each other as their stars formed, entraining metal
enriched material produced by other galaxies. A galactic wind may also
delay the onset of star formation in nearby low-mass halos by
delivering turbulent kinetic support to the gas
\cite{2004ApJ...613..159F}. In extreme cases, a wind may eradicate the
gaseous content of nearby low-mass halos through evaporation or
shock-driven stripping, suppressing star formation in these systems
entirely \cite{2000ApJ...536L..11S}.

Evidence that high redshift galaxies do impact on the IGM is provided
by \textcite{2005ApJ...629..636A}, who find that about a third of the
galaxies in their sample at $2\lta z\lta3$ show clearings of weak or
absent \HI\ absorption out to distances of $1h^{-1}$~Mpc
(comoving). Simulations suggest, however, that the clearings are
consistent with random fluctuations
\cite{2006MNRAS.367L..74D,2007ApJ...663...38K}.  The kinematic studies
of \textcite{2006AJ....131...24S} show that \CIV\ systems at $2<z<3.5$
with $N_{\rm CIV}>2\times10^{13}\,{\rm cm^{-2}}$ lie in complexes
having velocity widths of up to $300-400\kms$, suggestive of galactic
outflows. Lower column density systems lie in complexes with
substantially lower velocity widths of $\sim50\kms$, suggesting they
are not part of any galactic outflow, although they may be part of
residual outflows that were initiated at much higher redshifts
\cite{2004ApJ...613..159F}. The evidence for galactic outflows has
prompted an increasing interest in the impact of galactic winds on the
IGM, discussed next.

\subsection{Galactic winds and the IGM}
\label{subsec:winds}

Galactic winds occur throughout the Universe, in a variety of
environments, over a wide range of galactic masses, including
starbursts in the local Universe \cite{1998ApJ...503..646H} and Lyman
Break Galaxies at $z\simeq3$ \cite{2002ApJ...569..742P} and \Lya
emitting systems associated with them \cite{2005Natur.436..227W}. A
comprehensive review of galactic winds in the Universe is provided by
\textcite{2005ARA&A..43..769V}. Reviews specific to the IGM are
provided by \textcite{2007EAS....24..165A} and
\textcite{2008SSRv..134..295B}.

A natural mechanism for enriching the gas near galaxies with metals is
through winds. The possible enrichment by wind ejecta raises further
questions:\ What sort of galaxies ejected the metals? When did they do
it? How does the metal pollution relate to the star formation history
of the Universe?

Geometric estimates of the impact of galactic winds may be made on the
basis of the spatial correlations of the metals and galaxies.
\textcite{2005ApJ...625L..43P} find that metal-enriched bubbles of
comoving radius $\sim100$~kpc expelled at $z=9$ by a biased population
of dwarf galaxies with a halo mass of $10^9M_\odot$, polluting the
region around Lyman Break Galaxies, would account both for the metal
abundances measured at $z=3$ and the measured spatial
cross-correlation between metals and Lyman Break Galaxies. Using high
resolution, high signal-to-noise ratio spectra,
\textcite{2006MNRAS.365..615S, 2006MNRAS.366.1118S} study the
clustering properties of high (\CIVs and \SiIV) and low (\MgIIs and
\FeII) ionization metal absorption systems. In conjunction with
numerical simulations, they find that a uniform distribution of metals
fails to reproduce the correlations measured. Modulating the
metallicity by the density field similarly fails. They do find,
however, that the \CIVs correlation function is well-reproduced by
assigning the metals to clumps confined to bubbles of comoving radius
2~Mpc around systems with masses exceeding $10^{12}M_\odot$ at
$z=3$. A similar model with a slightly larger bubble radius recovers
the measured correlations of \MgIIs and \FeIIs at $z=0.5-2$. Both
models are degenerate with a biased high-redshift enrichment model
with the metals confined to 2.5~Mpc (comoving) radius bubbles around
$3\times10^9M_\odot$ systems at $z=7.5$. The results are consistent
with two scenarios, metals deposited by winds from massive galaxies at
$z\simeq3$, or by winds from dwarf galaxies at high redshifts,
$z>7$. An early wind model is also favored by the near constant mass
density associated with \CIVs systems over the broad redshift range
$2<z<6$ \cite{2005AJ....130.1996S, 2006MNRAS.371L..78R}. Whether any
of these scenarios are viable depends on whether wind models that
support them may be physically realized.

Correlations between low redshift absorption systems and galaxies
provide a potentially powerful means of inferring the origins of the
metals. \textcite{2005ApJ...623L..97T} find that dwarf galaxies are
able to reproduce the line-of-sight number density of \OVIs systems if
they produce winds extending out to about 200~kpc, while luminous
galaxies would need to produce winds extending to $0.5-1$~Mpc.
Similar results are obtained from the \OVIs survey analysis of
\textcite{2006ApJ...641..217S}, who match \OVIs systems at $z<0.15$
with galaxies in the field. They find the metal systems have median
distances of 350--500~kpc (for $h=0.7$) from $L_*$ galaxies, and
200--270~kpc around $0.1L_*$ galaxies. A contribution from $0.1L_*$
galaxies is required to reproduce the line-of-sight number density of
\OVIs absorbers.

While these results are consistent with the ejection of metals from
the galaxies detected, this is not the only possible picture. The
metals may have arisen at earlier times in smaller systems that
eventually merged into the galaxies seen. Alternatively, they may have
been ejected from much smaller systems, perhaps as small as globular
star clusters or very faint dwarf galaxies, that are spatially
correlated with the visible galaxies. The metals may also have been
deposited very early in the history of the Universe, perhaps by
blowout from the projenitors of present day galaxies in superwinds or
possibly driven out by the impact of AGN that may have been active in
the projenitors.

As a galactic wind expands into the IGM, it will drive a shockfront
that sweeps up the surrounding gas into a thin shell just behind the
shock. In this limit, the thin-shell approximation may be made to
describe the advance of the shockfront. Denoting the distance of the
shockfront from the center of the galaxy by $R_s$, the mass of the
expanding gas interior to $R_s$ by $M_g$, the mean interior pressure
by $P_b$ and enthalpy density by $h_b$, the external pressure and
density by $P_0$ and $\rho_0$, respectively, the velocity of the
external gas relative to the galaxy by $v_H$, the gravitational
potential energy of the wind gas by $W$, and assuming mass injection
from the center at the rate $\dot M_{\rm in}$ with velocity $v_{\rm
in}$, the equations describing the evolution of the shockfront radius
and the interior thermal energy $E_b$ of the bubble powered by energy
injected at the rate $L_{\rm in}(t)$ and cooling at the rate $L_{\rm
cool}(t)$ are, in the thin-shell approximation
\cite{1988RvMP...60....1O, 2001ApJ...555...92M},
\begin{eqnarray}
\frac{d}{dt}(M_g{\dot R_s})&=&4\pi R_s^2(1-f_{\rm cl})[(P_b - P_0)+
\rho_0v_H({\dot R_s}-v_H)]\nonumber\\
&&-M_g{\dot R_s}^2/\lambda_{\rm cl}+W/R_s+{\dot M}_{\rm in}v_{\rm in},
\label{eq:bw-mom}
\end{eqnarray}
\begin{eqnarray}
\frac{d}{dt}E_b &=& L_{\rm in}(t) - 4\pi R_s^2P_b{\dot R}_s
+4\pi R_s^2h_b({\dot R}_s-v_H)\nonumber\\
&& - P_b\frac{dV_{\rm cl}}{dt} - L_{\rm cool}(t),
\label{eq:bw-energy}
\end{eqnarray}
where isothermal jump conditions across the shock front have been
assumed for the energy equation, presuming the wind expands into a
photoionized medium. A clumpy interior is allowed for due to the
presence of clouds with a volume filling factor $f_{\rm cl}$ and
spatial density and cross section corresponding to a mean free path
$\lambda_{\rm cl}$, which enters as a drag term on the advance of the
shockfront in Eq.~(\ref{eq:bw-mom}). The interior net cooling $L_{\rm
cool}$ will be dominated by Compton cooling at high redshifts and two
body atomic processes, which in turn depend on the gas clumping
factor, and radiative heating from the galaxy or UV metagalactic
background at lower redshifts. The second and fourth terms on the RHS
of Eq.~(\ref{eq:bw-energy}) account for energy loss due to work done
by the shock on the surrounding gas and on crushing interior clouds,
respectively.

The factors in Eqs.~(\ref{eq:bw-mom})-(\ref{eq:bw-energy}) highlight
the various uncertainties in any galactic wind calculation. The rate
of mass and energy injection by the stars and supernovae depend on the
rate of star formation and the Initial Mass Function of the stars,
which may have a lower upper mass cutoff in starbursts compared with
normal galaxies \cite{1995ApJS...96....9L}, or may be dominated by
massive stars if the metallicity is low
\cite{2003Natur.425..812B}. The cooling rate will depend on the
metallicity of the bubble interior, its clumpiness, and on how well
mixed the metals are. It will also depend on the temperature of the
interior gas, which requires a specification of the amount of interior
mass even though it is negligible compared with the mass in the
shell. Most of the interior mass results from evaporative losses from
the contact surface between the hot interior and the cool shell
\cite{2001ApJ...555...92M}. The rate of evaporation depends on the
degree to which thermal heat conduction may be suppressed by interior
magnetic fields, although mechanical heat input balancing adiabatic
expansion losses dominates the interior thermodynamics, so this may
not be critical. The strength of the shock front may be substantially
dissipated as the shock overtakes small clouds and crushes
them. Lastly, the progress of the shockfront depends on the
environment of the galaxy, which will differ in different directions,
so that the winds will in general be anisotropic. The acceleration of
dense gas by hot rarefied gas and the motion of surfaces past one
another as the bubble penetrates the IGM will also generally be
subject to Rayleigh-Taylor and Kelvin-Helmholtz instabilities, giving
rise to the mixing of gas components with vastly contrasting densities
and temperatures and the formation of eddies both at the surface of
the bubble and in its interior as the shell fragments. Thermal
instabilities may also play a role. Magnetic fields will generally
also affect the fragmentation of the expanding shell, and may affect
the crushing of interior clouds. No numerical simulation is able to
contend with all these factors:\ the required spatial resolution,
range in timescales, and number of physical processes involved are far
too demanding. Instead, simulators must resort to the art of
approximation.

Several estimates of the possible impact of galactic winds on the IGM
have been made. An instantaneous stellar burst of $10^6M_\odot$,
typical of a dwarf galaxy with total halo mass $M_h\simeq10^9M_\odot$
and allowing for a conversion efficiency $f_*\simeq0.01$ of gas mass
into stars, produces a mechanical luminosity provided by supernovae
and stellar winds of $5\times10^{33}$~W \cite{Leitherer99}, assuming a
Salpeter Initial Mass Function between $1-100M_\odot$. The
contribution of supernovae dies off after about $4\times10^7$~yr, and
the mechanical luminosity plummets, having yielded a total mechanical
energy input of $2\times10^{48}$~J. If the wind is allowed to expand
until it reaches pressure equilibrium with the surrounding IGM, then a
maximum distance the wind travels may be estimated. For example, if a
fraction $\nu_m$ of the mechanical energy is available to power the
wind (taking into account losses internal to the galaxy and as the
wind expands), balancing the mean internal pressure
$P_b=(2/3)E_b/V_s=E_b/2\pi R_s^3$ (assuming $\gamma=5/3$ for the
ionized interior gas), where $V_s$ is the bubble volume, with the
pressure of the external IGM $P_{\rm IGM}\simeq
5\times10^{-20}T_4(1+z)^3\,{\rm N\,m^{-2}}$, gives a maximum comoving
bubble radius of $R_{s, c}\simeq 300(\nu_m/0.1)^{1/3} T_4^{-1/3}$~kpc.

The maximum radius is comparable to the geometric estimates discussed
above. Taking account of radiative losses, however, stalls the
expansion before it reaches the maximum value
\cite{2001ApJ...555...92M}. More massive galaxies driving stronger
winds might be appealed to, but this is problematic on two grounds:\
they are much rarer than dwarf galaxies at high redshifts and the ram
pressure of the surrounding infalling IGM can smother the expansion of
the wind \cite{2004ApJ...613..159F}.

An estimate of the maximum mass halo able to produce an escaping wind
may be made as follows. Galaxies have nearly flat rotation curves,
corresponding to envelope density profiles slightly steeper than
$\rho\sim1/r^2$ \cite{Peebles93}. The infall velocity of the gas is
then $v_H\simeq(2GM_V/r_V)^{1/2}$, where $M_V=(4\pi/3)\rho_V r_V^3$ is
the mass of the virialized halo of mean density
$\rho_V=2[3(3\pi+2)/4]^2\rho_{\rm crit}(0)(1+z_V)^3$ and radius $r_V$
(\S~\ref{subsec:minihalos}). Assuming the gaseous component follows
the profile of the envelope of dark matter, the ram pressure at the
surface of the bubble is $P_{\rm ram}=\rho
v_H^2=(3/2\pi)(\Omega_b/\Omega_m) GM_V^2/(r_V^2R_s^2)$ for $R_s\gg
r_V$. If the ram pressure dominates the IGM pressure, then it will
determine the stalling radius where it matches $P_b$, according to
Eq.~(\ref{eq:bw-mom}). The criterion $P_{\rm ram}<P_b<E_b/(2\pi r_V
R_s^2)$ (for $R_s>r_V$) imposes the upper limit on the virial mass of
the galaxies able to produce escaping winds of
\begin{equation}
M_V < (5\times10^7h^{-1}\,M_\odot)\biggl(\frac{\nu_m f_*}{0.001}\biggr)^{3/2}
\biggl(\frac{1+z_V}{10}\biggr)^{-3/2},
\label{eq:maxMh}
\end{equation}
in good agreement with hydrodynamical computations
\cite{2004ApJ...613..159F}. For a more complete treatment,
hydrodynamical simulations are required. Examples of recent wind
simulations include \textcite{2002ApJ...578L...5T},
\textcite{2005ApJ...620L..13A}, \textcite{2005ApJ...635...86C},
\textcite{2006ApJ...650..560C}, \textcite{2006MNRAS.373.1265O},
\textcite{2007MNRAS.376.1465K}, and \textcite{2007ApJ...663...38K}.

\section{Prospects for the Future}
\label{sec:future}

The future of IGM studies is entering a new phase in both observations
and computations that promises to settle a number of outstanding
questions. Prominent among these are the circumstances that led to the
reionization of the Universe. At the same time, the growing
recognition of the relationship between the IGM and galaxies is likely
to shift the subject increasingly toward a merger with the more
embracing subject of galaxy formation. Perhaps most spectacularly,
imminent and planned detectors and space missions may open up a new
era in IGM science:\ direct imaging of the structure of the IGM in the
infrared and radio.

Recent years have witnessed the detection of high redshift \Lya
emitting galaxies, long-sought as candidates for the first forming
galaxies \cite{1967ApJ...147..868P}. Measurements of the \Lya emitter
luminosity function and its evolution would in principle provide
evidence for the reionization epoch, as the \Lya\ radiative damping
edge of neutral patches would increasingly erode the lines and produce
a sharp decrease in the luminosity function. Evidence for such rapid
evolution over the redshift range $5.7<z<7$ has been obtained from the
Subaru Deep Field \cite{2006ApJ...648....7K, 2006Natur.443..186I}.
The interpretation of any decline, however, is complicated by
systematics, such as the unknown evolution in the dust content, the
structure of the local IGM that may produce a \Lya\ radiative damping
wing, and the mass of the systems producing star formation, as well as
the possible infall of the IGM towards the source and the impact of
winds produced by the source on the surrounding IGM, all of which
could affect the intensity and shape of the \Lya emission line and so
the evolution of the \Lya luminosity function
\cite{2004MNRAS.349.1137S, 2005ApJ...623..627H, 2007A&A...474..385N,
2006MNRAS.370..273D}. \textcite{2007MNRAS.381...75M} suggest these
uncertainties may be circumvented by using the angular correlation
function of the \Lya emitters. The angular correlations will be
strongly modulated by \HII\ regions during the EoR on large scales
where the dark matter fluctuations are weak, providing a distinctive
signature of reionization.
Much improved \Lya emitter statistics are anticipated with current or
upcoming narrow band surveys which may make it possible to realize
these tests \cite{2004SPIE.5492.1022H, 2006SSRv..123..485G,
2008ApJS..176..301O, 2007ApJ...668..627S}.

Anticipated CMB polarization measurements are expected to further
narrow the redshift range during which reionization
occurred. Polarization anisotropy measurements on arcminute scales by
the {\it Planck} satellite,\footnote{http://www.rssd.esa.int/Planck}
to be launched in 2008, may reveal the expected patchiness of
reionization, which may be used to discriminate between competing
reionization scenarios \cite{2003ApJ...598..756S, 2005ApJ...630..657Z,
2005ApJ...630..643M}. The Atacama Cosmology Telescope
(ACT)\footnote{http://www.physics.princeton.edu/act/} and the South
Pole Telescope (SPT)\footnote{http://spt.uchicago.edu/} have already
embarked on such measurements.

The relationship between the IGM and galaxies will be further
clarified by upcoming galaxy surveys, especially at low redshifts. UV
spectroscopy with the Cosmic Origins Spectrograph
(COS),\footnote{http://cos.colorado.edu} to be installed on {\it HST}
in 2008, will refine the measurements of low to moderate redshift
metal absorption systems, and increase sample sizes by an order of
magnitude, greatly improving the statistics on the correlations with
galaxies. Currently detections of the WHIM have been confined almost
entirely to \OVIs measurements. Detections of further high ionization
species would be possible using COS as well as the spectroscopic
capabilities of the proposed x-ray satellite {\it Constellation-X} or
a possible smaller scale satellite with high resolution x-ray
spectroscopic capability \cite{2006hrxs.confE..26N}, providing
important diagnostics of the physical properties of the WHIM. LOFAR is
expected to detect the synchrotron emission from $10^7$ starburst
galaxies, which with follow-up optical, UV and sub-millimeter
observations, will provide a wealth of data that may establish the
impact of winds on the IGM over a wide range of redshifts, including
the spreading of metals. Even greater numbers may be detected by a
Square Kilometer Array, with an expected detection rate of several
thousand at $z>5$ per square degree surveyed. Optical spectroscopy
using a Giant Segmented Mirror Telescope
(GSMT),\footnote{http://www.gsmt.noao.edu} such as the Thirty Meter
Telescope (TMT)\footnote{http://www.tmt.sorg} and the Giant Magellan
Telescope (GMT),\footnote{http://www.gmto.org} and an Extremely Large
Telescope (ELT)\footnote{http://www.eso.org/projects/e-elt/} could
measure the temperature of the moderate and even low density
components of the IGM by measuring the velocity widths of multiple
metal line absorption systems. Deep photometric surveys with such very
large telescopes could detect very low luminosity galaxies that may be
associated with low redshift metal absorption systems. The Ultra Deep
Survey of the {\it James Webb Space Telescope},
\footnote{http://www.jwst.nasa.gov/} to be launched in 2013, will
detect dwarf galaxies out to redshifts of $6<z<10$, and so may
positively identify the population of galaxies which reionized the IGM
\cite{2006SSRv..123..485G, 2006NewAR..50..113W}. These multiband
measurements, extending from the radio through x-rays, will help to
establish the metal pollution patterns throughout the IGM and their
relationship with star-forming galaxies.

These advances will enable simulations to provide a complete
description of the evolution of the IGM. Such a prospect would open up
new opportunities for other areas of astrophysics and cosmology to
exploit the IGM as a tool for testing models of interest. For example,
the abundances and distribution of metals as well as the line widths
of metal and hydrogen absorption features around galaxies could place
stringent constraints on galaxy formation models and the winds young
galaxies may produce, including the possible impact on the host galaxy
and its surroundings of an embedded AGN \cite{2007MNRAS.380..877S}.
The imaging of the IGM around QSO sources could directly reveal the
opening angles of their emission cones and relative orientation to the
observer, placing strong constraints on unification models of AGN.

Precise predictions for the absorption statistics of the IGM may
provide valuable constraints on modifications of standard cosmological
models such as those allowing for a ``dark energy'' equation of
state. Dark energy will affect the growth rate of structures in the
IGM, possibly at a detectable level in future surveys
\cite{2003MNRAS.340L..47V}. The density fluctuations in the IGM along
the line of sight and transvere directions will differ as a
consequence of the difference between radial and angular distances at
high redshifts \cite{1979Natur.281..358A}. With sufficient statistics,
the difference may be detectable in the correlation function or power
spectrum measurements of the transmitted \Lya\ flux through the IGM
\cite{1999ApJ...511L...5H, 1999ApJ...518...24M, 2008ApJS..175...29M,
2008ApJ...675..946M}. Even more ambitious would be the direct
detection of the expansion of the Universe through the monitoring of
absorption lines along fixed lines of sight over a period of several
years through spectroscopy on extremely large telescopes
\cite{2007PhRvD..75f2001C, 2008MNRAS.386.1192L}. Since realizing such
a measurement would require the development of an extraordinarily
stable detection system and adequate taming of systematics associated
with variations in the QSO brightness and spectral shape and their
effects on the noise statistics, such a project may spur efforts to
investigate the technical characteristics of large telescopes and the
behavior of QSOs at unprecedented levels of accuracy.

The next few years may produce the first images of the IGM. Ionizing
radiation reprocessed by the neutral gas in Lyman Limit Systems into
\Lya photons may soon be detectable by upcoming \Lya narrow-band
detectors \cite{1987MNRAS.225P...1H, 1996ApJ...468..462G,
2005ApJ...628...61C, 2007ApJ...657..135C}. It my even be possible to
detect the \Lya photons emitted by a central source like a QSO and
scattered by the surrounding neutral IGM prior to reionization
\cite{1999ApJ...524..527L} or \Lya photons produced by collisional
excitation within the rapidly expanding ionization fronts during the
EoR \cite{2008ApJ...672...48C}, providing \Lya tomographic imaging of
the neutral IGM. Such detections would complement the planned IGM
imaging by the LOFAR, MWA and SKA radio facilities using 21cm
tomography. Cross-correlations between the \Lya and radio images may
help to establish statistically significant detections.

At even higher redshifts, 21cm tomography prior to reionization may
become possible through the production of a 21cm signature via \Lya
and collisional decoupling of the spin temperature from the CMB. At
$z>30$, collisional decoupling alone is adequate, offering the
opportunity to explore the redshift desert $30<z<200$ and examine the
primordial density fluctuation power spectrum on scales smaller than
may be reached by CMB measurements. Such measurements could probe the
structure of an inflationary potential and induced non-gaussian
fluctuations \cite{2004PhRvL..92u1301L, 2004MNRAS.352..142B,
2005ApJ...624L..65B, 2006ApJ...653..815M,
2007ApJ...662....1P}. Astrophysical and instrumental systematics and
sources of radio frequency interference (RFI), both man made and
natural, such as arises from the ionosphere, likely preclude such
measurements using currently planned radio facilities
\cite{2007ApJ...661....1B}. These limitations, however, may be
circumvented by placing a radio telescope in space. Even more
adventurous would be to build a much larger facility on the far side
of the Moon, where it would be shielded from terrestrial RFI
\cite{2007astro.ph..2070C}. The exploration of long wavelengths
unavailable to terrestrial radio astronomy would open up a new
panorama on the Universe, introducing a new era of ``Astronomy on the
Far Side.'' While not all of these prospects may materialize, others
not currently envisaged will surely arise,
ensuring that IGM science will continue to flourish for many years to
come.

\section*{Acknowledgments}

This review benefitted from discussions with and comments by
J. Bland-Hawthorne, J. Bolton, A. Ferrara, J. Higgins, P. Madau,
J. Peterson, M. Rauch, S. Reynolds, J. Schaye, E. Tittley, D. Turnshek
and M. White.

\bibliography{meiksin}

\end{document}